\documentclass{JHEP3}
\usepackage{axodraw}
\usepackage{cite}
\usepackage{epsfig}
\usepackage{xspace}
\usepackage{graphics}
\usepackage[utf8]{inputenc}
 \usepackage[]{youngtab}
\usepackage{ifthen}

\def\showcommentsflag{0}
\newcommand{\showcomments}{\def\showcommentsflag{1}}

\newcounter{commentcounter}%
\newcommand{\comment}[1]{\ifnum\showcommentsflag > 0%
\addtocounter{commentcounter}{1}%
{{\Red{\ensuremath{\ddagger^{\arabic{commentcounter}}}}}}%
\marginpar{\raggedright\tiny\it{{\Red{\ensuremath{\ddagger^{\arabic{commentcounter}}}}} {#1}}}
\fi%
}
\newcommand{\commentdel}[2]{\ifnum\showcommentsflag > 0%
\Red{\sout{#1}}\comment{#2}%
\fi
}
\newcommand{\commentadd}[2]{\ifnum\showcommentsflag > 0%
\comment{#2}\Red{#1}%
\else
#1
\fi
}
\newcommand{\commentchange}[3]{\ifnum\showcommentsflag > 0%
\Red{\sout{#2}}\comment{#3}\Red{#1}%
\else
#1
\fi
}
\newcommand{\nocomment}[1]{\ifnum\showcommentsflag > 0%
{\tiny\it\Red{\{#1}\}}
\fi%
}
\newcommand{\nocommentdel}[1]{\ifnum\showcommentsflag > 0%
\Red{\sout{#1}}%
\fi
}
\newcommand{\nocommentadd}[1]{\ifnum\showcommentsflag > 0%
\Red{#1}%
\else
#1
\fi
}
\newcommand{\nocommentchange}[2]{\ifnum\showcommentsflag > 0%
\Red{\sout{#2}}\Red{#1}%
\else
#1
\fi
}

\newcommand{\pythia}{P\protect\scalebox{0.8}{YTHIA}\xspace}
\newcommand{\dipsy}{\protect\scalebox{0.8}{DIPSY}\xspace}
\newcommand{\herwig}{\protect\scalebox{0.8}{HERWIG}\xspace}
\newcommand{\pytppp}{P\protect\scalebox{0.8}{YTHIA}8\xspace}
\newcommand{\tee}{\ensuremath{e^+e^-}\xspace}
\providecommand{\eqref}[1]{eq.~(\ref{#1})\xspace}
\newcommand{\eq}[1]{(\ref{#1})\xspace}
\renewcommand{\eqref}[1]{eq.~(\ref{#1})\xspace}
\newcommand{\Eqref}[1]{Eq.~(\ref{#1})\xspace}
\newcommand{\eqsref}[1]{eqs.~(\ref{#1})\xspace}

\newcommand{\fig}[1]{\ref{#1}}
\newcommand{\figref}[1]{figure~\fig{#1}}
\newcommand{\figrefs}[1]{figures~\fig{#1}}
\newcommand{\Figref}[1]{Figure~\fig{#1}}

\newcommand{\sect}[1]{\ref{#1}}
\newcommand{\sectref}[1]{section~\sect{#1}}
\newcommand{\sectrefs}[1]{sections~\sect{#1}}

\newcommand{\appref}[1]{appendix~\sect{#1}}
\newcommand{\apprefs}[1]{appendices~\sect{#1}}

\newcommand{\ariadne}{A\protect\scalebox{0.8}{RIADNE}\xspace}
\newcommand{\thepeg}{T\protect\scalebox{0.8}{HE}PEG\xspace}

\def\done#1{}

\newcommand{\gtaeq}{\raisebox{-0.8mm}%
{\hspace{1mm}$\stackrel{>}{\sim}$\hspace{1mm}}}

\newcommand{\shiftleft}[2]{\makebox[0pt][r]{\makebox[#1][l]{#2}}}
\newcommand{\shiftright}[2]{\makebox[#1][r]{\makebox[0pt][l]{#2}}}
\def\as{\ensuremath{\alpha_S}\xspace}
\def\pmb#1{{\mbox{\boldmath$#1$}}}
\def\eg{\emph{e.g.}}
\def\ie{\emph{i.e.}}
\def\cf{\emph{c.f.}}
\def\etc{\emph{etc.}}
\def\plet#1{\ensuremath{\mathbf{#1}}}
\def\antiplet#1{\ensuremath{\mathbf{\overline{#1}}}}
\def\kapeff{\ensuremath{\tilde{\kappa}}}
\def\kaprope{\ensuremath{\kappa_{\mathrm{rope}}}}

\keywords{Hadronization, QCD, Jets, Parton Model, Phenomenological Models, Colour Reconnection}
\showcomments
\preprint{arXiv:1412.6259 [hep-ph]\\
MCnet-14-27\\
LU-TP 14-44\\
JLAB-THY-15-1991}

\skip\footins = 1\bigskipamount plus 2pt minus 4pt                              

\title{Effects of Overlapping Strings in \boldmath$pp$
  Collisions
  \footnote{
    Work supported in part by the MCnetITN FP7 Marie Curie Initial
    Training Network, contract PITN-GA-2012-315877, the Swedish Research
    Council (contracts 621-2012-2283 and 621-2013-4287), and contract
    DE-AC05-06OR23177 under which the Jefferson Science Associates, LLC
    operate the Thomas Jefferson National Accelerator Facility.}}
\author{Christian Bierlich$^1$, Gösta Gustafson$^1$, Leif Lönnblad$^{1}$ and
  Andrey Tarasov$^{2}$\\
  $^1$Dept.~of Astronomy and Theoretical Physics, Lund University, Sweden\\
  $^2$Theory Group, Jefferson Lab, Newport News, Virginia, USA}
  
\abstract{In models for hadron collisions based on string
  hadronization, the strings are usually treated as independent,
  allowing no interaction between the confined colour fields. In
  studies of nucleus collisions it has been suggested that strings
  close in space can fuse to form ``colour ropes''. Such ropes are
  expected to give more strange particles and baryons, which also has
  been suggested as a signal for plasma formation. Overlapping strings
  can also be expected in $pp$ collisions, where usually no phase
  transition is expected.  In particular at the high LHC energies the
  expected density of strings is quite high. To investigate possible
  effects of rope formation, we present a model in which strings are
  allowed to combine into higher multiplets, giving rise to increased
  production of baryons and strangeness, or recombine into singlet
  structures and vanish.  Also a crude model for strings recombining
  into junction structures is considered, again giving rise to
  increased baryon production. The models are implemented in the
  \dipsy MC event generator, using \pytppp for hadronization, and
  comparison to $pp$ minimum bias data, reveals improvement in the
  description of identified particle spectra.}

\begin{document}
 
\sloppy
 
\section{Introduction}
\label{sec:intro}

In most models for high energy collisions, like the popular \pythia
\cite{Sjostrand:2006za} or \herwig \cite{Bahr:2008pv} models, the
hadronization mechanism is described via strings or cluster
chains. The strings are often treated as independent, but in
connection with nucleus collisions it was early suggested that the
many strings produced within a limited space may interact and form
"colour ropes" \cite{Biro:1984,Bialas:1984ye}. Such ropes have
subsequently been studied by many authors with applications to high
energy nucleus collisions
\cite{Kerman:1985tj,Gyulassy:1986jq,Andersson:1991er,Braun:1991dg,Braun:1993xw,Amelin:1994mc,Armesto:1994yg}.
The stronger field in a rope is expected to give larger rates for
strangeness and baryons. The effect on the multiplicity is more
difficult to predict, and here the results depend on simplified
assumptions, often without a real motivation. Usually the result is
either a decreased or an unmodified particle multiplicity.

As rope formation is expected to give increased rates of strange
particles and baryons, which may mimic effects of plasma formation, it
makes signals for a phase transition more difficult to interpret.  It
has also been suggested that ropes may initiate the formation of a
quark--gluon plasma
\cite{Kajantie:1985jh,Gyulassy:1986jq,Gatoff:1987uf,Braun:1997ch}. At
LHC energies many overlapping strings are also expected in $pp$
scattering, where plasma formation normally is not expected. In this
paper we want to study string interference effects in $pp$ scattering,
with the aim to get a better understanding of the dynamics in $pp$
collisions, and simultaneously get a tool to estimate possible rope
effects in nucleus collisions.
  
For a quantitative estimate of interaction between neighbouring
strings, we believe it is essential to have a description formulated
in transverse coordinate space. Such a formulation is also suitable
for including effects of saturation for small $x$ and high gluon
densities. Here we will use the Lund dipole cascade model implemented
in the event generator \dipsy
\cite{Avsar:2005iz,Flensburg:2011kk}. This model is based on a
formulation of BFKL dynamics in transverse coordinate space, including
non-leading-log and saturation effects, and also taking fluctuations
and correlations into account.\footnote{String interaction effects in
  $pp$ collisions have also earlier been included in the event
  generator DTUJET \cite{Merino:1991nq}, formulated in momentum
  space. This was generalized to nucleus collisions, including a
  geometric distribution of nucleons within a nucleus
  \cite{Mohring:1992wm}.  Rope effects are also included, together
  with hadron rescattering, in the RQMD model, with applications in
  the SPS fixed target and RHIC energy ranges
  \cite{Sorge:1992ej,Bleicher:2000us,Soff:2002bn}.}

The Lund string hadronization model \cite{Andersson:1983jt,
  Andersson:1979ij}\footnote{For a review of the Lund hadronization
  model see ref.~\cite{Andersson:1983ia}, or a more recent summary in
  ref.~\cite{Buckley:2011ms}.}, which has been particularly successful
in describing data from $e^+e^-$ annihilation at LEP
\cite{Hamacher:1995df,Sjostrand:2014zea}, is based on the assumption
that a confined colour field between a quark and an antiquark is
compressed to a linear flux tube, similar to a vortex line in a
superconductor. When the string is stretched between separated colour
charges, it can break by the production of $q\bar{q}$ pairs
\cite{Casher:1978wy,Andersson:1980vj,Gurvich:1979nq,
  Glendenning:1983qq} in a process similar to the production of
$e^+e^-$ pairs in a homogeneous electric field
\cite{Schwinger:1951nm}. As demonstrated in ref.~\cite{Brezin:1970xf},
this can be interpreted as the effect of a quantum-mechanical
tunneling process. In the Lund model the dynamical motion of the flux
tube is approximated by an infinitely thin \textit{massless
  relativistic string}, and gluons are identified with transverse
excitations on such a string \cite{Andersson:1979ij}, which also has
the effect that the model is infrared stable, \ie\ insensitive to soft
or collinear gluons.

It was early observed that when the string hadronization model is
tuned to LEP data, it slightly underestimates the production of
strange particles in hadronic collisions
\cite{Alner:1985ra}. Similarly in DIS, the LEP tune works well in the
current fragmentation region (in the Breit frame), while in the proton
fragmentation end, one again observes an enhanced strange quark
fraction \cite{Aaron:2008ck}. This effect is enhanced in data from
LHC, where a rather dramatic increase is observed in the fractions of strange
particles and baryons, most notably that of strange baryons
\cite{Khachatryan:2011tm}.
These observations should not be surprising. The colour flux tubes are
expected to have a transverse width determined by the confinement
scale, of $\sim 1$~fm. In $pp$ collisions there can be several strings
close to each other, and it should actually be rather surprising that
models neglecting mutual string interaction are working as well as
they do.

Biro \emph{et al.} noted in ref.~\cite{Biro:1984}, that the colour
charge at the endpoint of a rope formed by strings with random colour
charges, is given by a random walk in colour space. The rope can break
up in a stepwise manner by repeated production of $q\bar{q}$ pairs, as
expected from a local interaction $\propto j_\mu A_\mu \sim
\bar{\psi}\gamma_\mu \psi A_\mu$. This process is analogous to the
production of $e^+e^-$ pairs in an electric field. It was pointed out
in ref.~\cite{Biro:1984} that for a rope formed by random charges, the
number of pairs produced before a total breakup of the rope is in
general smaller than the number of initial strings, and also that the
total time for this successive split is approximately the same as for
a single string.

Although such a stepwise breakup of a rope is assumed in most studies,
also an immediate breakup by production of multi-quark--antiquark
systems has been advocated by Amelin, Braun, and Pajares
\cite{Amelin:1994mc} (also mentioned as a possibility in
ref.~\cite{Biro:1984}), and a breakup by the production of gluon pairs
has been studied by Gyulassy and Iwazaki~\cite{Gyulassy:1986jq}.  It
has also been suggested that a rapid decay of the ropes into
elementary partons is facilitating the formation of a quark--gluon
plasma \cite{Gyulassy:1986jq,Gatoff:1987uf,Braun:1997ch}.

It was early suggested that, if the charges correspond to a specific
SU(3) multiplet, the tension (or energy density) in the rope is given
by the second Casimir operator \cite{Ambjorn:1984dp}. For an isolated
rope this conjecture has later been supported by lattice calculations
\cite{Bali:2000un}. However, if the rope is surrounded by other
strings or ropes, we expect that the transverse area, and thus the
rope tension, will be affected by the presence of the neighbouring
ropes, which exert a pressure keeping the radius small. Such a
pressure might also cause a collective expansion contributing to extra
transverse momentum for the hadrons. It ought to be kept in mind, that
this feature contributes to the necessary uncertainties in estimating
the effects of rope formation.

As mentioned above we will here use the event generator \dipsy to
study the effects of string interaction and rope formation in more
detail, beyond qualitative results such as increased strangeness and
baryon production.  The \dipsy model is a generalization and extension
of Mueller's dipole model
\cite{Mueller:1993rr,Mueller:1994jq,Mueller:1994gb}, which describes
BFKL evolution in transverse coordinate space.  At high energies the
high density of soft gluons effectively screens colour charges, thus
suppressing gluons with $p_\perp$ below a saturation scale $Q_s$.  As
in the Color Glass Condensate formalism for nucleus collisions
\cite{McLerran:1993ni, McLerran:1993ka}, we argue that this is a
motivation for a perturbative treatment of the initial phase in terms
of quarks and gluons. (This is also assumed in other models for soft
interactions, like \pythia and \herwig.)

While Mueller's model reproduces leading log BFKL evolution, the
\dipsy model includes essential non-leading corrections to BFKL, as
well as saturation within the cascade evolution, and confinement. It
reproduces total, elastic, and diffractive cross sections in $pp$
collisions and DIS, and gives a good description of particle
distributions in minimum-bias final
states\cite{Flensburg:2011kk}. However, as parton distributions are
generated within the model, and therefore not tuned to data, and are
in addition limited to gluons, the model is naturally less accurate
than \eg\ the \pythia and \herwig models. Our aim has instead been
focused on understanding the dynamics of small-$x$ evolution and
saturation, including correlations and fluctuations, \eg\ in
connection with multiple parton interactions \cite{Flensburg:2011kj}
and diffraction \cite{Flensburg:2010kq,Flensburg:2012zy}. The
formulation in transverse coordinate space makes the \dipsy model
particularly suited for studies of string interference and rope
formation. Although it has also been applied to collisions with nuclei
(see \eg\ refs.~\cite{Flensburg:2011wx, Flensburg:2012zz}), we will in
this paper limit our study to proton--proton collisions.

We will here assume that colour ropes can form by coherent interaction
between a group of strings confined within a limited transverse
size. As in ref.~\cite{Biro:1984} we assume random colour charges for
the individual strings, leading to a random walk in colour space.  We
also assume that the rope breaks by successive production of new
$q\bar{q}$ pairs. The nature of the tunneling process implies here
that an ``effective string tension'' is determined by the
\emph{reduction} in rope tension in each individual breakup. As
mentioned in ref.~\cite{Biro:1984}, and discussed in detail below, the
result of the rope formation is then a smaller number of $q\bar{q}$
pairs needed to break the rope, but a larger effective string tension.
As is generally expected, this implies larger strangeness and baryon
fractions. In addition our model also gives nontrivial effects on the
$p_\perp$-dependence for different particle ratios, which to some
extent mimic effects of transverse flow. In this paper we present
results where the model (with some approximations) is compared to LHC
$pp$ scattering data, with encouraging qualitative agreement. In the
future we plan to also study effects of rope formation in collisions
with nuclei.  We begin this article with a recapitulation of the
relevant ingredients of the Lund string fragmentation in
\sectref{sec:stringfragmentation}, before we describe the basic idea
of the rope model in \sectref{sec:ropes}. Then we describe the
proposed ``final-state swing mechanism'' together with the
implementation of our rope model in the interface between \dipsy and
\pytppp in \sectref{sec:the-dipsy}. In \sectref{sec:results} we
present some first results, and finally in \sectref{sec:outlook} we
summarize our findings and give an outlook.

For completeness, we end the article with a number of
appendices. There we first summarize details in the \dipsy model
(\appref{app:dipsy}) that are important for our rope implementation,
and also summarize the relevant colour algebra needed
(\appref{sec:color-algebra}). Most of this can be found in various
different references, but we think it is valuable to have the relevant
issues collected here. Furthermore, we have collected some of the more
technical points in our implementation and in the tuning procedure in
\apprefs{sec:deta-descr-rope} and \sect{sec:tuning}
respectively. Although these are important for our results, they tend
to hamper the readability of the main text, and are therefore
presented separately.

\section{String fragmentation}
\label{sec:stringfragmentation}

In this section we will discuss the fragmentation of a single string.
We first discuss the basic tunneling process, then the space-time
picture describing how the produced quarks and antiquarks combine to
hadrons, followed by a discussion of baryon production and spin
effects. We end by a discussion of the effects of a modified effective
string tension.

\subsection{Tunneling}
\label{sec:lundstring}
A linear colour electric field stretched between a quark and an
antiquark, moving in opposite directions, can break up by the
production of new $q\bar{q}$ pairs, in a way similar to the production
of $e^+e^-$ pairs in a homogenous electric field. As discussed by
Schwinger \cite{Schwinger:1951nm} the electric field is unstable, and
the decay can be interpreted as the result of the production of new
$e^+e^-$ pairs with a rate per unit time and unit volume given by:
\begin{equation}
\mathcal{P} \propto (e\,\mathcal{E})^2 \exp\left(-\,
  \frac{\pi\,\mu^2}{e\,\mathcal{E}}\right). 
\label{eq:schwinger}
\end{equation}
Here $\mu$ is the electron mass and $\mathcal{E}$ the electric field
strength.  Thus $e\mathcal{E}$ is the force acting on the produced
electron or positron.  As pointed out in ref.~\cite{Brezin:1970xf},
this result can be interpreted as the effect of a tunneling
process. Classically the electron and the positron cannot be produced
in a point, but only separated by a distance $2\mu/e\mathcal{E}$, such
that the reduction in the electric field energy can be transferred
into the mass of the pair. In quantum theory the particles are
produced locally by an interaction Lagrangian $\sim e \bar{\psi}(x)
\gamma_\mu \psi(x) A_\mu(x)$, and have to tunnel through the
classically forbidden region, where the wavefunctions can be estimated
by the WKB method.

When generalizing this result to $q\bar{q}$ pair production
\cite{Casher:1978wy,Andersson:1980vj,Gurvich:1979nq,
  Glendenning:1983qq} in a linear confined colour field, the tunneling
mechanism implies that $e\mathcal{E}$ has to be replaced by the force
acting on a quark, \ie\ by the string tension\footnote{In
  ref.~\cite{Casher:1978wy} $e\mathcal{E}$ was (for a quark with
  charge $g/2$) replaced by $g\mathcal{E}/2=2\kappa$, where the string
  tension $\kappa$ was estimated from the energy in the
  colour-electric field. However, ref.~\cite{Glendenning:1983qq} noted
  that, although the force on the electron is given by $e\mathcal{E}$
  in a classical (macroscopic) electric field, in case the flux
  corresponds to only a single charge quantum, an extra contribution
  comes from the decreased field between the created quark and
  antiquark, giving $e\mathcal{E}\mapsto g\mathcal{E}/4$, just
  corresponding to the energy in the colour-electric field. In
  ref.~\cite{Glendenning:1983qq} it was also noted that if the flux
  tube is embedded in a vacuum condensate, a further contribution to
  the string tension is given by the response from the condensate. In
  the bag model (similar to a type I superconductor) this contribution
  equals the energy in the colour-electric field.}  $\kappa\sim
1$~GeV/fm.  For the production of a pair with opposite transverse
momenta $p_\perp$ in this process, the mass $\mu$ in
\eqref{eq:schwinger} will be replaced by the transverse mass
$\sqrt{\mu^2+p_\perp^2}$. This gives
\begin{equation}
\frac{d\mathcal{P}}{d^2 p_\perp}\propto \kappa \exp\left(-\,
  \frac{\pi\,}{\kappa}(\mu^2+p_\perp^2)\right), 
\label{eq:schwinger2}
\end{equation}
which integrated over $p_\perp$ gives the result in
\eqref{eq:schwinger}, with the replacement $e\mathcal{E} \rightarrow \kappa$.

The result in \eqref{eq:schwinger2} can be used to estimate the
relative production of quarks with different flavour, and the
distribution in $p_\perp$.  As it is not possible to theoretically
determine the effective quark masses to be used in
\eqref{eq:schwinger2}, it is in practice necessary to tune the $s/u$
ratio to experimental data. Fits to LEP data give $s/u \approx 0.2$
\cite{Hamacher:1995df,Sjostrand:2014zea}, which is not inconsistent
with \eqref{eq:schwinger2} for reasonable quark masses.  This
mechanism also implies that charm and heavier quarks cannot be
produced in the soft hadronization process; they can only be produced
in an initial perturbative phase.  For the transverse momenta it gives
a Gaussian $p_\perp$-distribution with $\sqrt{\langle
  p_\perp^2\rangle}\approx 0.25\,\,\mathrm{GeV}$.  Phenomenological
fits to data will, however, also include effects from soft gluons
below a necessary $p_\perp$ cut in the perturbative parton
shower. Tunes to data therefore give a somewhat wider distribution,
with a width $\sigma_{p_\perp}\approx 0.32$~GeV.

\subsection{Space-time picture and longitudinal momentum distribution}
\label{sec:spacetime}
The strange quark fraction and the $p_\perp$-distribution are governed
by the tunneling mechanism, but for the longitudinal momentum
distribution it is essential to take into account how the produced
quarks and antiquarks can fit into a mesonic wavefunction, and combine
to form final-state hadrons. The Lund fragmentation model is here
inspired by the area law for Wilson loop integrals for a confining
theory \cite{Wilson:1974sk}, in analogy with the Nambu--Goto action
for a massless relativistic string. The boost invariance of the
relativistic string (and of a linear homogenous electric field) has to
result in a boost-invariant distribution of hadrons, produced around a
hyperbola in space-time.

Consider, for simplicity, a model in one space dimension with only one
quark flavour and a single mesonic state with mass $m$. In the Lund
hadronization model the probability, $\mathcal{P}$, for the production
of a specific state with $n$ mesons with momenta $p_i$ ($i=1,...,n$)
is given by the relation \cite{Andersson:1983jt}:
\begin{equation}
\mathcal{P} \propto \left\{\left[\prod_1^n Nd^2 p_i
    \delta(p_i^2-m^2)\right]\delta^{(2)}(\sum p_i-P_{tot})\right\} \exp(-bA).
\label{eq:Lund}
\end{equation}
Here the term in curly parenthesis is a phase space factor, where the
dimensionless constant $N$ determines the relative weight between
states with different number of mesons. The term $bA$ in the exponent
corresponds to the imaginary part of the action for the massless
string, which is responsible for the decay and finite lifetime of the
string. $A$ is a measure of the space-time area covered by the string
before the breakup, and $b$ is a constant. Conventionally the area $A$
is scaled by the square of the string tension $\kappa$:
\begin{equation}
A\equiv\mathcal{A}\, \kappa^2,
\label{eq:stringarea}
\end{equation}
where $\mathcal{A}$ is the area in space and time. Consequently the
dimension of $b$ is $(energy)^{-2}$. 

The result in \eqref{eq:Lund}
can be generated in a Monte Carlo simulation by producing the mesons
in an iterative way starting from one of the string ends, where each
meson takes a fraction $z$ of the remaining energy. In each step the
relevant $z$-value is given by the probability distribution or
splitting function:
\begin{equation}
f(z)=N\frac{(1-z)^a}{z}e^{-b m^2/z}.
\label{eq:split}
\end{equation}
Here the constant $a$ is related to $N$ and $b$ through the
normalization constraint $\int f(z)\ dz = 1$. 
The production points for the pairs will be located around a hyperbola in
space-time, with a typical proper time determined by 
\begin{equation}
\langle \tau^2\rangle = \frac{1+a}{b\,\kappa^2}.
\label{eq:averagetau}
\end{equation}
This timescale is also related to the particle multiplicity via the relation
\begin{equation}
d\,N/d\,y \sim \sqrt{\langle \tau^2\rangle}\,\kappa/m = \sqrt{\frac{1+a}{b\,m^2}}.
\label{eq:particledensity}
\end{equation}
We note that absorbing the string
tension in the definition of $b$, via the scaling in
\eqref{eq:stringarea}, implies that $\kappa$ does not appear
explicitly in this expression for the splitting function or the particle
density.  

In three dimensions the hadron mass $m$ in \eqsref{eq:Lund},
\eq{eq:split}, and \eq{eq:particledensity} has to be replaced by the
transverse mass 
$m_\perp=\sqrt{m^2+p_\perp^2}$, in accordance with
\eqref{eq:schwinger2}.  The parameters $a$ and $b$, determined by the
hadronic phase space and (the imaginary part of) the string action,
have been tuned to LEP data, which in our case gives $a=0.42$ and
$b=0.4$.

\subsection{Baryon production}
\label{sec:baryonprod}

Besides including different quark species, the relations in
\eqsref{eq:Lund} and \eq{eq:split} must also be generalized to include
baryon production and effects of spin interaction.  A quark and an
antiquark can combine to a total spin 1 or 0. Fits to data favour a
$\pi:\rho$ ratio about $1:1$ (rather than the $1:3$ expected from
naive spin counting), which also can be understood as a result of
normalization of the wavefunction in the tunneling process
\cite{Andersson:1982if,Andersson:1983ia}.

A baryon--antibaryon pair can be formed if the string can break by the
production of a diquark--antidiquark pair forming an antitriplet and a
triplet respectively \cite{Andersson:1981ce}. Such a process would be
suppressed by a larger effective diquark mass. In this case the
$\bar{B}B$ pair will always have two quark flavours in common, in
conflict with experimental data from $e^+e^-$ annihilation.  A
modified model with a stepwise production mechanism (called the
popcorn model) was presented in \cite{Andersson:1984af}, and has since
been incorporated in \pythia.\footnote{A stepwise production mechanism
  was suggested by Casher \emph{et al.} in ref.~\cite{Casher:1978wy}.}
In a red--antired ($r\bar{r}$) string-field a $g\bar{g}$ quark pair
can be produced as a vacuum fluctuation (see \figrefs{fig:popcorn}a
and \fig{fig:popcorn}b). If the $r$ and $g$ charges form a $\bar{b}$
antitriplet, a $\bar{b}b$ field can be formed between the new quarks,
which means that the net force on the green quark or antiquark is
zero. During this fluctuation a $b\bar{b}$ pair produced in the string
can split the system by an effective diquark--anti-diquark production
(see \figref{fig:popcorn}c).  In this way one (or more) mesons can be
produced between the baryon and antibaryon.

As hadronization is a non-perturbative process, estimating all
possible hadron species in an MC implementation necessarily implies a
set of additional phenomenological parameters, which all have to be
tuned to experimental data. Although \pythia has adopted the popcorn
model, it is reformulated in terms of diquark breakups with an
additional probability for having mesons produced in between baryon
pairs as in \figref{fig:popcorn}d.  Most important for the result is
therefore the diquark/quark ratio in the splitting process, together
with the extra suppression of strange diquarks beyond the $s/u$
suppression. A detailed description of the parameters involved in
baryon production within the Lund model is presented in
\appref{sec:deta-descr-rope}.

We here also want to point out that, besides the different parameters,
it is also very important to take into account that a produced baryon
must be symmetric in flavour and spin, in order to preserve SU(3)
flavour symmetry (as is done in the \pythia generator).

\FIGURE[ht]{
  \begin{picture}(300,100)(0,0)
  \ArrowLine(25,75)(125,75)
  \Text(75,65)[]{a}
  \Vertex(25,75){2}
  \Vertex(125,75){2}
  \Text(25,85)[]{$r$}
  \Text(125,85)[]{$\bar{r}$}
  
  \ArrowLine(175,75)(208,75)
  \ArrowLine(241,75)(208,75)
  \ArrowLine(241,75)(275,75)
  \Text(225,65)[]{b}
  \Vertex(175,75){2}
  \Vertex(208,75){2}
  \Vertex(241,75){2}
  \Vertex(275,75){2}
  \Text(175,85)[]{$r$}
  \Text(208,85)[]{$g$}
  \Text(241,85)[]{$\bar{g}$}
  \Text(275,85)[]{$\bar{r}$}
  
  \Text(75,15)[]{c}
  \ArrowLine(25,25)(50,25)
  \Line(50,25)(65,25)
  \Line(85,25)(100,25)
  \ArrowLine(100,25)(125,25)
  \Vertex(25,25){2}
  \Vertex(50,25){2}
  \Vertex(65,25){2}
  \Vertex(85,25){2}
  \Vertex(100,25){2}
  \Vertex(125,25){2}
  \Text(25,35)[]{$r$}
  \Text(50,35)[]{$g$}
  \Text(65,35)[]{$b$}
  \Text(85,35)[]{$\bar{b}$}
  \Text(100,35)[]{$\bar{g}$}
  \Text(125,35)[]{$\bar{r}$} 
  
  \Text(225,15)[]{d}
  \ArrowLine(175,25)(200,25)
  \Line(200,25)(210,25)
  \Line(220,25)(230,25)
  \Line(240,25)(250,25)
  \ArrowLine(250,25)(275,25)
  \Vertex(175,25){2}
  \Vertex(200,25){2}
  \Vertex(210,25){2}
  \Vertex(220,25){2}
  \Vertex(230,25){2}
  \Vertex(240,25){2}
  \Vertex(250,25){2}
  \Vertex(275,25){2}
  \Text(175,35)[]{$r$}
  \Text(200,35)[]{$g$}
  \Text(210,35)[]{$b$}
  \Text(220,35)[]{$\bar{b}$}
  \Text(230,35)[]{$b$}
  \Text(240,35)[]{$\bar{b}$}
  \Text(250,35)[]{$\bar{g}$}
  \Text(275,35)[]{$\bar{r}$}  

  \end{picture}
  \caption{ \label{fig:popcorn} Illustration of popcorn production of a
    diquark pair. In frame a) no fluctuation has occurred, and a full string is
  spanned between a red--antired $q\bar{q}$ pair. In frame b) a
  green--antigreen pair has 
  appeared on the string as a quantum fluctuation. If the red and green quarks
  form an antiblue triplet, this reverses the colour
  flow in this part of the string, and the net force acting on the green quark
  is zero. In frame c) the string breaks by the production of a $b\bar{b}$
  pair, resulting in two string pieces with diquark ends. In frame 
  d) another breakup in the blue triplet field results in an additional meson.}
}

\subsection{Effects of a modified string tension}  
\label{sec:modifiedtension}

We will in this paper assume that a rope breaks up by the repeated production
of $q\bar{q}$ pairs, as expected from an analogy with $e^+e^-$ pair production
in QED. As mentioned in the introduction, and discussed in detail in
\sectref{sec:ropes}, a rope formed by $n$ elementary strings with
random charges, can in general be fully extinguished by a number of
$q\bar{q}$ pairs smaller than or equal to $n$. Here $n$ breakups will be
needed in case the colour charges combine to the highest possible multiplet. 
As can be understood
from the tunneling mechanism, the "effective
string tension", to be inserted in \eqsref{eq:schwinger} and
\eq{eq:schwinger2} for each step, is determined by the energy released in the
step. This means the \textit{reduction} in rope tension when the new
$q\bar{q}$ pair is produced. We will therefore treat one step in
the breakup of the rope, as the breakup of an individual
string, with a modified effective string tension.

\subsubsection{Effects on particle ratios}
As discussed above, strangeness and baryon production is in the \pythia 
implementation determined by a set of phenomenological parameters, some of
which represent the relative tunneling probabilities for different
quarks and diquarks. 
Let the modification of the string tension be given by a simple scaling
with an enhancement factor $h$, such that $\kappa \mapsto
\kapeff = h\kappa$. The result in \eqref{eq:schwinger} or
\eq{eq:schwinger2} then implies that the  $s/u$ ratio (called $\rho$), will be
modified by the scaling relation
\begin{equation}
\rho \mapsto \tilde{\rho} = \rho^{1/h}.
\label{eq:rhoscaling}
\end{equation}

Baryon production in the popcorn model is somewhat more
complicated. Here we will assume that the production of the vacuum
fluctuation giving the first new pair in \figref{fig:popcorn}b is
insensitive to the string tension\footnote{Also if the fluctuation
  probability does depend on the string tension, it turns out that
  this effect can be compensated by a change in the effective
  coherence radius for the rope formation, described in
  \sectref{sec:implementation}.}, while the production of the second
pair shown in \figref{fig:popcorn}c, where the string breaks, is the
result of a tunneling process.  This means that we expect the same
kind of scaling as for the $s/u$ ratio in \eqref{eq:rhoscaling} for
the parameters which determine the extra suppression of diquarks with
strange quark content relative to diquarks without strange quarks, and
the suppression of spin 1 diquarks relative to spin 0 diquarks. As these
parameters go together with others that are not affected by string
overlapping in defining the final diquark/quark ratio (called $\xi$), the resulting expression for
the mapping becomes
\begin{equation}
  \label{eq:tildemapping}
 \xi \mapsto \tilde{\xi} = \tilde{\alpha}\beta\left(\frac{\xi}{\alpha\beta}\right)^{1/h}.
\end{equation}
The parameter $\alpha$ contains the parameters for all different types of diquark content, as mentioned above, and thus maps accordingly. The $\beta$-parameter is the popcorn fluctuation probability which in this work is assumed to be unaffected by changes in string tension. The complete mapping relation is derived in full in \appref{sec:deta-descr-rope}.

The effect of a modified string tension on the $s/u$ ratio and the net
diquark/quark ratio is presented in \figref{fig:effpar}. The range of $h$ chosen in \figref{fig:effpar} is much larger than the range relevant for the $pp$ collisions considered in this work (which generally have $h < 1.5$), but is chosen to show effects for large values of $h$ relevant for heavy ion collisions\footnote{We note that the \pythia implementation currently limits all supression parameters from above at a value of one, corresponding to the situation where the supression is gone. The parameter $\tilde{\xi}$ can in principle take on larger values and will, with $\beta = 0.25$, saturate at $\tilde{\xi} = 1.75$.}.

The tunneling probability in \eqref{eq:schwinger2} will
also give somewhat increased transverse momenta, but as the tunneling
effect is a minor contribution to the $p_\perp$-distribution, this
effect is rather small. A detailed description of the above
modifications is presented in \appref{sec:deta-descr-rope}.

\FIGURE[ht]{
  \includegraphics[width=0.7\textwidth, bb=40 180 550 580]{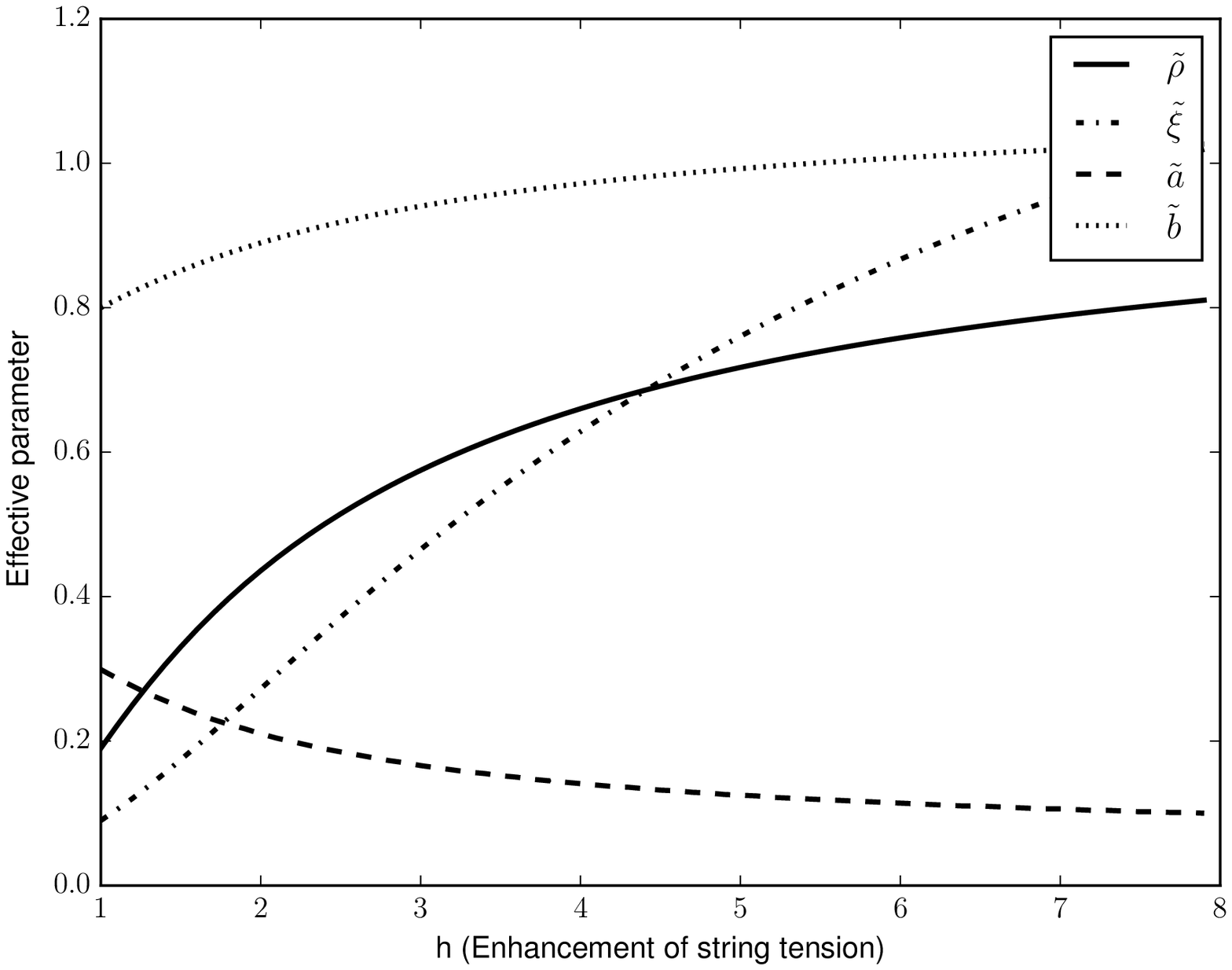}
  \caption{\label{fig:effpar} Effective parameters of the string model
    as a function of effective string tension. The parameters $\rho$
    and $\xi$ control the strangeness content and baryonic content
    respectively, $a$ and $b$ are related to multiplicity. A modified
    string tension has a sizeable effect on $\rho$ and $\xi$ in
    particular. The range of $h$ shown is much larger than relevant for $pp$ collisions, which typically have $h < 1.5$. Larger values of $h$ are, however, relevant for heavy ion collisions. The values of the parameters for $h=1$ comes from a
    tune to LEP data.}
}

\subsubsection{Effects on multiplicity}
\label{sec:eff-on-mult}
One could assume, that the factor $\kappa^2$ in the production rate in
\eqref{eq:schwinger} should imply a change in the $b$-parameter in
\eqsref{eq:Lund} and \eq{eq:split}, and thus significantly modify
hadron multiplicities. The quantity $\mathcal{P}$ in
\eqref{eq:schwinger} is the probability per unit volume and unit
time. In a one-dimensional string, it should be interpreted as
probability per unit length and unit time. Therefore the space-time
distance between break-up points is expected to be proportional to
$1/\kappa$. At the same time the ``yo-yo'' states representing mesons
have extension and oscillation times proportional to $1/\kappa$.  Thus
the earlier production time is compensated by the smaller string
length needed to form a meson. This is taken into account via the
scaling factor $\kappa^2$ in \eqref{eq:stringarea}, with the effect
that $b$ is essentially unchanged. (For $u$ and $d$ quarks, with
masses of the order 10~MeV, and a single string with tension $\kappa
\approx\, 0.2 \,\mathrm{GeV}^2$, the exponential factor in
\eqref{eq:schwinger} is very close to 1, and therefore not
significantly changed by an increased string tension.)  The second
free parameter in the model (when flavours and spin are neglected) is
the parameter $N$ in \eqref{eq:Lund}, which specifies the relative
weights between hadron states with different multiplicities. As the
density of hadronic states ought to be independent of the former
colour configuration, we do not expect any change in the
$N$-parameter, although we admit that such modifications cannot be
excluded when the mesons are produced in a stronger field.

Although the $b$-parameter would be unchanged if only light quarks
were produced, it will increase slightly due to the enhanced
production probability for strange quarks and diquarks, given by an
increased exponential factor in \eqref{eq:schwinger}. This gives a
shift
\begin{equation}
	b \mapsto \tilde{b} = \frac{2 + \tilde{\rho}}{2 + \rho}\,b.
\label{eq:hscaling3}
\end{equation}
As the $a$-parameter is calculated from the normalization constraint
for the splitting function in \eqref{eq:split}, it will get a
correspondingly moderate modification.  The effect on the parameters
$a$ and $b$ is also shown in \figref{fig:effpar}, and we see that
these parameters are less affected by an increased string tension than
the parameters determining strangeness and baryon ratios. As it will
be discussed in \sectref{sec:model-behaviour}, a typical value for the
string tension enhancement factor $h$ at LHC is around 1.2. Changes in
$a$ and $b$ could therefore naively account for $\sim 5$\% decrease in
multiplicity in $pp$, as particle density is approximately
proportional to $\sqrt{(1+a)/b}$ (see
\eqref{eq:particledensity}). Since some parameters must be retuned
after implementation of the rope model, this effect will not appear in
final state observables (note that the string hadronization parameters
are still tuned to LEP data, see \appref{sec:tuning} for a detailed
account of the tuning procedure.)  The model should, however, not be
further retuned when it is applied to nucleus collisions at LHC
energies. Here $h$ is expected to be so large, that a decrease of the
order $15 - 20$\% in multiplicity, due to changes in $\tilde{a}$ and
$\tilde{b}$, is a prediction of the model.

\section{Ropes}
\label{sec:ropes}
In this section an ideal situation is considered, where separated
colour charges within a limited area in transverse space act
coherently to form a colour rope (assuming that the total system is a
colour singlet.) As mentioned in the introduction, lattice
calculations show that if the endpoint charges correspond to a
specific SU(3) multiplet, the tension (or energy density) in the rope
is given by the second Casimir operator \cite{Bali:2000un}. This
result is valid for an isolated rope, modifications must be expected
in a situation where the rope is surrounded by other ropes or
strings. Here we discuss the formation of a rope, its tension, and its
eventual decay.

\subsection{Rope formation}
\label{sec:ropeformation}
We study a situation, where a rope is formed by a group of ordinary
triplet--antitriplet strings, where the net colour charge is obtained
from the addition of $m$ colour triplets and $n$ antitriplets with
random colours.  As pointed out first in ref.~\cite{Biro:1984}, the
result corresponds to a kind of random walk in colour space. A
detailed discussion of this process can be found \eg\ in
ref.~\cite{Jeon:2004rk} or in \appref{sec:pq-calc}. Here we only
present the main features essential for the later discussion.

While in SU(2) a multiplet is specified by the quantum number $j$, or
its multiplicity $(2j+1)$, an SU(3) multiplet can be specified by two
quantum numbers $p$ and $q$. A specific state then corresponds to $p$
coherent triplets (\eg\ all red) and $q$ coherent antitriplets (\eg\
all antigreen). In addition the triplet and the antitriplet must be in
an octet state (as is the case for red--antigreen), and not a
singlet. The multiplicity, $N$, of the multiplet $\{p,q\}$ is then
given by:
\begin{equation}
N = \frac{1}{2} (p+1)(q+1)(p+q+2).
\label{eq:multiplicityrev}
\end{equation}

The result of adding a set of triplets and antitriplets can be
calculated in an iterative way.  Starting from a multiplet $\{p,q\}$
adding one more triplet, with random colour, one obtains the
multiplets
\begin{equation}
  \{p+1,q\},\,\,\, \{p-1,q+1\},\,\, \mathrm{and}\,\, \{p,q-1\}, 
\label{eq:randomaddition}
\end{equation}
with weights proportional to the corresponding multiplicities given by
\eqref{eq:multiplicityrev}. From symmetry, the addition of an
antitriplet gives the multiplets $\{p,q+1\}$, $\{p+1,q-1\}$, and
$\{p-1,q\}$. Multiplets with negative values for $p$ or $q$ are not
allowed. In \appref{sec:pq-calc} the result for adding random colour
octets (corresponding to gluons) is also described.

The average walk can be easily calculated numerically, and in
\figref{fig:ranwalk} we show $\langle p+q\rangle$ obtained from
randomly adding triplets and antitriplets, with a fixed number, $m+n$,
of charges. (The individual numbers $m$ and $n$ are then given by a
binomial distribution. The same result was also presented in
ref.~\cite{Biro:1984}.) The width of the distribution is indicated by
the band showing $1\,\sigma$ variations.

\FIGURE[ht]{
  \includegraphics[width=0.75\textwidth]{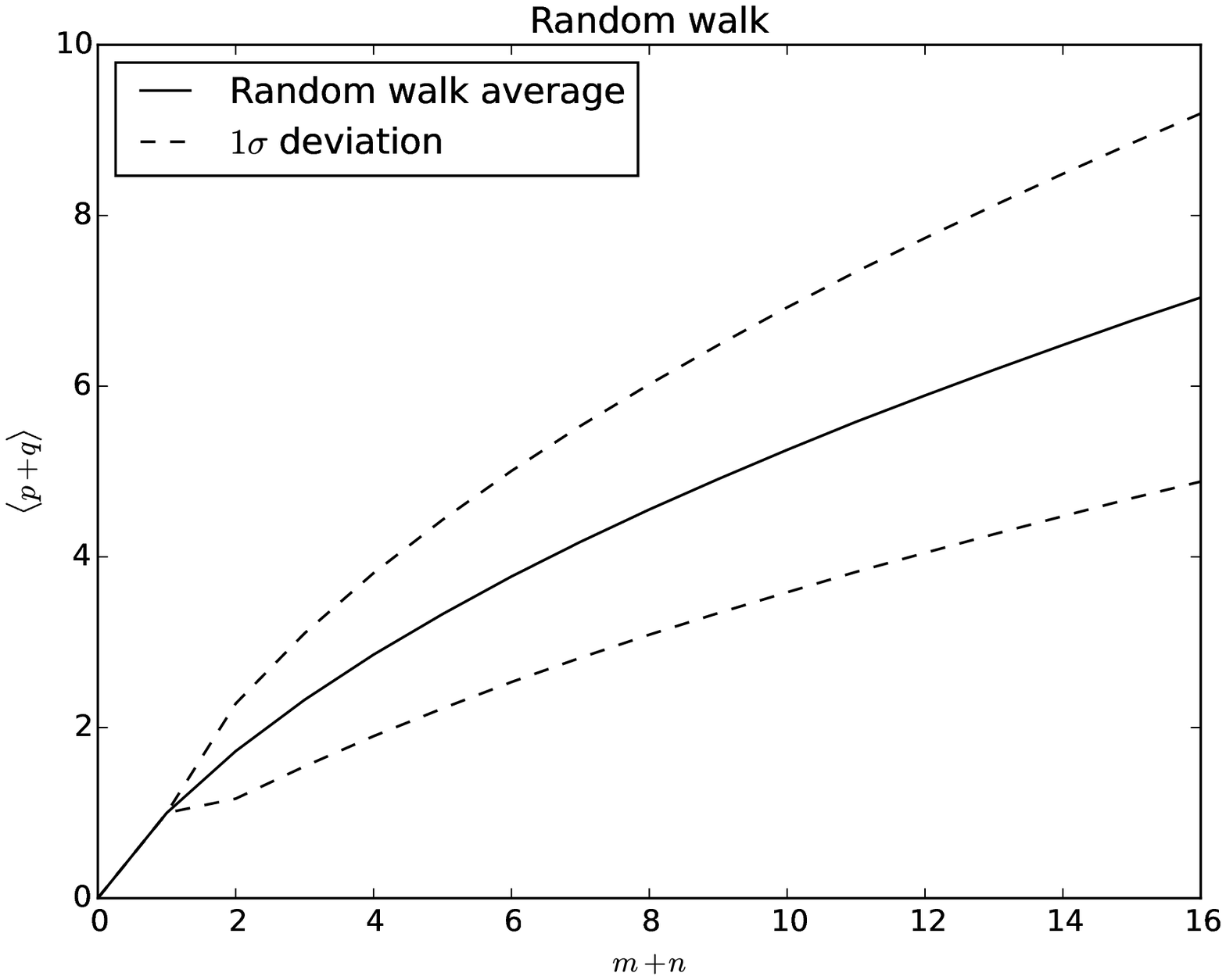}
  \caption{ \label{fig:ranwalk} Relation between $\langle p+q \rangle$
    and $m+n$ after a random walk in $p,q$-space. The shaded area
    corresponds to the standard deviation around the average $\langle
    p+q \rangle$.}
}

\subsection{Rope tension}
\label{sec:ropetension}

As mentioned in the introduction, lattice calculations show that the
tension in an isolated static rope is proportional to the quadratic
Casimir operator $C_2$ \cite{Bali:2000un}. The Casimir operator is
only defined up to a normalization constant, but for our purpose we
only need the value of $C_2$ normalized to a triplet (antitriplet)
$\{1,0\}$ ($\{0,1\}$), corresponding to the tension in a single
string. This is given by
\begin{equation}
C_2(\{p,q\})/C_2(\{1,0\})=\frac{1}{4}\left(p^2+pq+q^2 +3p+3q\right),
\label{eq:scaledcasimir}
\end{equation}
which thus can be regarded as the relative strength of the ``rope
tension''. 

It is, however, possible that the energy is increased if the
transverse area is constrained by neighbouring ropes and strings. This
effect is very hard to estimate, and contributes to the uncertainties
in the predictions from string interactions.  To illustrate this
problem we first look at the analogous situation in a normal
superconductor.  Here a longitudinal magnetic flux is confined by
currents in the surrounding condensate. In an extreme type~I
superconductor the surface energy is small, and a tube with two flux
quanta will have doubled cross section and doubled energy. If,
however, the flux tube would be constrained within the same transverse
area, the magnetic field energy will be multiplied by 4, but the
energy to annihilate the condensate within the tube (originally equal
to the field energy) will be unchanged. The net result is a change by
a factor 2.5. In a type II superconductor the magnetic flux and the
energy density are more concentrated close to the center of a flux
tube or vortex line, and there is a repulsive (attractive) interaction
between parallel (antiparallel) vortices, with a logarithmic
dependence on the separation.

The simplest model for a colour flux tube is the bag model
\cite{Johnson:1975zp, Johnson:1975sg}, which is analogous to a type I
superconductor. The tension in an unconstrained flux tube is here
proportional to $\sqrt{C_2}$ \cite{Johnson:1975sg}, in contradiction
to the lattice result. However, for a flux tube constrained to the
width of an elementary string, the energy is indeed proportional to
the lattice result $C_2$ \cite{Semay:2004br}.

Lattice calculations by Cardoso \emph{et al.} \cite{Cardoso:2009kz}
show that the energy in the longitudinal colour-electric field is
dominating over the transverse colour-magnetic field, with other
components small, and the result shows similarities with a type II
superconductor. However, recent calculations by Cea \emph{et al.}
\cite{Cea:2014uja} indicate, that the best analogy with a dual
superconductor is close to the transition between type I and type II,
but with the Ginzburg--Landau parameter within the type I region.

We conclude that there are considerable uncertainties in the
estimation of the tension in a rope. Fortunately it turns out that,
for the results presented here, the effects of a modified effective
rope tension can be compensated by a modified value for the effective
rope width $r_0$. We will therefore in the following assume that the
tension is proportional to the Casimir operator of the colour
multiplet, in accordance with the lattice calculations for an isolated
rope.

\subsection{Fragmentation of a rope}
\label{sec:ropefragmentation}

\subsubsection{Effective string tension and particle ratios}

As in most studies of rope fragmentation, we will assume that the
colour rope will break up in a stepwise manner, by the production of
quark--antiquark pairs. This is the result for an interaction
Lagrangian proportional to $\bar{\psi}\gamma_\mu\psi A_\mu$. In the
region between the newly produced quark and antiquark, the field
corresponding to the $\{p,q\}$ multiplet can be reduced to either a
$\{p-1,q\}$ or a $\{p,q-1\}$ multiplet. In the first case the
antiquark is pulled towards the $\{p,q\}$ charge (and the quark
towards the $\{q,p\}$ charge in the other end of the rope), and in the
second case it is pulled in the opposite direction. As discussed in
\sectref{sec:modifiedtension}, an essential point is that the energy
released in the breakup is what enters in the tunneling process
discussed in \sectref{sec:lundstring}, and thus determines the
production probabilities in \eqsref{eq:schwinger} and
\eq{eq:schwinger2}. This is given by the \emph{difference} between the
Casimir operators for the multiplets $\{p,q\}$ and $\{p-1,q\}$ or
$\{p,q-1\}$ respectively. This difference thus represents an
\textit{effective string tension}, or $\kapeff$, introduced in
\sectref{sec:modifiedtension}.

For a breakup via the transition $\{p,q\} \rightarrow \{p-1,q\}$ we
get from \eqref{eq:scaledcasimir} the effective string tension
\begin{equation}
\kapeff=\frac{2p+q+2}{4}  \kappa.
\label{eq:kappaeff}
\end{equation}
We note that for large charges $p$ this result grows $\sim
(p/2)\kappa$, (\ie\ more slowly than $\propto p\kappa$).  As an
example we have for a rope consisting of two parallel strings in a
\{2,0\}-state the Casimir operator $C_2(\{2,0\})=5/2\times
C_2(\{1,0\})$. In the first breakup the effective string tension will
then be $\kapeff=(5/2-1)\kappa=(3/2)\kappa$, while for the second
break-up we will have the normal tension $\kapeff=\kappa$.  A more
complicated example is presented in \appref{sec:example}.

We want to emphasize that this result is significantly smaller than
what is assumed in most studies of rope effects. Although the change
in rope tension is used \eg\ in ref.~\cite{Andersson:1991er}, it is
quite common to use Schwinger's result in \eqref{eq:schwinger}, with a
constant field $\mathcal{E}$ proportional to the charge $Q$. This
would correspond to a situation where the field is confined within a
tube with constant cross section $A$, and that the contribution from
the bag pressure or confining currents is neglected \footnote{In some
  cases the transverse area is taken as an undetermined parameter,
  \eg\ in refs.~\cite{Bialas:1984ye,Andersson:1991er}.}.  The field
energy per unit length is then given by $\frac{1}{2} A
\,\mathcal{E}^2=\frac{1}{2}Q\,\mathcal{E}$. We here used that the
total flux, $A\, \mathcal{E}$, is determined by the charge $Q$
spanning the rope. When a pair with elementary charge $g/2$ is
produced, the field is reduced to $\mathcal{E}-\epsilon$, where
$g/2=A\,\epsilon$ is the flux from the elementary charge $g/2$. (The
factor 1/2 is due to the conventional definition of $g$.) Neglecting
the contribution to the tension from the bag pressure, and assuming
that $A$ is not modified, we obtain the ``effective string tension''
from the difference in field energy:
\begin{equation}
\kapeff = \frac{1}{2}A \{\mathcal{E}^2 - (\mathcal{E}-\epsilon)^2\}=
A\epsilon(\mathcal{E}-\epsilon/2)=\frac{1}{2} g(\mathcal{E} -
\epsilon/2)
\label{eq:kappaeff2}
\end{equation}
For a classical macroscopic field $\mathcal{E}$, the term $\epsilon/2$
can be neglected in the parenthesis $(\mathcal{E} -\epsilon/2)$, and
the expression in \eqref{eq:kappaeff2} therefore looks like
\eqref{eq:schwinger}.  A problem shows up, however, if this formula is
used to relate the effective string tension to the tension in an
elementary string.  Here the term $-\epsilon/2$ can not be neglected,
and it corresponds to the correction to the string tension due to the
field produced by the new pair, as pointed out in
ref.~\cite{Glendenning:1983qq}. Taking this into account we get for an
elementary string, with $\mathcal{E} =\epsilon$, the tension
$\kappa=g\epsilon/4$, and the result
\begin{equation}
\kapeff = (2p-1)\kappa. 
\label{eq:kappaeff3}
\end{equation}
We note that for strong ropes, \ie\ large values of $p$, this result
is a factor 4 larger than the result in \eqref{eq:kappaeff}.  Here a
factor 2 comes from the correction term in the elementary string, and
another factor 2 is due to the relatively slow increase of the rope
tension in \eqref{eq:scaledcasimir} for large $p$.

\subsubsection{Particle multiplicity}

For a rope stretched by ($m+n$) random colour charges, the rope can break up
totally by only ($p+q$) $q\bar{q}$ pairs. The random walk in colour space here
gives $p+q \sim\sqrt{m+n}$ (see \figref{fig:ranwalk}), and this effect
therefore contributes to a reduction in the density of produced particles. 
In many references it is assumed that this is the dominant effect determining
the multiplicity. Lacking a physical picture, it is often assumed that  
the hadrons are produced with the same separation in rapidity, see \eg\
refs.~\cite{Biro:1984,Kerman:1985tj,Braun:1991dg}. 

Andersson and Henning \cite{Andersson:1991er}
argue, however, 
that the early breakup can give a large energy to the leading particle(s),
leaving less energy to produce softer particles.
Amelin \emph{et al.} \cite{Amelin:1994mc} also argue
that, as the breakups according to 
\eqref{eq:averagetau} occur earlier for a stronger tension,
the multiplicity should be lower. 
The arguments in refs.~\cite{Andersson:1991er,Amelin:1994mc} do, however not
take into account that not only the space-time distance between breakup points
becomes smaller for a higher effective string tension; also the size of the
string pieces making up a hadron becomes smaller with the same scale factor.
Therefore we argued in \sectref{sec:eff-on-mult} that the early breakup 
only gives a minor correction to the multiplicity, owing to the
increased production of strange quarks (also admitting that unknown effects
could influence the 
possibility for a $q\bar{q}$ pair to fit into a final state hadron in case of
a stronger effective string tension). Thus
in our approach, the dominant effect on the total multiplicity is due to
the random walk in colour space discussed above.

\section{Implementation of ropes in the \dipsy Generator}
\label{sec:the-dipsy}
The rope model outlined above has been implemented in the \dipsy
\cite{Avsar:2005iz,Avsar:2006jy,Avsar:2007xg,Flensburg:2011kk} event
generator. \dipsy is based on Mueller's dipole cascade model
\cite{Mueller:1993rr,Mueller:1994jq,Mueller:1994gb}, which is a
formulation of leading-log BFKL evolution
\cite{Kuraev:1977fs,Balitsky:1978ic} in transverse coordinate space,
making it very well suited for a study of coherence effects based on
spatial overlap of strings.\par
The \dipsy model introduces many sub-leading effects to Mueller's
dipole model (see \appref{app:dipsy} for an introduction to the \dipsy
model). Important to emphasize here is the initial state ``swing'',
which is a finite $N_c$ effect. Two dipoles with the same colour form
a colour quadrupole, where the colour field is better approximated by
dipoles formed by the closest charge--anticharge pairs (\cf\
\figref{fig:initial-swing}). As smaller dipoles have smaller cross
sections, this effect contributes to the saturation at small $x$.

\FIGURE[ht]{
\begin{picture}(300,100)(-65,0)
\Line(10,10)(20,25)
\Line(25,40)(20,25)
\Line(50,40)(80,30)
\Line(75,60)(80,30)
\Line(75,60)(55,50)
\Line(30,50)(15,70)
\Line(15,70)(10,90)
\Text(95,45)[c]{$\longleftrightarrow$}
\Line(100,10)(110,25)
\Line(115,40)(110,25)
\Line(140,40)(170,30)
\Line(165,60)(170,30)
\Line(165,60)(145,50)
\Line(120,50)(105,70)
\Line(105,70)(100,90)
\Vertex(10,10){1}
\Vertex(20,25){1}
\Vertex(25,40){2}
\Vertex(50,40){2}
\Vertex(80,30){1}
\Vertex(75,60){1}
\Vertex(55,50){2} \Vertex(30,50){2} \Vertex(15,70){1} \Vertex(10,90){1}
\Vertex(100,10){1} \Vertex(110,25){1} \Vertex(115,40){2} \Vertex(120,50){2}
\Vertex(140,40){2} \Vertex(170,30){1} \Vertex(165,60){1} \Vertex(145,50){2}
\Vertex(105,70){1} \Vertex(100,90){1}

\ArrowLine(50,40)(25,40)
\ArrowLine(30,50)(55,50)
\ArrowLine(120,50)(115,40)
\ArrowLine(140,40)(145,50)
\end{picture}

\caption{ \label{fig:initial-swing} Sketch of how the initial state swing could
  reconnect colours between two dipoles in impact parameter space.}
}

By introducing saturation in the cascade using the colour swing, saturation
directly becomes a finite-$N_c$ effect, as the effect obviously vanishes in
the large-$N_c$ approximation. In \sectref{sec:final-state-swing} we will argue
how adding a similar reconnection effect in the final-state cascade, is a
convenient way to introduce the singlet from the $\plet{3} \otimes
\antiplet{3} = \plet{8} \oplus \plet{1}$ decomposition.

The gluons produced in the initial state in \dipsy are in the end
ordered in both positive and negative light-cone momenta, and are
allowed to continue radiating final-state bremsstrahlung according to
the time-like dipole radiation model in momentum space
\cite{Gustafson:1986db,Gustafson:1987rq} implemented in the \ariadne
program\cite{Lonnblad:1992tz}.  Before forming the ropes, which will
be sent to \pytppp \cite{Sjostrand:2007gs} for hadronization as
described in \sectref{sec:implementation}, the strings may also
reconnect via a ``final state swing'' between dipoles with identical
colours. This is introduced to account for the singlet from the $3
\otimes \bar{3} = 8 \oplus 1$ decomposition of interacting string
colours, and described in more details in
\sectref{sec:final-state-swing}.

\subsection{The Final-state Swing}
\label{sec:final-state-swing}

The swing mechanism was inspired by the colour reconnection mechanism
proposed in \cite{Lonnblad:1995yk} for the time-like dipole shower in
\ariadne \cite{Lonnblad:1992tz}. The need for colour reconnections in
the final state had already been noted in the context of multiple
interactions \cite{Sjostrand:1987su}.  Several other investigations
were performed in the eighties and nineties looking at recoupling
effects both in \tee, DIS and hadronic collisions, see \eg\
\cite{Gustafson:1988fs,Sjostrand:1993rb,Edin:1995gi,Enberg:2001vq}.

\FIGURE[ht]{
  \begin{picture}(300,100)(0,0)
    \DashLine(0,35)(20,40){2}
    \DashLine(0,65)(20,60){2}
    \DashLine(100,40)(120,35){2}
    \DashLine(100,60)(120,65){2}
    \Vertex(20,40){1}
    \Vertex(20,60){1}
    \Vertex(100,40){1}
    \Vertex(100,60){1}
    \ArrowLine(20,60)(100,60)
    \ArrowLine(100,40)(20,40)
    \Text(140,50)[c]{$\longleftrightarrow$}
    \DashLine(160,35)(180,40){2}
    \DashLine(160,65)(180,60){2}
    \DashLine(260,40)(280,35){2}
    \DashLine(260,60)(280,65){2}
    \Vertex(180,40){1}
    \Vertex(180,60){1}
    \Vertex(260,40){1}
    \Vertex(260,60){1}
    \ArrowLine(180,60)(180,40)
    \ArrowLine(260,40)(260,60)

  \end{picture}
  \caption{ \label{fig:final-swing} Sketch of how the final
    state-swing could reconnect colours between two dipoles in
    momentum space.}
}

These reconnection models were based on the principle of minimizing
``effective string lengths''. This was inspired by the string
fragmentation, where the string action is given by the area law for
the Wilson loop \cite{Wilson:1974sk}. For a simple $q\bar{q}$-string,
which does not break, the action for one period is proportional to the
total invariant squared mass $s$. For a string which breaks up into
hadrons, the string area is instead proportional to $\ln(s/m_0^2)$,
where the scale parameter $m_0$ is a typical hadronic scale,
$\sim1$~GeV. This area also determines the average hadron
multiplicity. For a more complicated string configuration it is also
possible to generalize this area to the so-called
\textit{lambda}-measure \cite{Andersson:1985qr,Andersson:1988ee}. The
lambda-measure is infrared stable, but in cases without soft or
collinear gluons, it can be approximated by
\begin{equation}
  \label{eq:lambda}
  \lambda_s\propto\sum_{i=1}^{n-1}\ln{\frac{(p_i + p_{i+1})^2}{m_0^2}}.
\end{equation}
Hence, when
looking at the re-coupling of individual dipoles in a string it seems
natural to try to minimise the sum of the logarithms of the dipole
masses, or equivalently, the product of dipole masses.

In the original implementation in \ariadne, only reconnections
which decreased the total $\lambda_s$ were allowed, but now we have
reimplemented it in a way very similar to the swing in \dipsy. Between
every final-state dipole radiation there is a possibility to
recouple two dipoles, $(12)(34)\to(14)(32)$, if they are in the same
colour state (using the same colour indices as in \dipsy). Again we
treat emission and swing as competing processes, and while the
emission of a gluon is simply given by the dipole formula
\begin{equation}
  \label{eq:fsdiprad}
  \frac{d\mathcal{P}_g}{d\rho}\approx dy\frac{C_F\as}{2\pi}
\end{equation}
(for a $q\bar{q}-$ dipole. For a gluon dipole $C_F \mapsto C_A/2$).
Here $\rho=\ln(p_\perp^2)$ is the evolution parameter.
We define the relative probability for the swing as
\begin{equation}
  \label{eq:fsswingprob}
  \frac{d\mathcal{P}_s}{d\rho}=\lambda
  \frac{(\pmb{p}_1+\pmb{p}_2)^2(\pmb{p}_3+\pmb{p}_4)^2}
  {(\pmb{p}_1+\pmb{p}_4)^2(\pmb{p}_3+\pmb{p}_2)^2},
\end{equation}
with a free parameter $\lambda$ governing the relative strength
of the swing.

Dipoles spanning large distances in impact parameter space are heavily
suppressed in \dipsy due to the confinement effects imposed by the
introduction of a small gluon mass, $m_g$ (see \appref{app:dipsy}). Even
though the final-state swing in \eqref{eq:fsswingprob} is formulated
in momentum space and does not take into account any impact parameter
dependence we still need to preserve these confinement effects. This
is done by only considering dipoles which are closer in impact
parameter space than a distance $\propto 1/m_g$ as candidates for a
swing.

It is interesting to note that if we have two completely anti-parallel
dipoles, the probability of them having the same colour index is
simply $1/N_c^2$, and if this is so, they will always reconnect,
effectively breaking the ``rope''. This corresponds to the singlet
term which arises when combining the triplet and anti-triplet string,
$\plet{3}\otimes\antiplet{3}=\plet{8}\oplus\plet{1}$ (see
\eqref{eq:trip-atrip}). This means that this effect in principle has
already been taken care of before the formation of the rope, and when
we later perform the random walk in colour space for overlapping
strings, we need to constrain it to take this into account.

\FIGURE[ht]{
  \begin{picture}(300,100)(0,0)
    \DashLine(0,35)(20,40){2}
    \DashLine(0,65)(20,60){2}
    \DashLine(100,40)(120,35){2}
    \DashLine(100,60)(120,65){2}
    \Vertex(20,40){1}
    \Vertex(20,60){1}
    \Vertex(100,40){1}
    \Vertex(100,60){1}
    \ArrowLine(20,60)(100,60)
    \ArrowLine(20,40)(100,40)
    \Text(140,50)[c]{$\longleftrightarrow$}
    \DashLine(160,35)(180,40){2}
    \DashLine(160,65)(180,60){2}
    \DashLine(260,40)(280,35){2}
    \DashLine(260,60)(280,65){2}
    \Vertex(180,40){1}
    \Vertex(180,60){1}
    \Vertex(260,40){1}
    \Vertex(260,60){1}
    \ArrowLine(180,60)(195,50)
    \ArrowLine(180,40)(195,50)
    \ArrowLine(245,50)(260,60)
    \ArrowLine(245,50)(260,40)
    \ArrowLine(245,50)(195,50)
    \BCirc(245,50){3}
    \BCirc(195,50){3}
  \end{picture}
  \caption{ \label{fig:sextetswing} Sketch of how a antitriplet swing
    could reconnect colours between two dipoles by introducing two
    string junctions  (denoted by circles).}
}

One could also imagine introducing another swing mechanism for the
case of parallel strings, which would then imply that two (parallel)
dipoles could swing into a single string according to the sketch in
\figref{fig:sextetswing}. This would involve the formation of two
so-called \textit{junctions} where three colour lines join. Although
there is a mechanism for hadronizing junction strings in \pytppp, it
has some technical limitations\footnote{While preparing this
  manuscript a slightly improved version of the junction fragmentation
  was implemented in \pythia together with a colour reconnection
  mechanism producing junctions in the way depicted in
  \figref{fig:sextetswing} \cite{Sjostrand:2014zea}.}. Also, the
treatment of radiation from junction topologies in \ariadne requires
additional work, and we will thus defer the treatment of this new kind
of swing to a future paper. Instead we will treat the corresponding
multiplet configurations in the rope model below.

\subsection{Estimates of overlap region}
\label{sec:implementation}
As discussed in more detail in \appref{app:dipsy}, the \dipsy Monte Carlo
describes two colliding parton cascades, producing colour-connected
partons located in transverse coordinate space and rapidity. This is
followed by final-state radiation in momentum space from \ariadne
together with colour reconnection by the final-state swing. The
location in the transverse plane is the basis for the interaction
between the strings and the formation of ropes. As described in
\sectref{sec:modifiedtension}, we expect that the dominant effect of
rope formation is an enhanced production of strangeness and baryons,
and in this paper we concentrate on these effects. 

\FIGURE[ht]{
\includegraphics[width=0.7\textwidth]{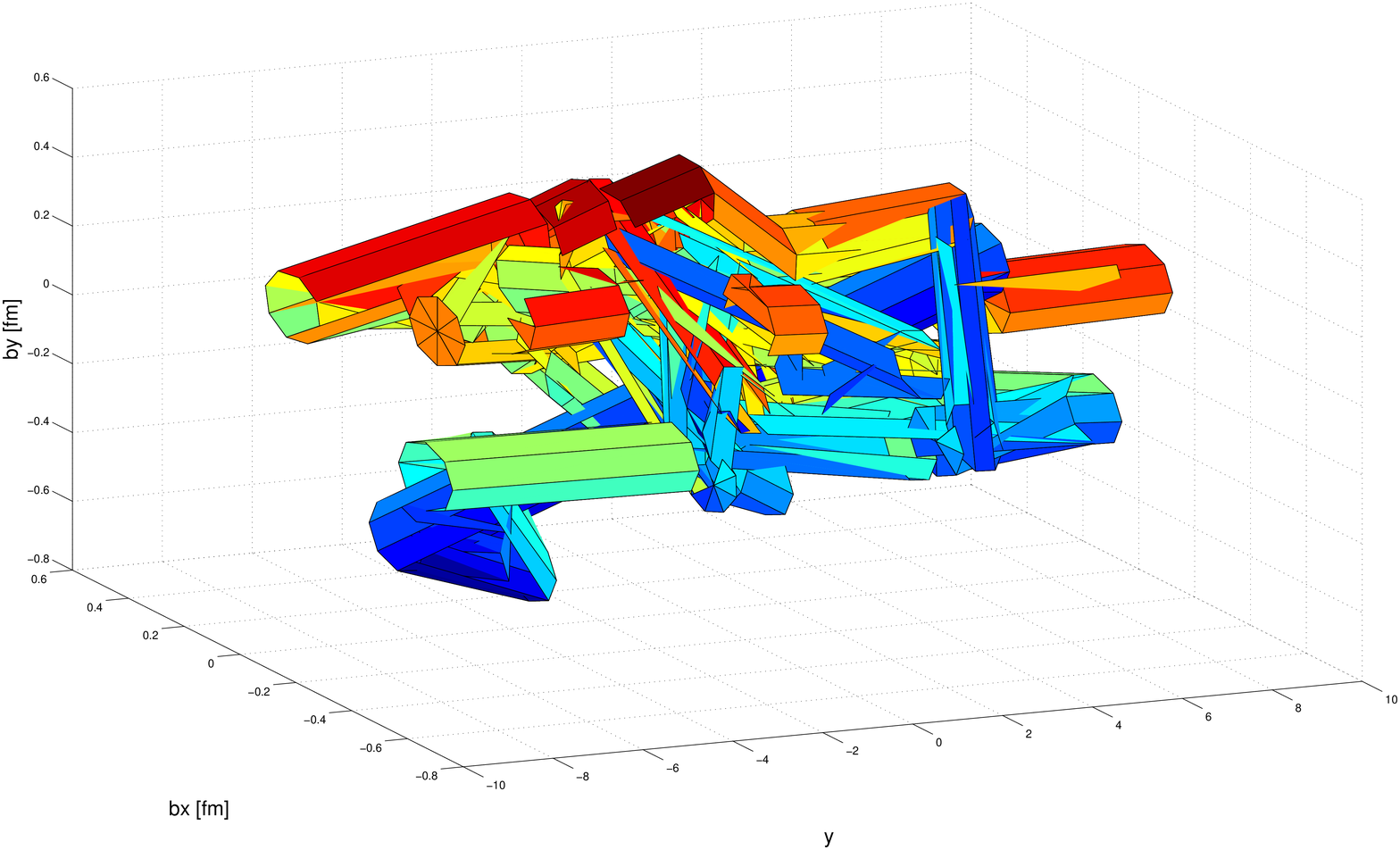}
\caption{ \label{fig:tubes} Illustration of strings from a $pp$ event
  at $\sqrt{s} = 7$~TeV in $(\vec{b}_\perp,Y)$-space before
  hadronization. Notice that the string radius is set at 0.1~fm -- an
  order of magnitude less than in the calculation -- in order to
  improve readability of the figure.}
}

In \sectref{sec:ropes} we discussed an ideal situation with strings or
colour flux tubes, which overlap completely in transverse space, and
interfere constructively or destructively to form a coherent rope. A
typical event, ready for hadronization, is shown in impact parameter
space and rapidity in \figref{fig:tubes}. The tubes in the figure
represents colour connections between partons, and it is easy to see
that real events are far from similar to the ideal situation, and we
therefore need a way to estimate the amount of interaction.  We
naturally expect that strings close in transverse space should
interfere more strongly than strings further away, with a typical
interaction range of the order of the confinement scale. Our main
assumption is therefore, that the degree of coherence between the
strings is determined by the overlap between the corresponding flux
tubes.

Since the \dipsy-generated events provide access to space-time
information of strings, it is natural to alter the effect depending on
the amount of overlap. Space-time information is usually not available in
generators for $pp$ collisions. It is, however, normally
  accesable in Molecular Dynamics Monte Carlo generators aimed for heavy ion
  collisions, where a similar approach  (not including all fluctuations), has
  been studied in ref.~\cite{Sorge:1992ej}. We expect the
coherence range (the radius of the flux tubes) to be of the order of
the confinement scale, and put it to 1~fm. One could treat it as a
completely free parameter, and tune it to data together with other
free parameters, in order to give the most accurate description of
data.  Since neither the method of calculation of overlap between
strings, nor the connection between this overlap and $m$ and $n$ (the
number of uncorrelated colour and anticolour charges in one end of the
rope) is obvious from first principles, we will present two different
approaches for calculating the overlap.  The first method is very
crude, and approximates all strings as straight flux tubes ("pipes")
parallel to the rapidity axis.  In this pipe-based approach, a string
will be given values for $m$ and $n$ that are averages over the whole
string, in fact an average over the full area in transverse space
covered by the string. The string is subsequently hadronized with a
single average value for $\kapeff$, determined from $m$ and $n$ by the
random walk procedure.

The second method is more detailed and takes into account more
fluctuations along the string. In this dipole-based approach, a string
is viewed as a chain of dipoles, connected by gluons, either with
quarks at the endpoints, or as closed gluonic loops. Overlaps are
then calculated and a random walk in colour space is performed for
each dipole, at a specific rapidity value, in order to change the
value for $\kapeff$ at each breakup in the hadronization.

In both treatments the random walk is constrained to take into account
that steps corresponding to $\{1,0\}\otimes\{0,1\}\mapsto\{0,0\}$ have
already been treated in the final-state swing mechanism. Nevertheless
it is clear that for $m+n$ overlapping strings or dipoles, may end up
in multiplets $\{p,q\}$, where $p<m$ and/or $q<n$. This poses a
problem, since for technical reasons, each of the $n+m$ strings are
hadronized separately, so we cannot break them with only $p+q$
break-ups. Since the net effect is to reduce the multiplicity by a
factor $\propto(p+q)/(m+n)$, we instead emulate this by simply
randomly discarding strings in the pipe-based treatment with a
probability $1-(p+q)/(m+n)$. In the dipole-based treatment the
approach is somewhat more sophisticated and we instead discard
individual dipoles in a procedure inspired by the suggested
\emph{junction swing} in \figref{fig:sextetswing}.

The details of the implementation of the two treatments are fairly
technical and a full description is therefore deferred to
\apprefs{sec:pipe-based-treatment} and
\sect{sec:dipole-based-treatment}.

The aim of introducing two different approaches for calculating
overlap, is to demonstrate that even the very crude pipe-based
approach catches the gist of the model and improves the description
of strangeness and baryon production, as described in \sectref{sec:results}.
Since further
sophistication in the dipole-based approach can improve description of
strange and baryonic content even further, we argue that a direct
mapping from overlap in tranverse space to $m$ and $n$ is indeed
sensible. Further sophistication of this calculation of overlap is
left for future publications.

\subsection{Exclusive observables with \dipsy}

The \dipsy Monte Carlo is implemented in the event generator framework
called \thepeg \cite{Lonnblad:2006pt}. Also the \ariadne program for
final-state parton showers has been implemented in this framework and
there we have now added the final-state swing mechanism described in
\sectref{sec:final-state-swing}. In \thepeg we have also written an
interface to the hadronization routines of the \pytppp event
generator, and it is here we have implemented our rope hadronization
models described in the previous section. The whole code is available
from the authors upon request.\footnote{See also
  \texttt{http://home.thep.lu.se/DIPSY} for installation
  instructions.}

The full code can generate full, exclusive final states for $pp$, $pA$
and $AA$.  The goal driving event generators such as \dipsy is to be
able to describe all collider physics with the same models, using the
same parameters. As event generators in general have a quite large
number of parameters, which parameterize the uncertainties in the
models implemented, these parameters need to be estimated from data in
a "tuning" process. It is important to state that tuning does not mean
fitting of individual spectra to data. It rather means that one set of
parameter values are estimated such that the same models can describe
anything from \tee over $pp$ to $pA$ and $AA$ collisions.

We have already discussed some of the parameters on the Lund string
fragmentation level, but there are many others, such as the
non-perturbative cutoff in the parton cascade and scale factors used
in the running of \as (to emulate untreated higher orders). The tuning
of these parameters is quite a complicated task, and for each event
generator there are typically several different tunes available (see
\eg\ \texttt{mcplots.cern.ch} \cite{Karneyeu:2013aha} for comparisons
between different tunes of different programs).

The most common strategy is to first tune parameters associated with
the final-state showers and hadronization to data from
\tee colliders. Assuming jet universality these are then fixed when
tuning further parameters related to initial-state showers and
multiple parton interactions to data from hadron colliders.

As we have tried to argue in this paper, the concept of jet
universality is not quite straightforward, and the hadronization may
very well behave differently in hadronic and \tee collisions. To see
the effects of our new model it is therefore necessary to take some
care and make sure that the description of the flavour-dependent
observables we wish to study is not dominated by a general change of
multiplicity distributions for all particles, as such global effects
would normally be removed in a tuning procedure. Therefore we have
made a careful tuning both for the cases with rope effects and
without, as detailed in \appref{sec:tuning}.

Since the \dipsy event generator does not yet include a model for
diffractive events, care also must be taken to only compare to
observables that are not sensitive to diffraction. For that reason, we
will primarily look at particle ratios. This point is also expanded
upon in \appref{sec:tuning}.

\section{Results}
\label{sec:results}

In this section we will present some results from applying the
introduced rope model. We will concentrate on flavour observables in
minimum bias events in hadronic collisions in the energy range where
we believe the small-$x$ approximation in the \dipsy model is valid,
$\sqrt{s}\gtaeq 100$~GeV. After presenting comparisons to experimental
data in \sectref{sec:data-cmp}, we will look at the model's
sensitivity to parameters and its behaviour at higher energies in
\sectref{sec:model-behaviour}. In \sectref{sec:des-flow} we discuss
the flow-like effects shown in the results.

\subsection{Comparison to data}
\label{sec:data-cmp}
Results from \dipsy including the rope model in the dipole scheme
(labelled 'Rope') are here compared to CMS data
\cite{Khachatryan:2011tm} at 900 and 7000~GeV as well as STAR data
\cite{Adams:2006nd} at 200~GeV. We also show \dipsy with no rope
effects on, labelled '\dipsy', and finally a \pytppp
reference\footnote{Version 8.180, tune 4C.} labelled
'Pythia'. Comparison to data for more energies, and other kinematic
variables \etc\ can be found on the project home page at
\texttt{http://home.thep.lu.se/DIPSY/}.  The parameters for rope
hadronization used are $r_0 = 1$~fm, $\beta = 0.25$ and $m_0 =
0.135$~GeV. These choices will be further discussed in
\sectref{sec:model-behaviour}.

In \figref{fig:pt-data} (left) we see the proton/pion ratio in bins of
$p_\perp$, as measured by STAR, compared to simulations. We clearly
see that \dipsy with no added effects fails to describe this ratio, in
the same way as \pytppp does. The proton/pion ratio is a good measure
of the relative amount of baryons with no strangeness, governed by the
$\xi$-parameter, and we see that \dipsy with added rope effects indeed
describes data better, both in terms of relative proton content and
$p_\perp$.

In \figref{fig:pt-data} (right) the $\Lambda/K^0_s$ ratio at
$\sqrt{s}=7$~TeV is shown in bins of $p_\perp$ as measured by CMS. As
both the meson and the baryon has strangeness, this should also be a
good measure of the influence of the $\xi$-parameter. We again see
that including rope effects improves the description, especially in
the low-$p_\perp$ end, where most of the multiplicity is
concentrated. The high $p_\perp$-tail is poorly described, and we will
discuss this further in \sectref{sec:des-flow}.

\FIGURE[ht]{
  \includegraphics[width=0.45\textwidth]{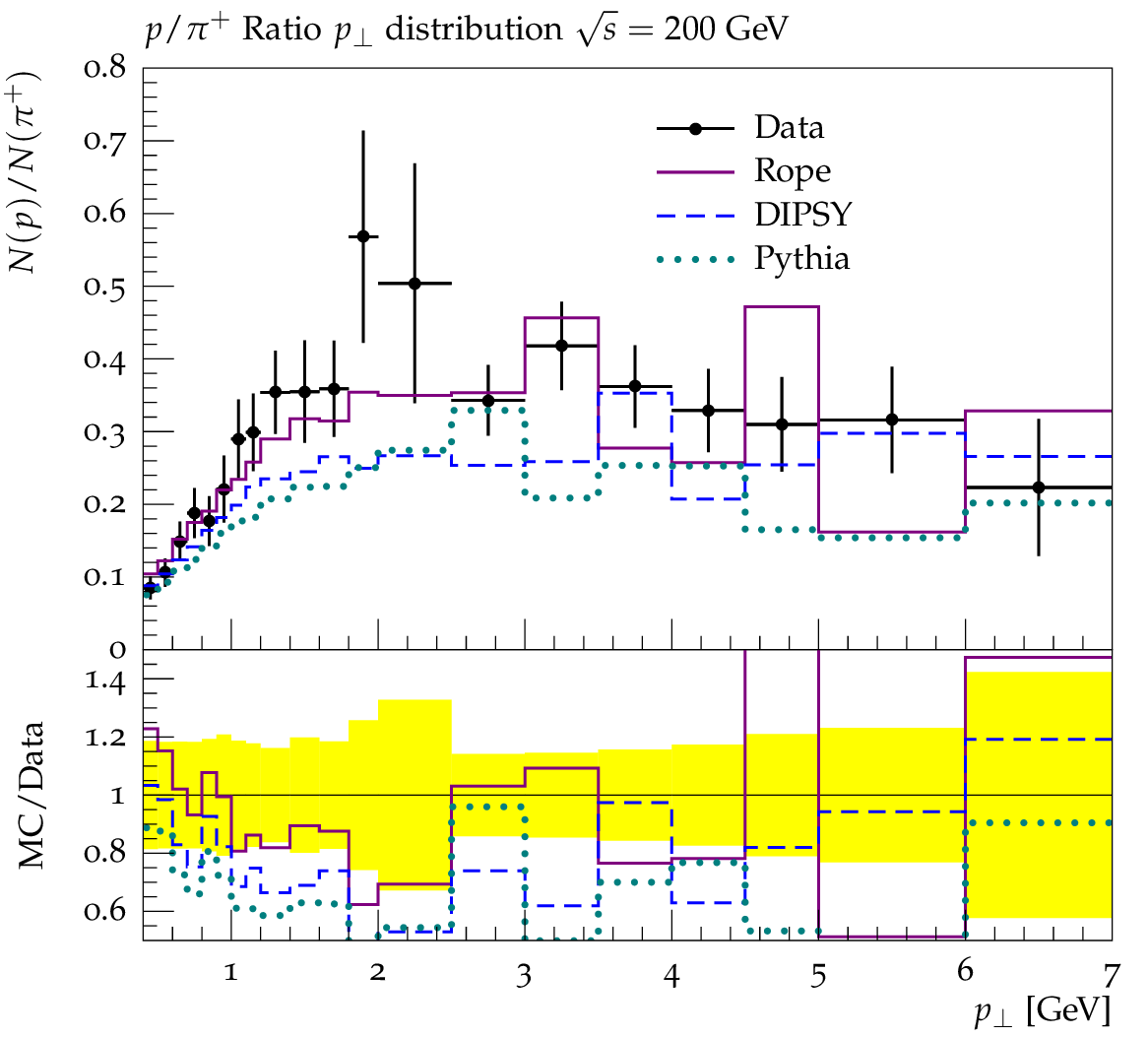}
  \includegraphics[width=0.45\textwidth]{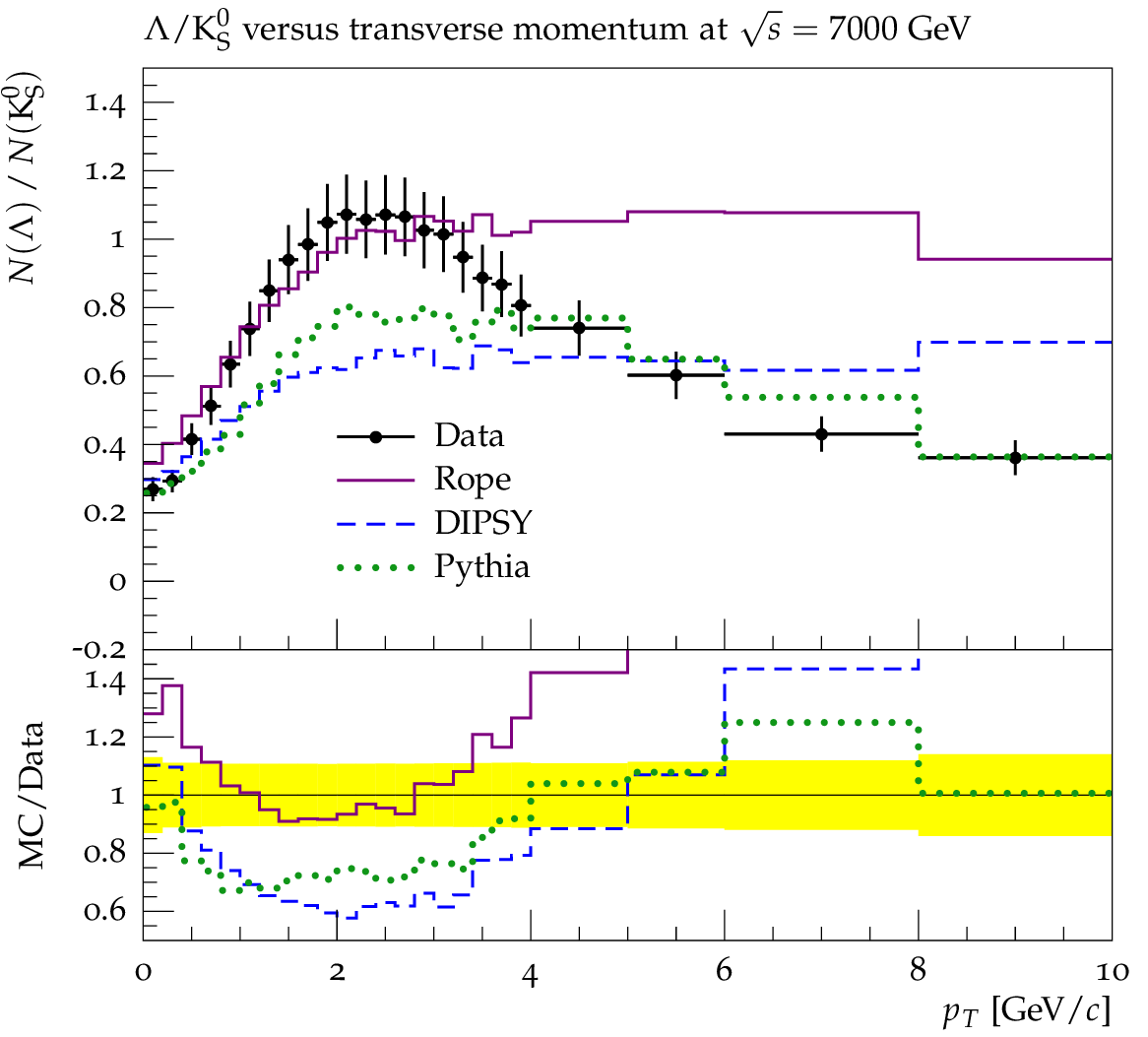}
  \caption{ \label{fig:pt-data} The proton to pion ratio in bins of $p_\perp$
    as measured by STAR at $\sqrt{s} = 200$~GeV (left) and $\Lambda/K^0_s$ at
    7000~GeV as measured by CMS (right). Both results are compared to \dipsy
    with and without rope, as well as with \pytppp.}
}

\Figref{fig:y-data-energy} shows the
$\Lambda/K^0_s$ ratio at $\sqrt{s}=900$~GeV (left) and $7$~TeV (right) in bins
of rapidity as measured by CMS. We see here that the relatively weak dependence
on energy is well described by the rope model,
with the same values for $r_0$, $\beta$ and $m_0$.

\FIGURE[ht]{
  \includegraphics[width=0.45\textwidth]{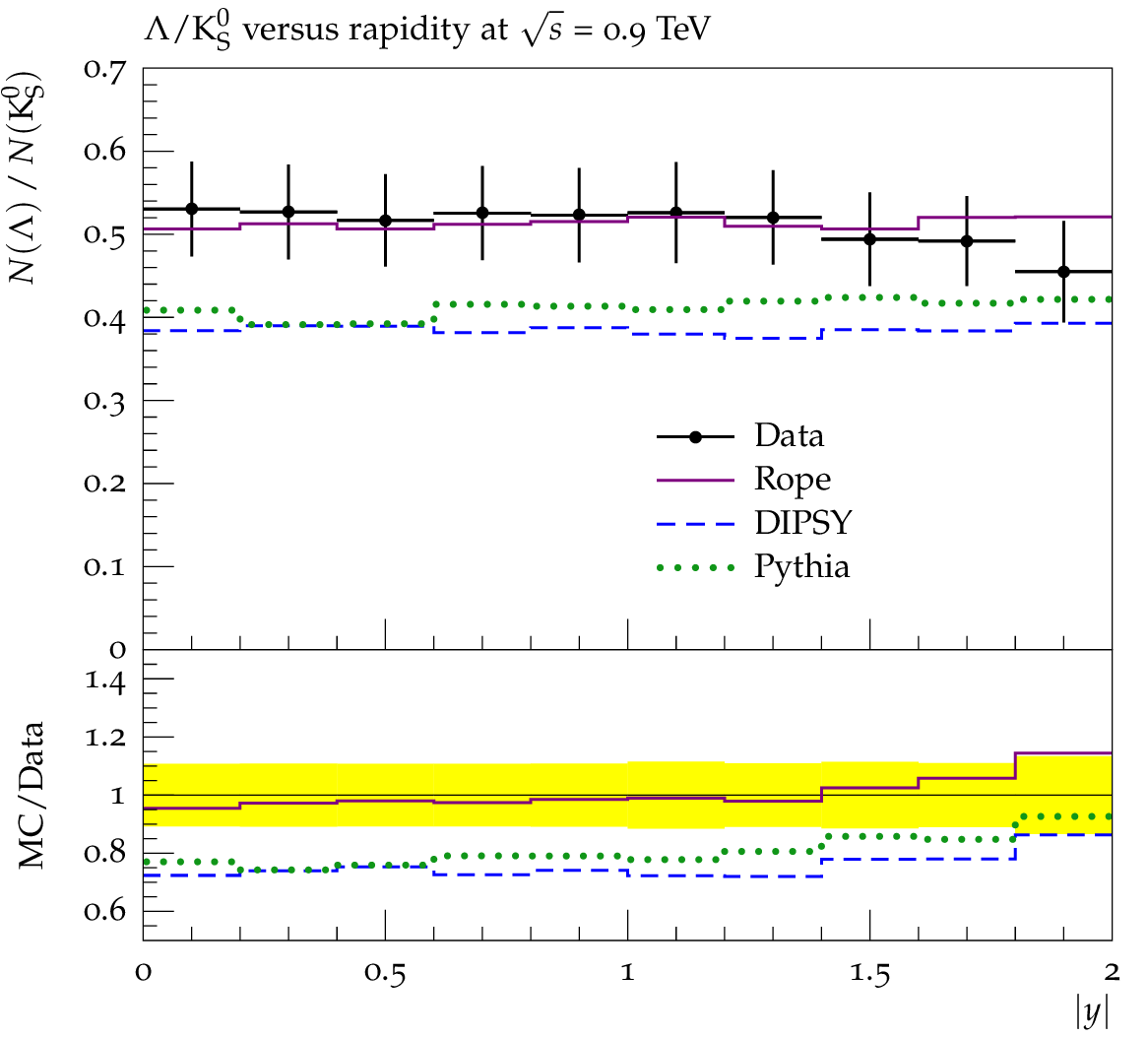}
  \includegraphics[width=0.45\textwidth]{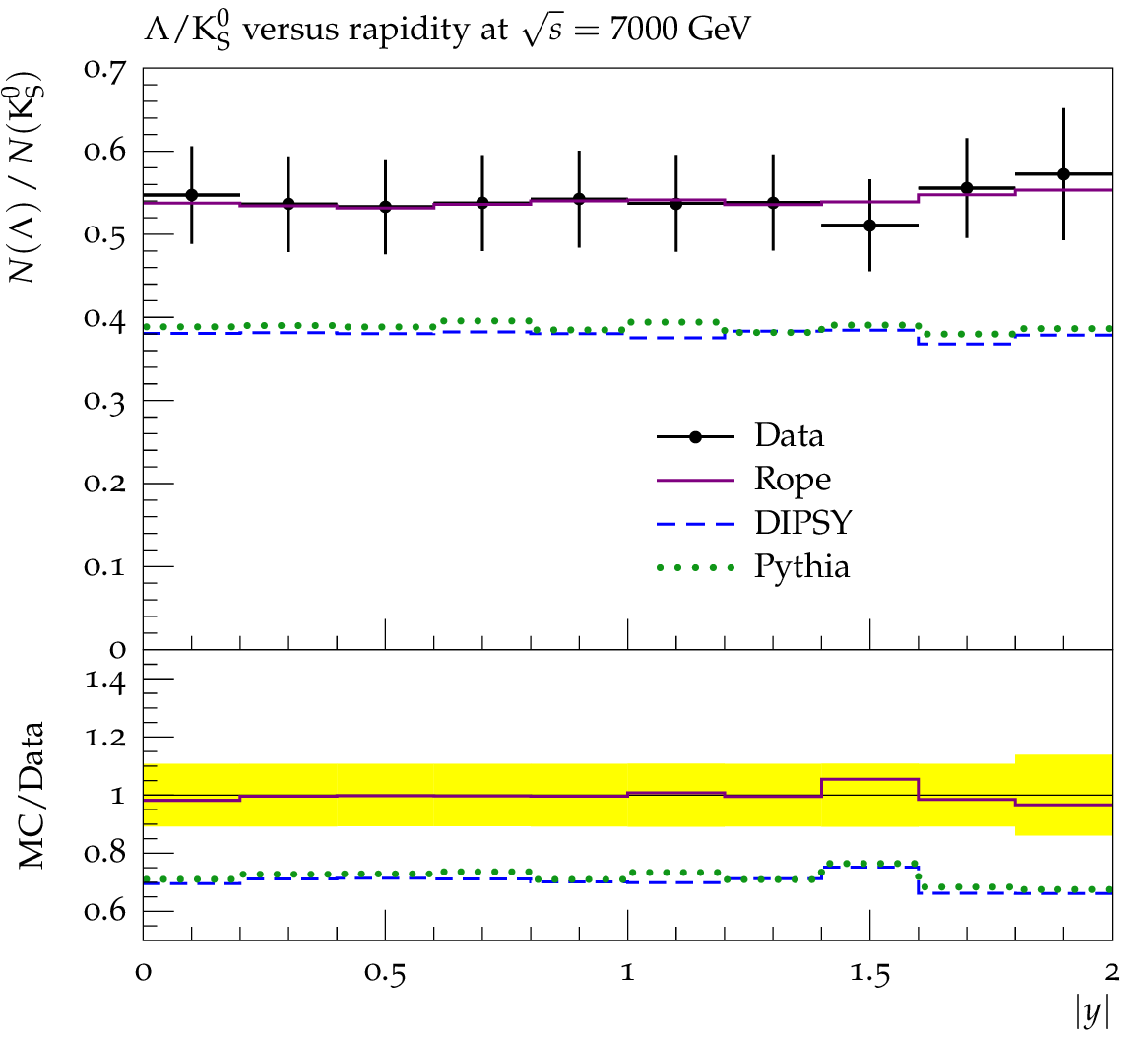}
  \caption{ \label{fig:y-data-energy} The $\Lambda/K^0_s$ ratio at
    900~GeV (left) and 7000~GeV (right) as measured by CMS in bins of
    rapidity. The figure shows that the rope model captures the
    (albeit weak) energy dependence of this ratio, while \dipsy
    without ropes, as well as \pytppp, shows no energy dependence.}
}

The $\Xi^-/\Lambda$ ratio is an observable, that is particularly sensitive to
the $\rho$ parameter. \Figref{fig:y-XL} shows this ratio in bins of
rapidity at 900~GeV and 7000~GeV, and 
we see that the rope model also
reproduces the behaviour of these data fairly well.

\FIGURE[ht]{
  \includegraphics[width=0.45\textwidth]{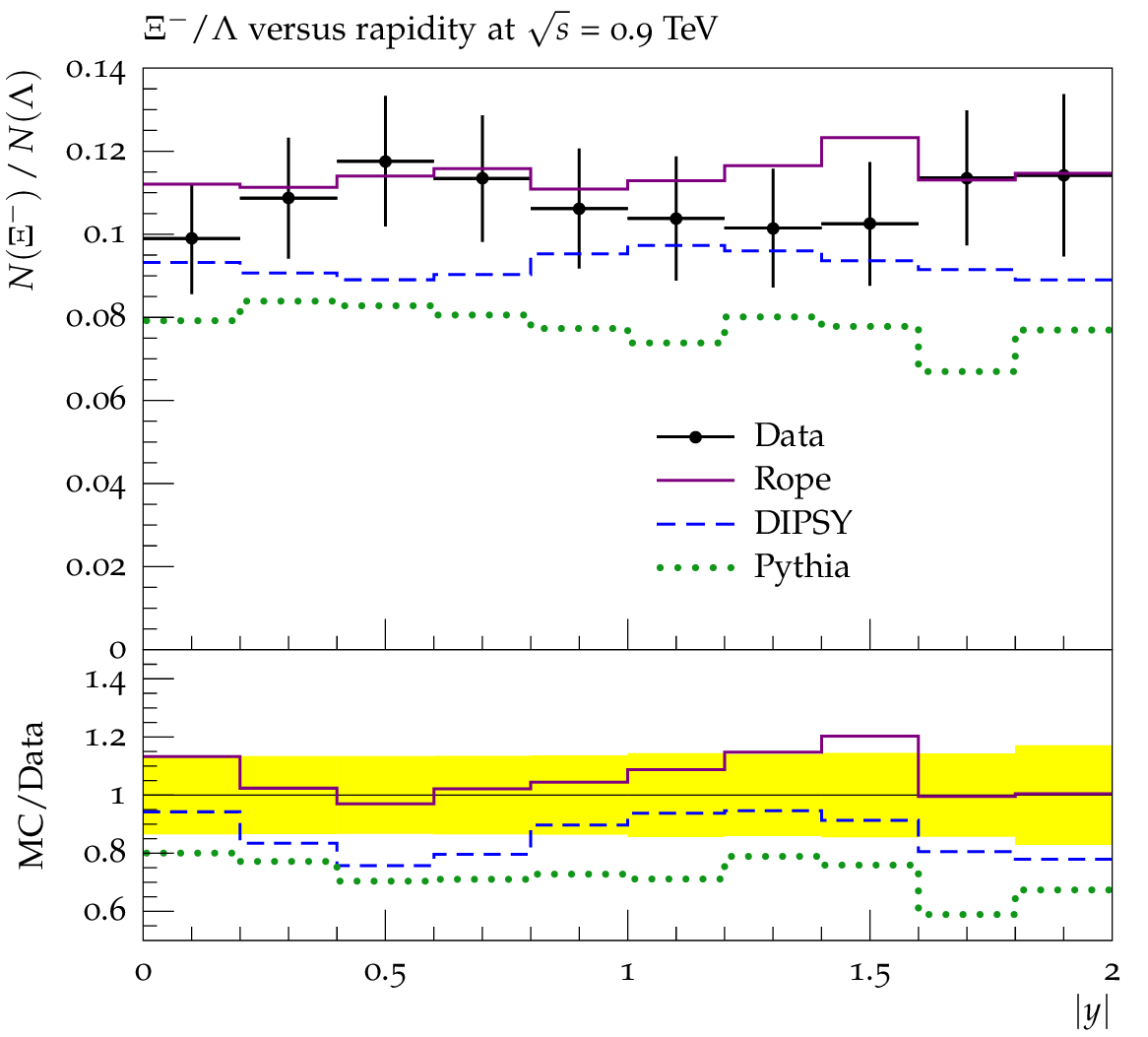}
  \includegraphics[width=0.45\textwidth]{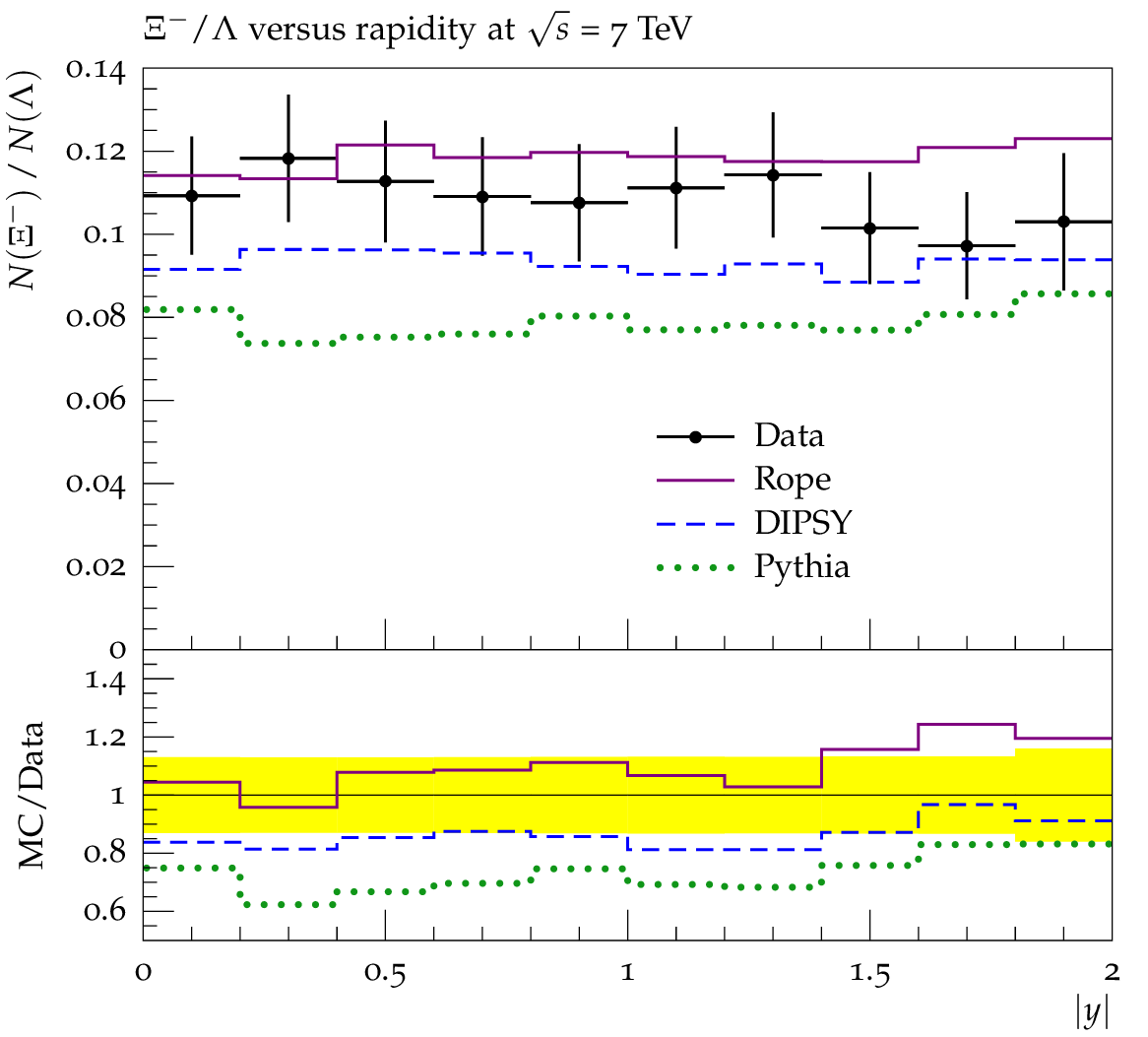}
  \caption{ \label{fig:y-XL} The ratio $\Xi^-/\Lambda$ at 900~GeV
    (left) and 7000~GeV (right) as measured by CMS in bins of
    rapidity.}

}Although the energy dependence shown in \figrefs{fig:y-data-energy}
and \fig{fig:y-XL} is fairly weak, the fact that it is well described
by the rope model is a very important point. At higher energies more
strings are confined within a small space, and with the amount of
overlap as a measure of the size of the rope effect, one could expect
a larger increase with energy. To further illustrate this point, and
to serve as qualitative predictions for the integrated particle
ratios, we show in \figref{fig:integrated-ratios} the total $K/\pi$,
$\Lambda/K^0_s$ and $\Xi^-/\Lambda$ ratios as a function of
$\sqrt{s}$. 
We note here that the combined effect of strangness and baryon suppression is
not factorizing in a simple way. As discussed in 
\appref{sec:deta-descr-rope}, the effects of an increased string
tension is quite involved, especially for baryons. The fact that
the relative abundances presented in \figrefs{fig:pt-data}-\ref{fig:y-XL}
are well described, also for different energies, gives us confidence
that our model has some physical relevance.

\FIGURE[ht]{
  \includegraphics[width=0.75\textwidth]{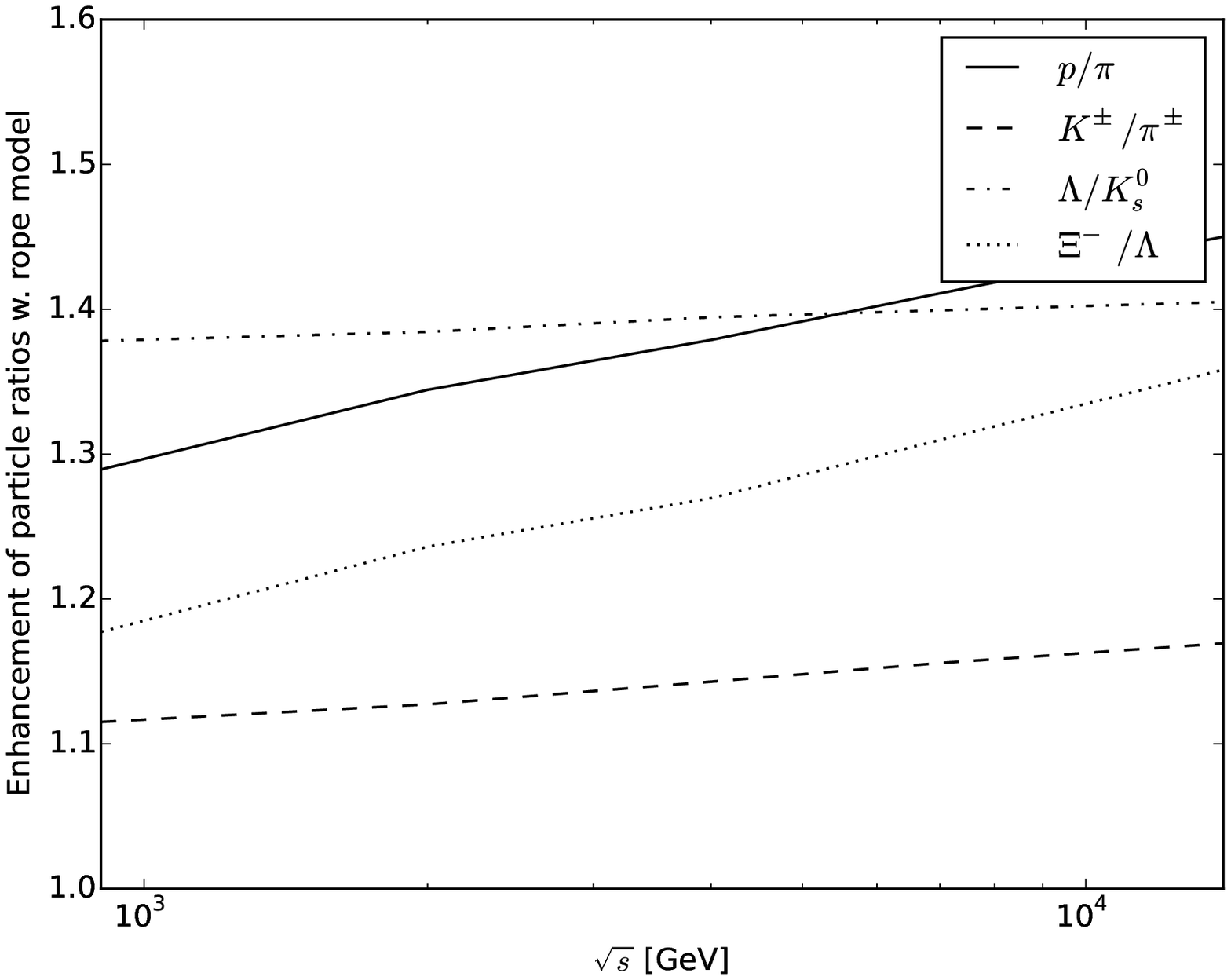}
  \caption{ \label{fig:integrated-ratios} Enhancement of particle
    ratios of function of $\sqrt{s}$. Integrated ratios of $p^\pm$ and
    $K^\pm$ to $\pi^\pm$, $\Lambda \bar{\Lambda}$ to $K^0_s$ and
    $\Xi^-$ to $\Lambda \bar{\Lambda}$ with the rope model (dipole
    approach) applied, normalized to the same ratio with ordinary
    string hadronization. All particles with $p_\perp>200$~MeV are
    included.}
}

\subsection{Model behaviour}
\label{sec:model-behaviour}

The model introduces three new parameters. The string radius $r_0$,
the popcorn-parameter $\beta$, and the parameter $m_0$ which is
specific to the dipole approach. The parameters have not been tuned to
data in the usual sense, but set to reasonable physical estimates. We
believe that the model in its current state is not mature enough to
warrant a tuning, but one should nevertheless get an intuition for the
uncertainties associated with the choice of parameters. We will here
motivate our choices, and show the sensitivity of the model to changes
in the parameter values, and how the results vary with $\sqrt{s}$.

To gauge the sensitivity we look at two quantities: the average string
tension and the number of junctions. Focusing on particle ratios, we
normalize to the $\lambda$-measure, which is a measure for the
hadronic multiplicity (see \eqref{eq:lambda}). Thus we study the event
averaged string tension, defined by the relation
\begin{equation}
  \langle \kapeff/\kappa \rangle \equiv \frac{\sum_{i}
    \lambda_i {\kapeff}_i/\kappa}{\sum_i \lambda_i}
\end{equation}
(where $i$ counts all dipoles and $\lambda_i = \ln(m_i^2/m_0^2)$), and
the number of junctions per unit $\lambda$, $\langle
n_j/\sum_i\lambda_i\rangle$.  These quantities can act as indicators
for the amount of string overlap, which grows with increasing energy,
but is also sensitive to the tunable parameters $r_0$ and $m_0$.

\Figref{fig:avg-enhancement} (left) shows the average string tension
as functions of $\sqrt{s}$, for $r_0 = 1$~fm and $m_0 =
0.135$~GeV. The dashed lines in \figref{fig:avg-enhancement} indicate
the event-by-event fluctuations, showing one standard deviation.  It
is clearly visible that the enhancement effect rises logarithmically
with $\sqrt{s}$. This is expected, as the number of gluons in the
BFKL-based \dipsy cascade has the same energy dependence.

\FIGURE[ht]{
  \includegraphics[width=0.45\textwidth]{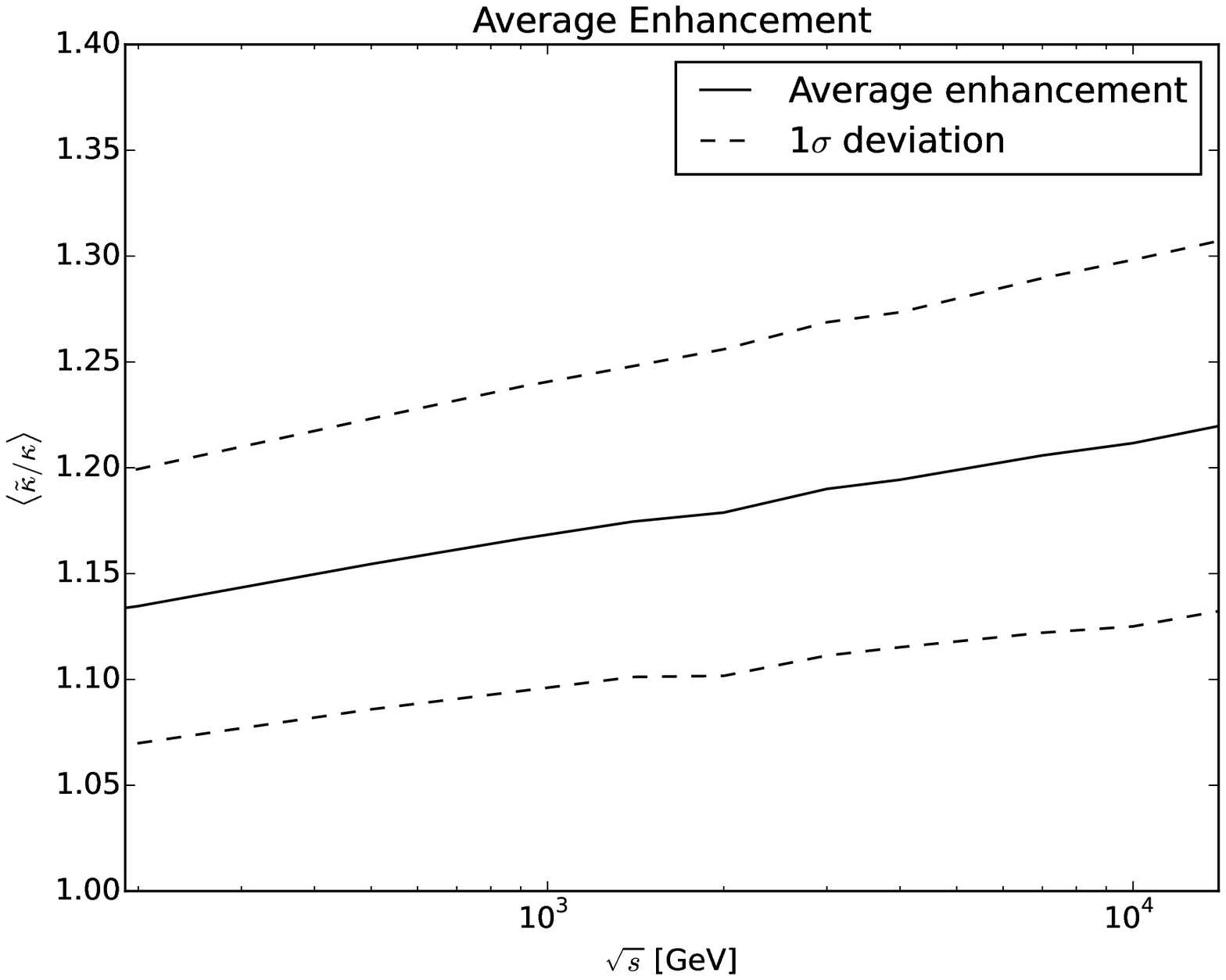}
    \includegraphics[width=0.45\textwidth]{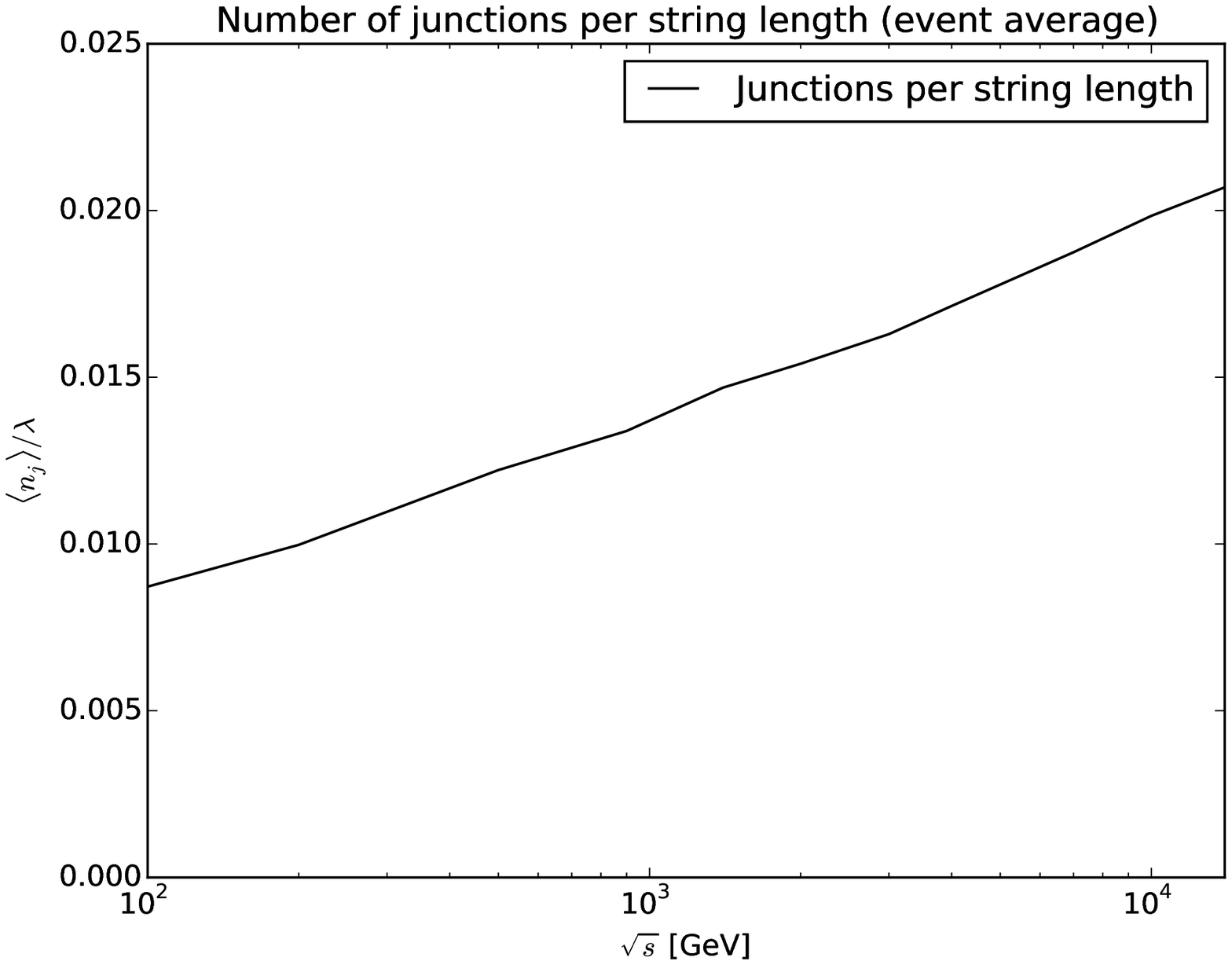}
  \caption{ \label{fig:avg-enhancement} (left) Average enhancement $\langle
    h\rangle = \langle \kapeff/\kappa\rangle$ as a function of
    $\sqrt{s}$ in $pp$ collisions. The band indicates one standard
    deviation.  \label{fig:njunctions} (right) Number of junctions per string
    length as a function of $\sqrt{s}$. }
}

The average value of the string tension will thus increase with
energy, and from \figref{fig:effpar} it is clear that this gives a
larger amount of strange and baryonic activity as $\sqrt{s}$ goes up,
as well as having a moderate effect on total multiplicity due to the
effect on the parameters $a$ and $b$ in the splitting function in
\eqref{eq:split}. However, the multiplicity in $pp$ collisions are
heavily influenced by the parameters controlling the initial-state
evolution in \dipsy, and as described in \appref{sec:tuning}, we tune
these (while keeping the hadronization parameters tuned to LEP data
fixed) to obtain the same multiplicity with and without rope
effects. In this way the only effects from our rope model are the
relative amounts of baryons and hadrons with a strange content; both
are expected to increase.

The increase in the number of junctions per string length is also
shown in \figref{fig:njunctions} (right).  The amount of baryons
emerging from the produced junctions is, as explained in
\appref{sec:dipole-based-treatment}, controlled by the popcorn
strength parameter $\beta$ (see \eqref{eq:hxiscaling}).  We note that
while increasing $\beta$ results in a stronger increase of baryons
produced through diquark break-ups in the strings, it also decreases
the probability that baryons are produced in junction structures, as
explained in \sectref{sec:dipole-based-treatment}. As our results thus
have very little sensitivity to variations in $\beta$, it is fixed to
0.25 throughout the article, and we do not expect large theoretical
uncertainty to be ascribed to this parameter should the model be
thoroughly tuned.

The amount of overlap in an event will also increase by increasing
$r_0$. In \figref{fig:avg-enhancement-r0} (left) the average
enhancement is shown as a function of $r_0$ at fixed energy $\sqrt{s}
= 900$~GeV and $m_0 = 0.135$~GeV.
\FIGURE[ht]{
  \includegraphics[width=0.45\textwidth]{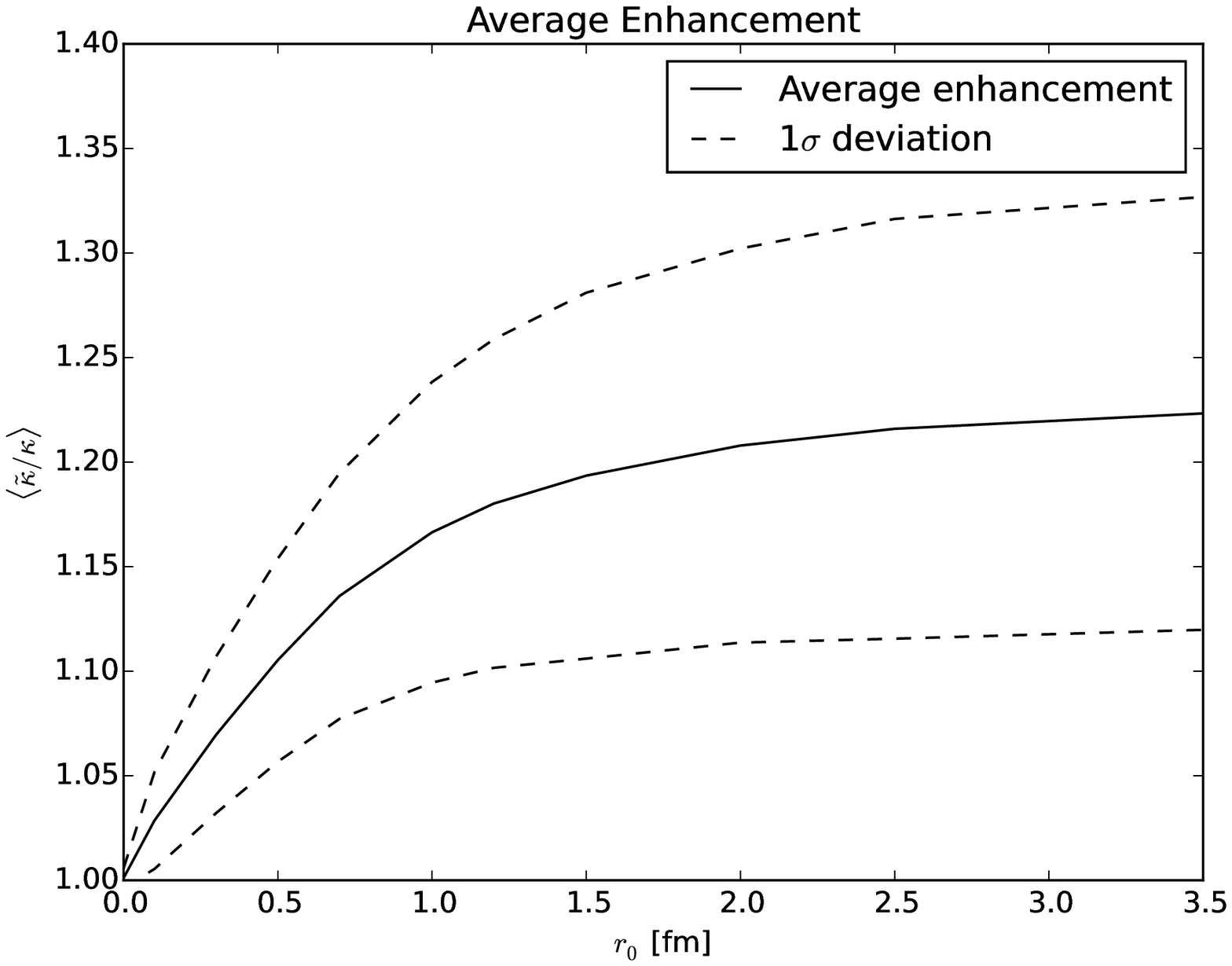}
  \includegraphics[width=0.45\textwidth]{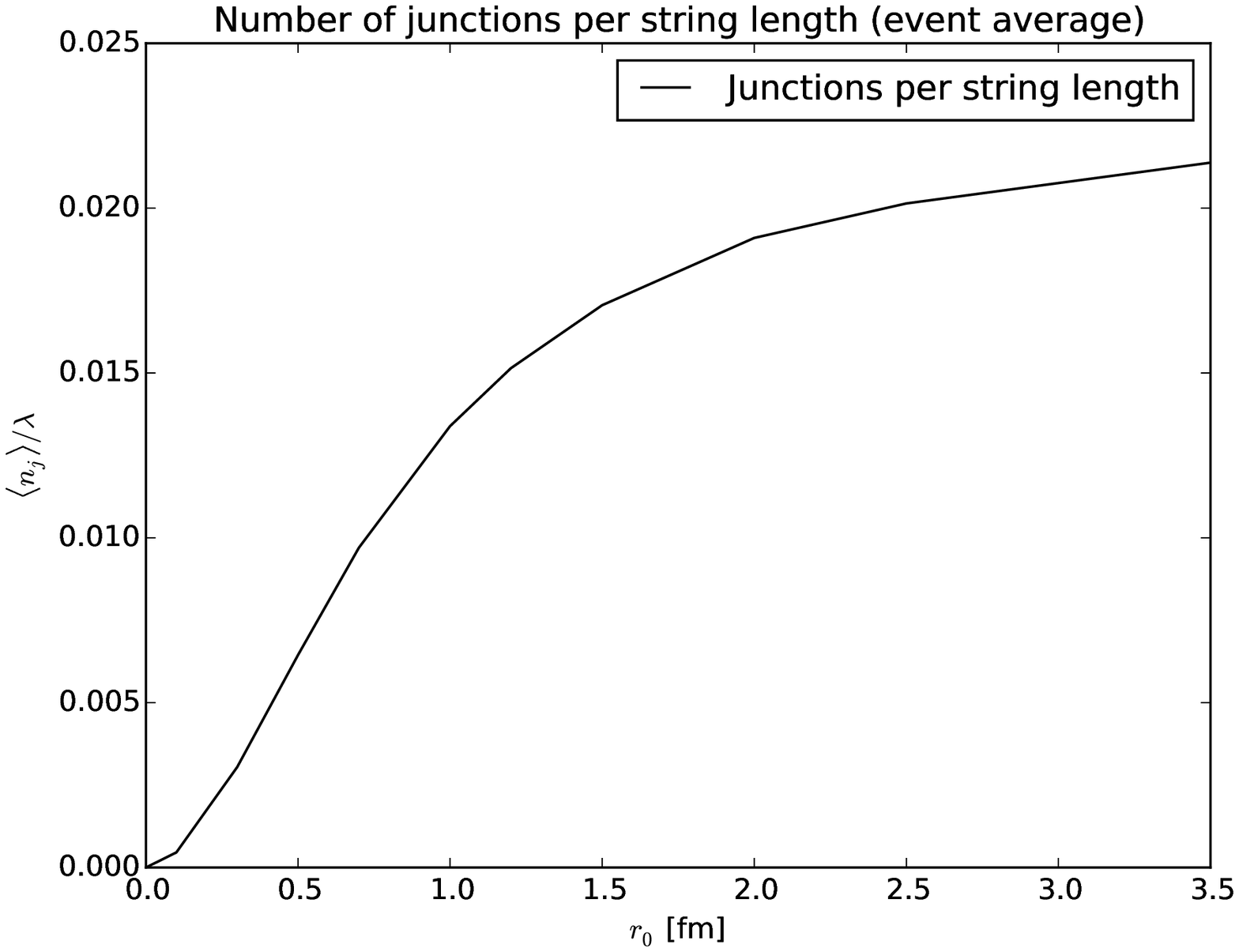}
  \caption{ \label{fig:avg-enhancement-r0} Average enhancement (left) $\langle
    h\rangle = \langle \kapeff/\kappa\rangle$ as a function of
    $r_0$  at $\sqrt{s} = 900$~GeV in $pp$ collisions. The band indicates one standard
    deviation. Number of junctions (right) per string
    length as a function of $r_0$.}
}
It is interesting to note that the overlap saturates at
$r_0\sim1.5$~fm, as the size of the strings becomes larger than the
proton. This behaviour is almost independent of collision energy, as
the cross section only increases logarithmically with energy, although
the value at saturation is higher for higher energies. The same type
of saturation effect is found in the number of junctions, shown in
\figref{fig:avg-enhancement-r0} (right). Throughout the article, $r_0$
is set to 1~fm, which is taken as a typical hadronic length scale.  We
expect variations in the parameter $r_0$ to be the largest source of
theoretical uncertainty should the model be tuned. In the region
around $r_0\sim1$~fm, small changes in $r_0$ can give up to 5 \%
change in average effective string tension, which will of course be
reflected in the results.

Finally the parameter $m_0$ serves as a characteristic scale for the
dipoles. This has both the effect of a cut-off in the rapidity span
(gluons in a dipole at rest would otherwise give infinite rapidity),
and as a propagation time ($1/m_0$), to let the gluons propagate a
finite distance (determined by their $p_\perp$) before the overlap is
calculated and hadronization takes place (see
\appref{sec:dipole-based-treatment} for further explanation).  The
average enhancement factor and the density of junctions as function of
$m_0$, is shown in \figref{fig:avg-enhancement-m0}.

\FIGURE[ht]{
  \includegraphics[width=0.45\textwidth]{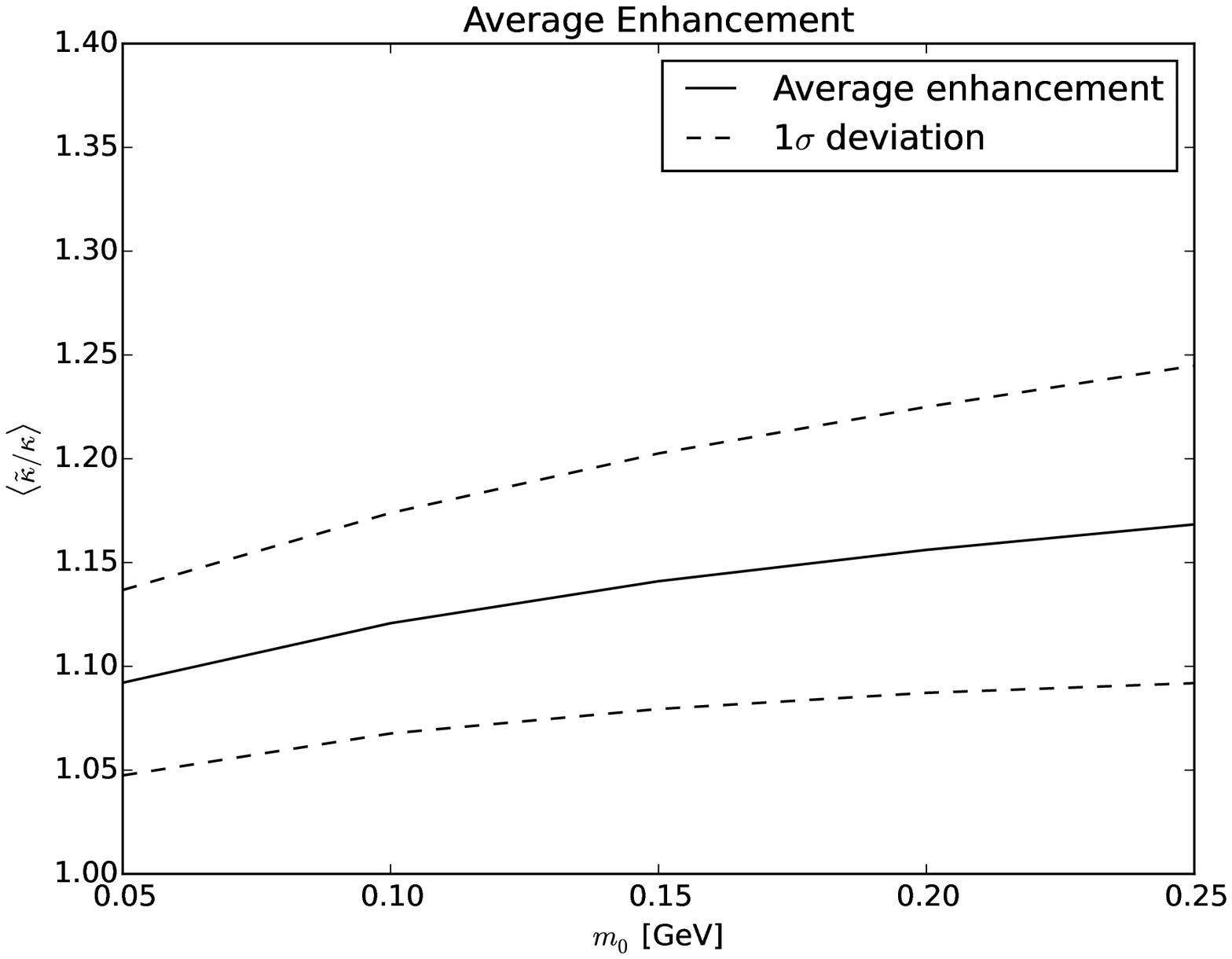}
  \includegraphics[width=0.45\textwidth]{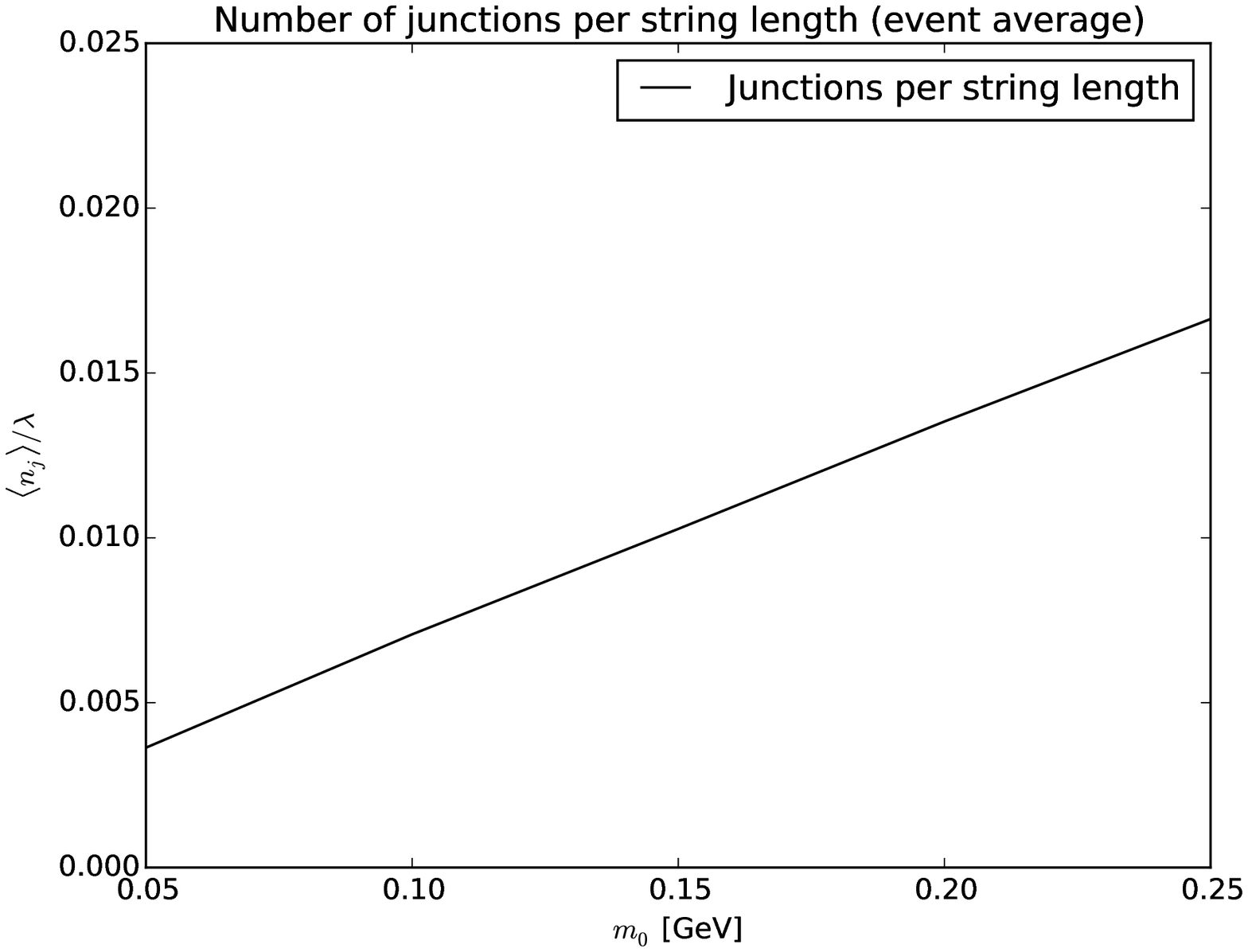}
  \caption{ \label{fig:avg-enhancement-m0} Average enhancement (left) $\langle
    h\rangle = \langle \kapeff/\kappa\rangle$ as a function of
    $m_0$  at $\sqrt{s} = 900$~GeV in $pp$ collisions. The band indicates one standard
    deviation. Number of junctions (right) per string
    length as a function of $r_0$.}
}

The model is not as sensitive to $m_0$ as to $r_0$, and we expect that the
uncertainty after a tuning, which can be ascribed to $m_0$, will be only on
the order of a few percent. In this article we have chosen to set the
parameter $m_0$ to the pion mass, $m_0 = 0.135$~GeV, as no hadron can have a
larger rapidity than the pion. The pion formation time will then be defining
for the dipole propagation time. We believe that a tuning of this parameter
will not give large deviations from the pion mass, as we have also tried
hadronic scale $1/r_0\approx0.2$~GeV. This is numerically close to the
pion mass, but does not give an equally good energy dependence.   

\subsection{Particle ratios and flow-like effects}
\label{sec:des-flow}

As seen, rope effects introduce a $\sqrt{s}$-dependence of flavour
ratios and baryon ratios in the fragmentation, and we note in particular that 
the rise at small $p_\perp$ is well described. This effect is often seen
as an indication for the formation of a quark--gluon plasma phase, also
in $pp$-collisions \cite{Werner:2014xoa,Abreu:2007kv}, as the pressure in the
hot plasma would push large mass particles to higher
$p_\perp$ (compared to low mass ones). In our rope model it is mainly
the result of the colour reconnections induced by the final state swing
mechanism, which originate from the formation of lower colour multiplets.
Ortiz \textit{et al.}\ have previously noted that the colour reconnection
model implemented in \pythia, gives rise to a flow-like 
effect in $pp$ collisions at LHC \cite{Ortiz:2013yxa}.

It is also seen that, although the rise at small $p_\perp$ is well
described, the experimentally observed fall at higher $p_\perp$, in
\eg\ $\Lambda/K^0_s$ shown in \figref{fig:pt-data}, is not reproduced
by the present implementation of rope effects. In studies of plasma
effects the hadrons with larger $p_\perp$ are expected to originate
from high-$p_\perp$ jets fragmenting outside the plasma (see \eg\
ref.~\cite{Werner:2014xoa}). Such an effect should also be expected in
the rope picture, where high-$p_\perp$ gluons are expected to
hadronize outside the region where strings interfere, illustrated in
\figref{fig:tubes}. This effect is, however, not taken into account in
the present implementation.

As a means to approximately account for the reduced rope effect for
high-$p_\perp$ jets, we have studied a modified version of the
``pipe'' implementation described in
\appref{sec:pipe-based-treatment}. Here, in case any parton in the
string, or any hadron arising from the string, obeys the criteria $|y|
< 2.0$ and $p_\perp > 4$~GeV, the string will not feel any
enhancement, but be hadronized with $\kapeff/\kappa = 1$. As this
happens in only a small fraction of the events, we believe that this
crude measure gives a qualitatively correct estimate of the effect.

\FIGURE[ht]{
  \includegraphics[width=0.45\textwidth]{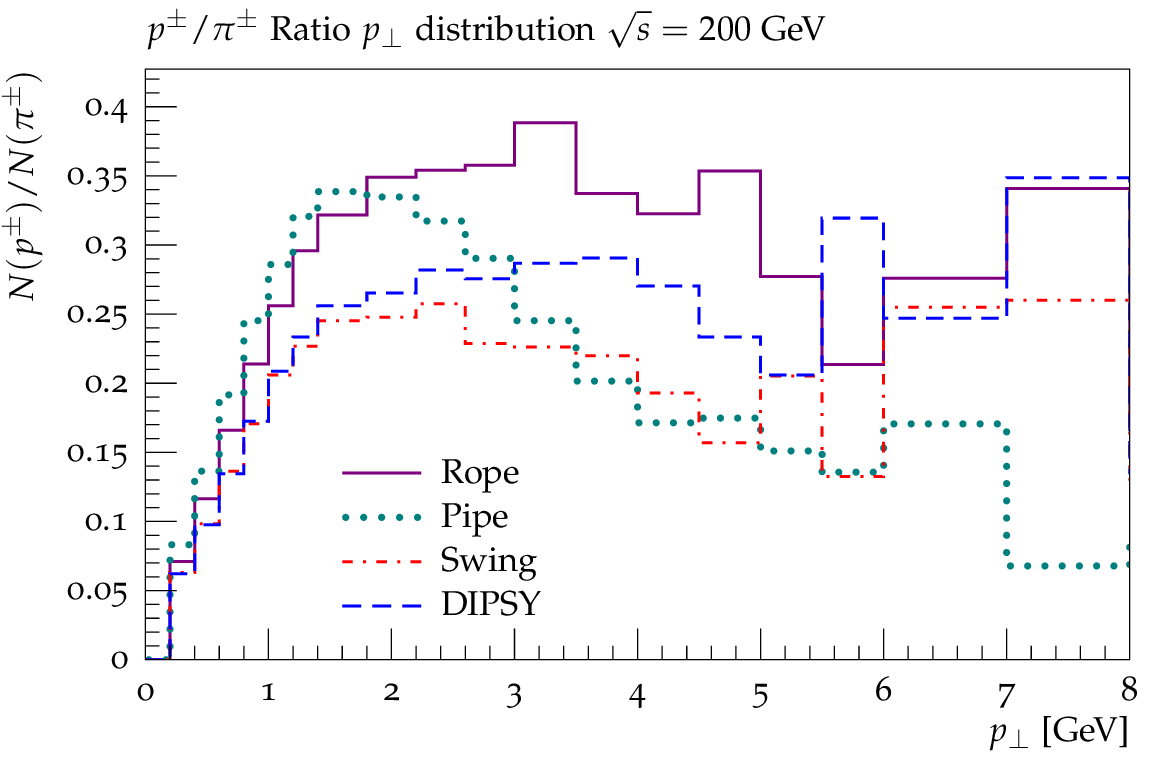}
  \includegraphics[width=0.45\textwidth]{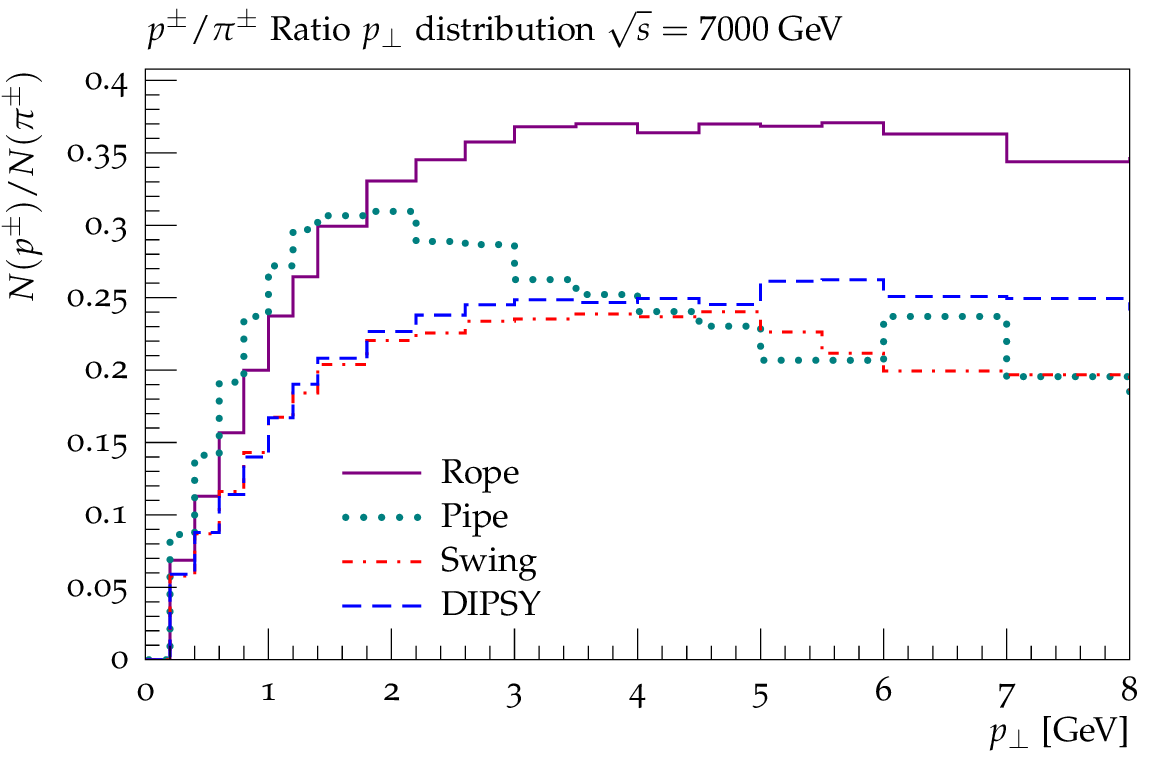}
  \caption{ \label{fig:proton-pion-pT} Proton ($p+\bar{p}$) to pion ($\pi^++\pi^-$) ratio in bins of $p_\perp$ ($|y| < 2.0$) at $\sqrt{s} = 200$~GeV (left) and 7000~GeV (right).}
}

In \figref{fig:proton-pion-pT} we show the $p^{\pm}/\pi^{\pm}$ rato at
$\sqrt{s} = 200$ and 7000~GeV. The curve marked '\dipsy' shows
simulation with no rope effects, but as the final state swing
(labelled 'Swing') is added, we already see how the high-$p_\perp$
tail of the ratio distribution (at both energies) falls off a bit.
Adding all rope effects (labelled 'Rope'), we see how the integrated
ratio increases (as shown before), but since the major effect on
$p_\perp$, in the rope model, comes from the final state swing, the
shape is not altered much. The curve labelled 'Pipe' shows the
modified pipe-based approach to estimate the overlap, as discussed
above.  We see that in this version the high-$p_\perp$ tails are more
suppressed, thus following the data better.  The 'Pipe'-curves also
show that the ratios for low to intermediate $p_\perp$ (where most of
the multiplicity is) is affected roughly as expected, even with this
very simple way of counting overlap\footnote{We remind the reader that
  not even additional junctions are added in the 'Pipe'-approach, all
  is due to changes in the $\xi$ parameter.}. The fact that even a
simple treatment like the pipe based one can catch the gist of the
rope model, is an encouraging indication that the interesting physics
lies in the model itself, and not in a more or less arbitrary choice
of how to estimate the numbers $m$ and $n$ denoting the number of
interacting strings.  We do, however, see that the
$\sqrt{s}$-dependence for the pipe based approach is not nearly as
good as the dipole based one. For this reason, we believe the dipole
approach to bear more physical sense, and in \sectrefs{sec:data-cmp}
and \sect{sec:model-behaviour}, we have thus only shown the dipole
approach.

In a future work we will address the issue of the high-$p_\perp$ tails
in ratios. A more sophisticated version of the cut applied to the pipe
based approach must be added to ensure that hadronization takes place
with local parameters suitable for the actual location of the process,
and not just let the dipoles propagate a fixed length.

\FIGURE[ht]{
  \includegraphics[width=0.45\textwidth]{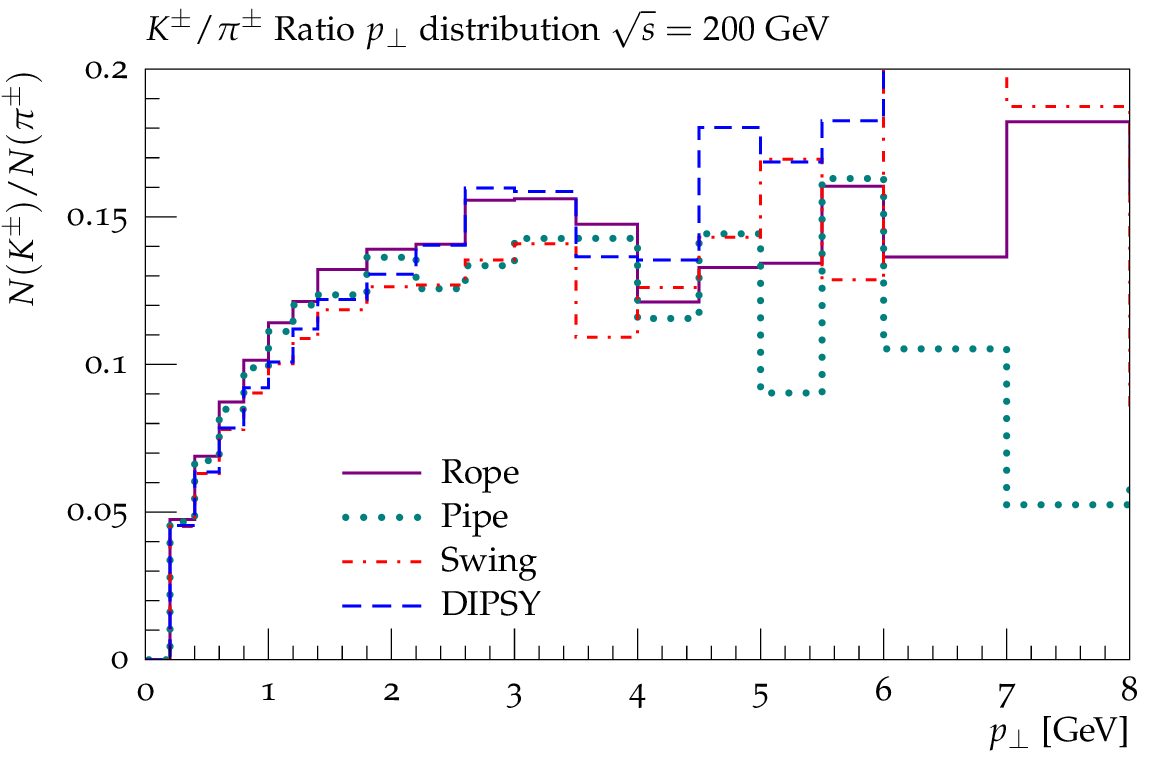}
  \includegraphics[width=0.45\textwidth]{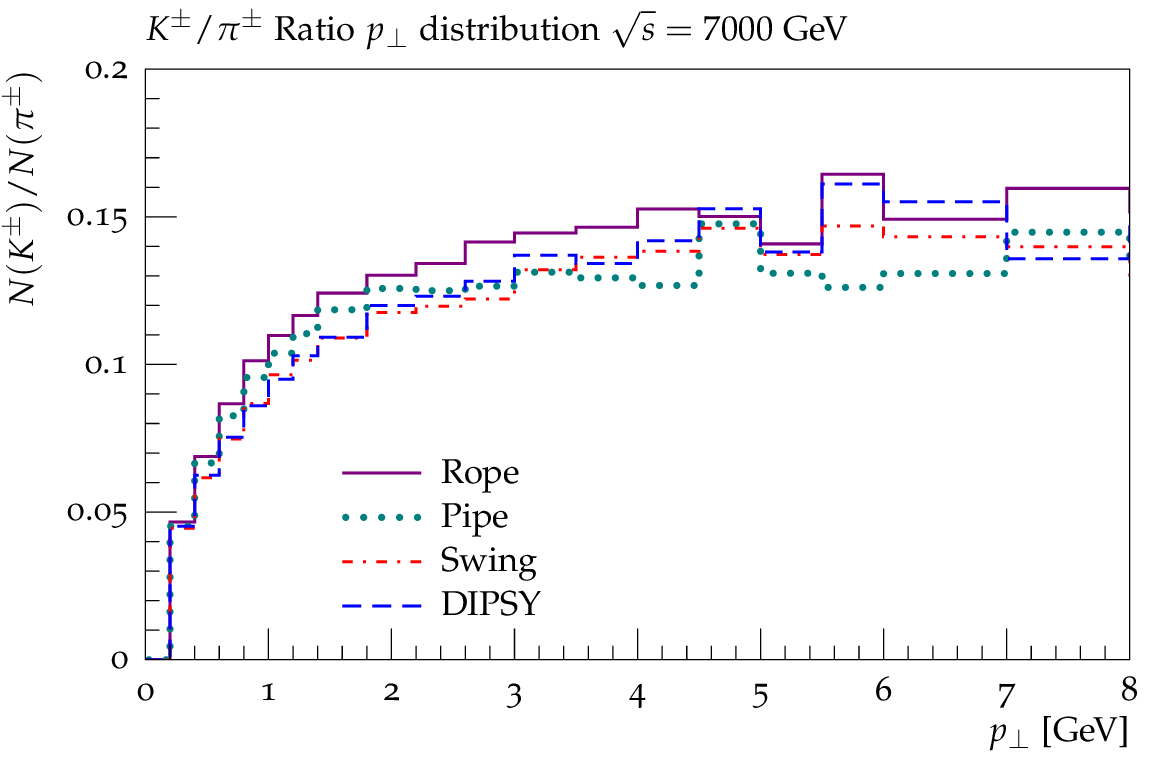}
  \caption{ \label{fig:kaon-pion-pT} Kaon ($K^++K^-$) to pion
    ($\pi^++\pi^-$) ratio in bins of $p_\perp$ at $\sqrt{s} = 200$~GeV
    (left) and 7000~GeV (right).}
}

In \figref{fig:kaon-pion-pT} we see the $K^{\pm}/\pi^{\pm}$ ratio in
bins of $p_\perp$ at $\sqrt{s} = 200$~GeV and 7000~GeV. This ratio
does not show the same type of intermediate-$p_\perp$ "bump" (also not
present in data, see \eg\ ref.~\cite{Abelev:2014laa}), but rather a
more smooth rise. The rope model (in both dipole and pipe approaches)
shows some, but not much, effect, in accordance with our expectations.

\section{Conclusions and outlook}
\label{sec:outlook}

It was early observed that string hadronization models, when tuned to
$e^+e^-$ annihilation data at LEP, underestimates the production of
strange quarks in $pp$ collisions. At the higher LHC energies the
experiments show significantly enhanced production of strangeness and
baryons, in particular strange baryons are strongly enhanced. In $pp$
collisions the strings or cluster chains are usually assumed to
hadronize independently, although the density of strings becomes quite
high at LHC energies, and interaction between the strings therefore
ought to be expected.  Interaction between strings have been discussed
by many authors in connection with nucleus collisions, where very high
string densities are also expected. Here the formation of ``ropes''
are generally predicted to give higher ratios of strange particles and
baryons.  Although the geometrical distribution of nucleons within a
nucleus can give a good estimate of the density of strings in nucleus
collisions, for a quantitative description of string interaction in
$pp$ collisions, a description of the parton distribution in impact
parameter space must be essential.

In this work we use the \dipsy model, which is a formulation of BFKL
evolution in transverse coordinate space, including NLL effects and
effects of saturation and confinement, taking also fluctuations and
correlations into account.  Within this model it is possible to
calculate the distribution of strings in the transverse plane, and
thus estimate the amount of interaction. For the actual hadronization
process we use \pytppp.

Following the early work by Biro \emph{et al.}\ \cite{Biro:1984}, we
assume that a set of strings within a limited transverse size can
interact coherently, forming a colour rope. If the strings are
stretched between random colour charges, the net charge at the end of
a rope is obtained by a random walk in colour space.  Results from
lattice calculations show that the tension in a rope is proportional
to the corresponding quadratic Casimir operator. If the rope breaks up
in a step-wise manner by the production of $q\bar{q}$ pairs, then the
number of such pairs needed to break the rope, is in general smaller
than the initial number of strings.  More energy will, however, be
released in the production of the individual pairs, thus simulating a
higher effective string tension. An important point is here that it is
the decrease in rope tension following the $q\bar{q}$ pair production,
which specifies the ``effective string tension'', and we note here
that this leads to a significantly smaller increase, compared to what
is usually assumed.

Besides higher fractions of strange particles and baryons, a higher string
tension also implies that the string breaks earlier. Early breakups usually
imply lower multiplicity, but we argue here that for rope hadronization 
this effect is compensated by
the fact that shorter string pieces are needed to form a final state hadron.

Special attention is needed for handling colour singlets, which can be
formed \eg\ when a triplet and an antitriplet combines as
$\plet{3}\otimes\antiplet{3}=\plet{8}\oplus\plet{1}$. We treat this by
colour reconnection via a ``final state swing'', described in
\sectref{sec:final-state-swing}. This idea could potentially also be
used to reconnect parallel strings into the anti-triplet in
$\plet{3}\otimes\plet{3}=\plet{6}\oplus\antiplet{3}$ (see
\eqref{eq:trip-trip}), as sketched in \figref{fig:sextetswing}. As the
hadronization model in \pytppp is currently being improved to better
handle complicated junction topologies, we expect to be able to
implement such a sextet swing mechanism in \ariadne in the near
future.

Naturally the range within which the strings can act coherently cannot
be calculated from basic principles. It ought to be of the hadronic
scale $\sim$1~fm, but might be treated as a tunable parameter. Partly
overlapping strings also give rise to uncertainties. We have here
studied two different schemes for estimating the effects of rope
formation, which both attempt to account for the actual overlap of
strings in impact parameter space and rapidity. The schemes differ in
the level of detail considered; the ``pipe-based'' scheme is only
looking at the average enhancement of the tension in a string, while
the ``dipole-based'' version estimates the increased string tension in
each individual string break-up. The dipole scheme also introduces a
simple junction model, both formation and breaking.  In spite of the
differences between the schemes, the results are fairly similar. In
both cases observables sensitive to the increased relative abundance
of baryons and strange hadrons are much better described by the rope
models, compared to conventional string fragmentation.

The fact that the model reproduces the increase both for several
collision energies and for several different hadron species, is a
strong indication that our picture of the increased string tension in
overlapping strings, and its effect on the fragmentation process, is
reasonable. Although our model introduces a couple of new parameters,
we have shown that these mainly affects the overall strength of the
effect, while the influence of the string tension on individual hadron
species is fixed by the model and by the tuning of parameters in
\pytppp to single-string data from \tee-experiments. Also the energy
dependence is fairly well constrained by the comparison to data
presented in this article, and our implementation in the \dipsy
generator can therefore make rather firm predictions, \eg\ for relevant
observables to be measured at Run~2 of the LHC.

A particularly interesting result is that the model reproduces the
increase in the ratios $p/\pi$ and $\Lambda/K$ with $p_\perp$ in the
range $p_\perp<2$~GeV, in a way mimicking a hydrodynamic transverse
flow. This effect is frequently interpreted as caused by a transition
to a quark--gluon plasma. It was also pointed out in
ref.~\cite{Ortiz:2013yxa} that colour reconnection, as implemented in
\pythia, gives rise to a flow-like effect in $pp$ collisions. Thus the
results presented in \sectref{sec:results} originate partly from the
increased tension in ropes with high colour multiplets, and partly
from colour reconnections in cases where strings combine to colour
singlets or other small multiplets.

Our model does, however, not reproduce the drop in the $p/\pi$ ratio
for $p_\perp > 2$~GeV. In analyses based on flow, it is frequently
assumed that high-$p_\perp$ particles result from fragmentation of
jets not participating in the thermalisation, and hadronizing outside
the plasma (see \eg\ ref.~\cite{Werner:2014xoa}). A similar effect
should be expected in our rope model. High-$p_\perp$ hadrons may be
predominantly formed outside the overlap regions, and therefore not
feel the increased tension in the rope. This effect is not included in
the present implementation of the model, giving the results presented
in \sectref{sec:results}. A crude modification of our pipe-based
scheme indicates that the effect may be qualitatively accounted for,
but further studies are needed of the formation times and the
transverse propagation in space within our rope model. In this context
we also need to revisit the description of high-$p_\perp$ gluons,
which currently are not well modelled in \dipsy.

We have in our analyses also neglected a possibly increased pressure
exerted by the ropes. In the bag model the pressure from a high colour
flux tends to expand the transverse size of a flux tube, in a way
which also could contribute to flow-like effects. An estimate of this
effect also needs a better understanding of the relative time-scales
for rope formation and the hadronization process.

We conclude that several mechanisms can contribute to the flow-like
behaviour in high energy collisions: besides a phase transition to a
plasma, also increased string tension in colour ropes, colour
reconnection in low colour multiplets, and transverse expansion due to
high pressure inside the ropes. To estimate the relative contributions
from these sources, it is important to study different reactions,
$pp$, $pA$, and $AA$, and also all possible observables, besides those
discussed in this paper also \eg\ angular flow, fluctuations, and
correlations of different kinds. The time-scales for the different
processes is here very important. We want to return with results of
such studies in forthcoming publications.

\section*{Acknowledgments}
\label{sec:acknowledgements}

Work supported in part by the MCnetITN FP7 Marie Curie Initial
Training Network, contract PITN-GA-2012-315877, the Swedish Research
Council (contracts 621-2012-2283 and 621-2013-4287), and contract
DE-AC05-06OR23177 under which the Jefferson Science Associates, LLC
operate the Thomas Jefferson National Accelerator Facility.

\clearpage
\appendix

\section{The \dipsy model}
\label{app:dipsy}
It has since long been clear
that a proper description of the multi-particle final states in high
energy hadron collisions requires some kind of multi-parton interaction
model. The most successful such model to date is the one developed by
Sjöstrand and van Zijl \cite{Sjostrand:1987su}, but also other models
have been proposed (see \eg\ \cite{Engel:1995yda} and
\cite{Butterworth:1996zw}).

For the purpose of our investigations, however, it is important that
not only the momentum distribution of the produced partons is
described; to estimate the degree to which strings overlap we also
need to understand the impact-parameter distribution of partons. For
this reason we have used \dipsy event generator
\cite{Flensburg:2011kk}, which will be described briefly in this
appendix.

\dipsy is based on Mueller's dipole cascade model
\cite{Mueller:1993rr,Mueller:1994jq,Mueller:1994gb}, which is a
formulation of leading-log BFKL
evolution \cite{Kuraev:1977fs,Balitsky:1978ic} in transverse coordinate
space. This model relies on the fact that initial-state radiation from
a colour charge (quark or a gluon) in a hadron is screened at large
transverse distances by an accompanying anticharge, and that gluon
emissions therefore can be described in terms of colour-dipole
radiation. Thus the partonic state is described in terms of dipoles in
impact-parameter space and rapidity, which is evolved in rapidity when
an emitted gluon splits a dipole into two.  We here note that the
suppression of large dipoles in transverse coordinate space is
equivalent to the suppression of small $k_\perp$ in the conventional
BFKL evolution in momentum space.

For a dipole with charges at the transverse points $\pmb{x}_1$ and $\pmb{x}_2$,
the probability to emit a gluon at $\pmb{x}_g$  is given by
\begin{equation}
  \label{eq:dipoleradiation}
  \frac{d\mathcal{P}_g}{dY}=\frac{\bar{\alpha}}{2\pi}d^2\pmb{x}_g
\frac{(\pmb{x}_1-\pmb{x}_2)^2}{(\pmb{x}_1-\pmb{x}_g)^2 (\pmb{x}_g-\pmb{x}_2)^2}\,,
\,\,\,\,\,\,\, \mathrm{with}\,\,\, \bar{\alpha} = \frac{N_c\alpha_s}{\pi}.
\end{equation}
The emission produces two new dipoles, $(\pmb{x}_1,\pmb{x}_g)$ and
$(\pmb{x}_g,\pmb{x}_2)$, which can split independently by further
gluon emissions. Repeated emissions form a cascade, with dipoles
connected in a chain. When two cascades collide, a dipole
$(\pmb{x}_1,\pmb{x}_2)$ in a right-moving cascade can interact with a
left-moving dipole $(\pmb{x}_3,\pmb{x}_4)$, with probability
\begin{equation}
  \label{eq:dipoleinteraction}
P=\frac{\alpha_s^2}{4}\left[\ln\left(
\frac{(\pmb{x}_1-\pmb{x}_3)^2(\pmb{x}_2-\pmb{x}_4)^2}
{(\pmb{x}_1-\pmb{x}_4)^2(\pmb{x}_2-\pmb{x}_3)^2}\right)\right]^2.
\end{equation}

In a series of papers
\cite{Avsar:2005iz,Avsar:2006jy,Avsar:2007xg,Flensburg:2011kk} a
generalization of Mueller's model, implemented in the Monte Carlo
event generator \dipsy, has been described in detail. Here we will
only discuss the main points.  The basic idea behind the model is to
include important non-leading effects in the BFKL evolution,
saturation effects in the evolution, and confinement.

The full next-to-leading logarithmic corrections have been calculated
and have been found to be very large
\cite{Fadin:1998py,Ciafaloni:1998gs}.  A physical interpretation of
these corrections has been presented by Salam \cite{Salam:1999cn}, and
a dominant part is related to energy--momentum conservation. In the
\dipsy model this is achieved by equating the emission of a gluon at
small transverse distances with high transverse momenta of the emitted
and recoiling gluons. In this way the gluons emitted in the evolution
are traced in both momentum and coordinate space, allowing us to
generate the final-state momentum distribution of gluons. The
conservation of energy and momentum implies a dynamic cutoff for very
small gluons with correspondingly high transverse momenta. This
constraint has also important computational advantages; in the
standard Mueller model the number of small gluons diverges which,
although the cross section is still finite, gives computational
problems.

Other important non-leading effects are the running coupling,
$\alpha_s(p_\perp^2)$, and the ``energy scale terms'' (which correspond
to the consistency constraint discussed by Kwiecinski et
al. \cite{Kwiecinski:1996td}). The latter implies that the emissions
are ordered in both the positive and negative light-cone components
\cite{Andersson:1996ju}. Besides these perturbative corrections,
confinement effects are included via a small gluon mass, $m_g$, and
non-linear saturation effects through the so-called swing mechanism,
described in more detail below.

In a high energy collision, two hadrons are evolved from their
respective rest frames to a Lorentz frame in which they collide.  In
its own rest system a proton is currently modelled by a simple triangle
of gluons connected by dipoles, and the gluonic Fock state is built by
successive dipole emissions of virtual gluons. (The small-$x$ gluons
are rather insensitive to the initial parton configuration, apart from
the overall size, and valence quarks are later introduced by hand in
the final state). The two evolved systems are then made to collide,
allowing some of the dipoles in the left-moving system to interact
with some in the right-moving ones.  This enables the gluons in these
dipoles to come on-shell, together with all parent dipoles, while
non-interacting dipoles must be regarded as virtual and thus be
reabsorbed.

\subsection{The initial-state Swing mechanism}
\label{app:initial-swing}
The swing mechanism in \dipsy is a saturation effect within the
evolution, which is conceptually interesting in connection with the
rope formalism in this article.  Mueller's dipole evolution is derived
in the large $N_c$ limit, where each colour charge is uniquely
connected to an anticharge within a dipole. Saturation effects are
here included as a result of multiple dipole collisions, in the frame
used for the analysis. Such multiple interactions give dipole chains
forming loops, and are related to multiple pomeron exchange. Loops
formed within the evolution are, however, not included.\footnote{This
  is also the case \eg\ for the non-linear BK equation
  \cite{Balitsky:1995ub,Kovchegov:1999yj}, which describes the
  interaction between a relatively dilute cascade and a dense target.}
Besides missing parts of the saturation, this also makes the result
dependent on the frame used for the calculation. As dipole interaction
in \eqref{eq:dipoleinteraction} is colour suppressed compared to the
dipole splitting, loop formation is related to the possibility that
two dipoles have identical colours.

If two dipoles happen to have the same colour, we have actually a
colour quadrupole, where a colour charge is effectively screened by
the nearest anticolour charge.  Thus approximating the field by a sum
of two dipoles, one should preferentially combine a colour charge with
a nearby anticharge.  This interference effect is taken into account
in \dipsy in an approximate way, by allowing two dipoles with the same
colour to recouple forming the new dipoles, in a way that favours small
dipoles.

In the simulation this is handled by assigning all dipoles a colour
index running from 1 to $N_c^2$, not allowing two dipoles connected to
the same gluon to have the same index.  A pair of two dipoles,
$(\pmb{x}_1,\pmb{x}_2)$ and $(\pmb{x}_3,\pmb{x}_4)$, with the same
colour may be better approximated by the combination
$(\pmb{x}_1,\pmb{x}_4)$ and $(\pmb{x}_3,\pmb{x}_2)$, if these dipoles
are smaller.  In the evolution the pair is allowed to ``swing'' back
and forth between the two possible configurations as indicated in
\figref{fig:initial-swing}. The swing mechanism is adjusted to give the
relative probabilities
\begin{equation}
\frac{P_{(12)(34)}}{P_{(14)(32)}}=\frac{(\pmb{x}_1-\pmb{x}_4)^2(\pmb{x}_3-\pmb{x}_2)^2}{(\pmb{x}_1-\pmb{x}_2)^2(\pmb{x}_3-\pmb{x}_4)^2},
\end{equation}
thus favouring the
configuration with smallest dipoles. (In the implementation of the 
cascade evolution, the
swing is competing with the gluon emission in
\eqref{eq:dipoleradiation}, where a Sudakov-veto algorithm can be
used to choose which of the two happens next.)

The dipole interaction in \eqref{eq:dipoleinteraction} is smaller for
smaller dipoles, which reproduces the colour transparency effect.
As the swing leads to smaller average dipole size, the probability for
interactions is reduced, and thus the swing represents 
a saturation effect within the evolution. This reduced interaction probability
is equivalent to the
$2\rightarrow 1$ and $2\rightarrow 0$ vertices in \eg\ the BK evolution
equation.

\section{Colour algebra}
\label{sec:color-algebra}

\subsection{Calculation of $p$ and $q$}
\label{sec:pq-calc}
In this section the recursion relations presented in
\eqref{eq:randomaddition} for calculating all possible $\{p,q\}$
multiplets arising from the combination of $m$ triplets and $n$
antitriplets will be presented in detail.\par
It is worth noting that the combination of any SU(3) multiplets can be
carried out using Young tableaux (just as the more familiar case of
SU(2)). In the notation of this article, an SU(3) multiplet is denoted
by the quantum numbers $p$ and $q$, which can be directly a Young
tableaux, as $\{1,0\} = \tiny{\Yvcentermath1 \yng(1)}$ and $\{0,1\} =
\tiny{\Yvcentermath1 \yng(1,1)}$, and so the number of boxes in the
top row is $p + q$. and the number of boxes in the bottom row is
$q$.\par
Now the usual rules of manipulating Young tableaux can be used to
review the simple cases of combining a single triplet with an
(anti-)triplet:
\begin{eqnarray}
\label{eq:trip-trip}
\{1,0\} \otimes \{1,0\} = \tiny{\Yvcentermath1 \yng(1)} \otimes \tiny{\Yvcentermath1 \yng(1)} = \tiny{\Yvcentermath1 \yng(2)} \oplus \tiny{\Yvcentermath1 \yng(1,1)} = \{2,0\} \oplus \{0,1\}, \\
\label{eq:trip-atrip}
\{1,0\} \otimes \{0,1\} = \tiny{\Yvcentermath1 \yng(1)} \otimes \tiny{\Yvcentermath1 \yng(1,1)} = \tiny{\Yvcentermath1 \yng(2,1)} \oplus I = \{1,1\} \oplus \{0,0\},
\end{eqnarray}
where $I = \{0,0\}$ denotes a singlet. Note that symmetry ensures that
$\{1,0\} \otimes \{1,0\} = \{0,1\} \otimes \{0,1\}$. Physically
\eqref{eq:trip-trip} corresponds to the situation where two colour
strings, with colour flow in the same direction, merge to a rope. In
the case where the colour pairs are equal, the resulting rope will be
a sextet ($\{2,0\}$), and in all other cases an antitriplet
($\{0,1\}$). \Eqref{eq:trip-atrip} corresponds to two strings with
opposite colour flow merging. This can either result in a gluon-like
octet rope (\{1.1\}) or no colour flow at all, in the singlet case.\par
The recursion relations of \eqref{eq:randomaddition} can be derived
using a similar procedure. Adding a single triplet to an existing
multiplet $\{p,q\}$ gives \eqref{eq:randomaddition} directly,
as:
\begin{eqnarray}
\{p,q\} \otimes \{1,0\} = \underbrace{\tiny{\Yvcentermath1\yng(2,2)} ... \tiny{\Yvcentermath1\yng(1,1)}}_{q} \shiftleft{0.5ex}{\raisebox{0.75ex}{$\underbrace{\tiny{\Yvcentermath1\yng(2)} ... \tiny{\Yvcentermath1 \yng(1)}}_{p}$}} \shiftright{6.5ex}{$ \otimes \tiny{\Yvcentermath1\yng(1)} = $} \\ \nonumber
 \{p,q-1\} \oplus \{p-1,q+1\} \oplus \{p+1,q\}.
\label{eq:rec1}
\end{eqnarray}
Combining the general $\{p,q\}$ multiplet with an antitriplet proves
directly symmetry is ensured as:
\begin{eqnarray}
\{p,q\} \otimes \{0,1\} &=& \underbrace{\tiny{\Yvcentermath1\yng(2,2)} ... \tiny{\Yvcentermath1\yng(1,1)}}_{q} \shiftleft{0.5ex}{\raisebox{0.75ex}{$\underbrace{\tiny{\Yvcentermath1\yng(2)} ... \tiny{\Yvcentermath1 \yng(1)}}_{p}$}} \shiftright{6.5ex}{$ \otimes \tiny{\Yvcentermath1\yng(1,1)} $}\\ \nonumber
&=&\{p-1,q\} \oplus \{p,q+1\} \oplus \{p+1,q-1\}.
\label{eq:rec2}
\end{eqnarray}
Using the stated recursion relations, a random walk in colour space is
simulated, as in ref.~\cite{Biro:1984} by starting from a singlet $I =
\{0,0\}$. The resultant relation between $\langle p+q \rangle$ and
$m+n$ is shown in \figref{fig:ranwalk}.\par
In the same way the recursion relation for all multiplets arising from
combining the general $\{p,q\}$ multiplet with an octet can be
defined. The proof follows in a straightforward way from the same
considerations as above, and only the result is stated here:

\begin{eqnarray}
\{p,q\} \otimes \{1,1\} = \{p-1,q-1\} \oplus \{p+1,q+1\} \oplus 2 \cdot \{p,q\}  \oplus  \{p-1,q+2\}\\ \nonumber
  \oplus  \{p+1,q-2\} \oplus  \{p-2,q+1\} \oplus  \{p+2,q-1\}.
\end{eqnarray}

\subsection{An illustrative example}
\label{sec:example}
As a simple illustrative example we can look at a rope spanned between
three quarks with random colours in one end, matched by three
antiquarks in the other end, illustrated in
\figref{fig:addtriplets}. The first quark is a triplet, \plet{3},
denoted \{1,0\}, and depicted to the left in
\figref{fig:addtriplets}. The addition of a second quark can give a
\plet{6} (\{2,0\}) or an \antiplet{3} (\{0,1\}), as shown in the
central column of \figref{fig:addtriplets}, with probabilities 2/3 and
1/3 respectively. Adding the third quark to the sextet can give
\plet{10} (\{3,0\}) or \plet{8} (\{1,1\}), while adding it to the
\antiplet{3} gives \plet{8} (\{1,1\}) or a singlet (\{0,0\}), as shown
in the rightmost column in \figref{fig:addtriplets}. The result is
therefore a decuplet, two octets, and a singlet, with probabilities
proportional to their respective multiplicities (\ie\ 10/27, 8/27,
8/27, and 1/27).

\FIGURE[ht]{
\includegraphics[width=0.7\textwidth]{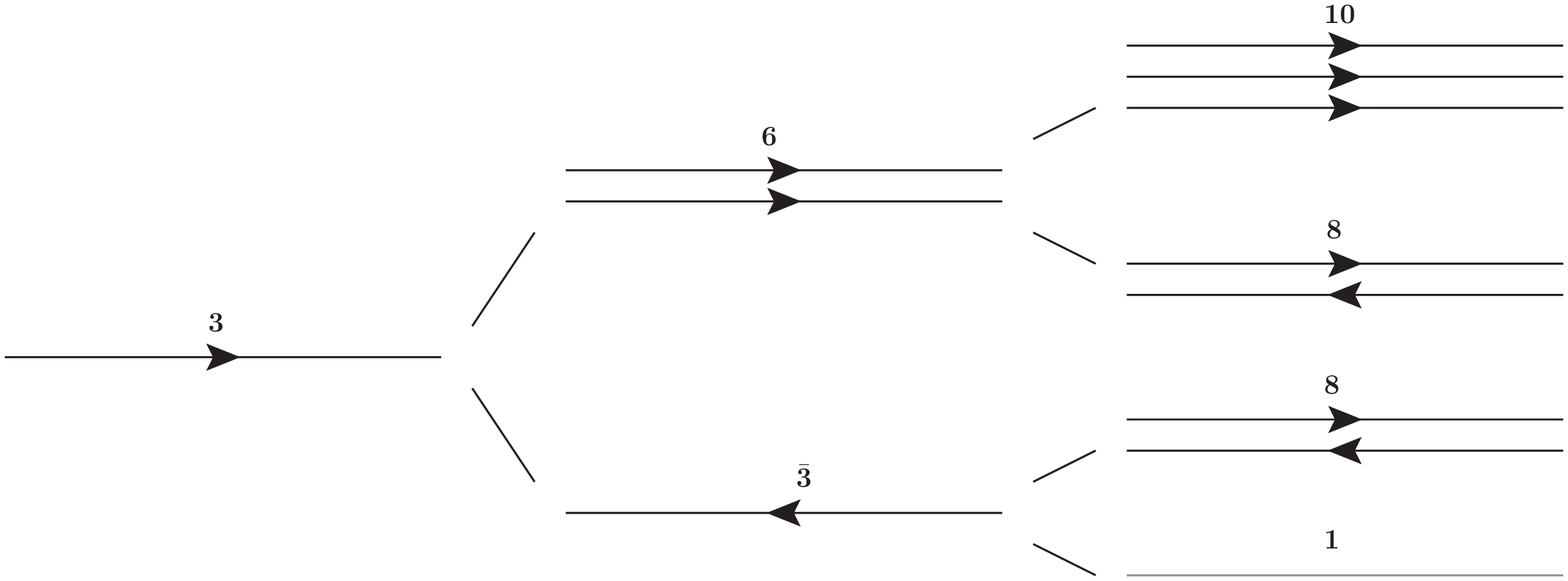}
\caption{ \label{fig:addtriplets} Illustration of the addition of
  triplets to an initial triplets, reads from left to right. By
  combining the initial \plet{3} in the left column with another
  \plet{3}, one gets either \plet{6} or \antiplet{3} and so on (see
  text).}
}

For the fragmentation of the rope, we find first that in the case of
the singlet, there is no colour field stretched. For the decuplet and
the octets the relative rope tension is 9/2 and 9/4 respectively. The
decuplet can fragment in three steps giving successively a sextet and
a triplet, before the final breakup. The relative effective string
tension, $\kapeff/\kappa$, in these steps are $9/2-5/2=2$,
$5/2-1=3/2$, and 1 respectively. Thus the first breakup will give the
highest effective string tension, and correspondingly higher
$s/u$-ratio and higher $p_\perp$. An initial octet will similarly
break in two steps via a triplet, with $\kapeff/\kappa$ equal
to $9/4-1=5/4$, followed by the break of the triplet with the tension
$\kappa$ for a single string.  If the $\{p,q\}$ multiplet is in the
left end of the rope, the antiquark is pulled to the left when $p$ is
reduced, and pulled to the right when $q$ is reduced.

\section{Detailed description of the rope models}
\label{sec:deta-descr-rope}

The models for rope formation and fragmentation presented in this
paper are similar in spirit, but as always when implementing models in
a Monte Carlo code, there are a number of choices to be made and
different levels of detail that can be chosen. One of our models is
fairly crude, using the average overlap for individual strings in an
event, while the other is very detailed in the treatment of the individual
string break-ups.

In common for the two models is that, for technical reasons, the
fragmentation of a rope is done one individual string at the time,
emulating the rope effects by modifying the parameters in the string
fragmentation implementation of \pytppp, thus taking into account the
effective string tension of the rope. We have concentrated on a
selection of parameters which should be particularly sensitive to rope
effects:
\begin{itemize}
\item $\rho$: the suppression of $s$ quark production relative to $u$
  or $d$ type production.
\item $\xi$: the suppression of diquark production relative to quark
  production, meaning\\ \emph{(all diquarks)/(all quarks)}.
\item $x$: the suppression of diquarks with strange quark content
  relative to diquarks without strange quarks (in addition to the
  factor $\rho$ for each extra $s$-quark).
\item $y$: the suppression of spin 1 diquarks relative to spin 0
  diquarks (not counting a factor three due to the number of spin
  states of spin 1 diquarks).
\item $\sigma$: the width of the transverse momentum distribution in
  string break-ups.
\end{itemize}
Of these we assume that $\rho$, $x$ and $y$ are directly related to
mass effects in the tunneling mechanism in \eqref{eq:schwinger2}, such
that if the modification of the string tension be given by a simple
scaling with an enhancement factor $h$, such that $\kappa \mapsto
\kapeff = h\kappa$, we obtain the following rescalings:
\begin{eqnarray}
	\rho &\mapsto& \tilde{\rho} = \rho^{1/h},\nonumber \\
	x &\mapsto& \tilde{x} = x^{1/h},\nonumber \\
	y &\mapsto& \tilde{y} = y^{1/h}.
\label{eq:hscaling}
\end{eqnarray}
Also the scaling of $\sigma$ is quite straight forward and is simply
given by $\sigma \mapsto \tilde{\sigma} = \sigma\sqrt{h}$.

The treatment of the $\xi$ parameter is somewhat more involved as it
gives a global probability of having a diquark break-up relative to a
simple quark break-up, which means it cannot be simply related to the
tunneling mechanism. Looking at the relation between the individual
probabilities for different quarks and diquarks, they are determined
by the relations: ${\cal P}_s=\rho{\cal P}_u$, ${\cal
  P}_{ud_1}=3y{\cal P}_{ud_0}$ and ${\cal P}_{us_1}=x\rho{\cal
  P}_{ud_1}$, {\etc} The total probability for diquark production
relative to quark production, can therefore be expressed in terms of
the ratio ${\cal P}_{ud_0}/{\cal P}_u.$
\begin{equation}
  \label{eq:xidefinition2}
  \xi\equiv\frac{\sum_{qq_s}{\cal P}_{qq_s}}{\sum_q{\cal P}_q}=\frac{1 +
    2x\rho + 9y + 6x\rho y + 3yx^2\rho^2}{2+\rho} 
  \frac{{\cal P}_{ud_0}}{{\cal P}_u}
  \equiv\alpha\frac{{\cal P}_{ud_0}}{{\cal P}_u},
\end{equation}
(where we have assumed that $u$ and $d$ quarks are equivalent).

However, it is not the case that ${\cal P}_{ud_0}/{\cal P}_u$ is
affected by the string tension in the same way as the
$\rho$-parameter. According to the popcorn mechanism described in
\sectref{sec:baryonprod} and \figref{fig:popcorn}, there is a two step
procedure where first a $q\bar{q}$ pair is produced as a fluctuation
that does not break the string, and then another pair is produced
which actually does, allowing a diquark--antidiquark pair to tunnel
out. We will therefore assume that the ratio is composed of two
factors, one is related to the probability to have a $q\bar{q}$
fluctuation in the first place, and one related to the differences in
masses. We will call these $\beta$ and $\gamma$ respectively, where we
assume that $\beta$ is independent of the string tension, while
$\gamma$ transforms as $\rho$.

Hence we have $\xi=\alpha\beta\gamma$ where
\begin{equation}
  \alpha \mapsto \tilde{\alpha}= 
  \frac{1 + 2\tilde{x}\tilde{\rho} + 9\tilde{y} + 6\tilde{x}\tilde{\rho}\tilde{y}
    + 3\tilde{y}\tilde{x}^2\tilde{\rho}^2}
 {2+\tilde{\rho}},
\label{eq:hscalingalpha}
\end{equation}
and
\begin{equation}
\gamma \mapsto \tilde{\gamma} = \gamma^{1/h},
\label{eq:hscaling2}
\end{equation}
and the total effect on $\xi$ from a modified string tension is given by
\begin{equation}
  \label{eq:hxiscaling}
  \xi=\alpha\beta\gamma \mapsto \tilde{\xi} =
  \tilde{\alpha}\beta\left(\frac{\xi}{\alpha\beta}\right)^{1/h}.
\end{equation}

As explained in \sectref{sec:eff-on-mult}, also the parameters $a$ and
$b$ are indirectly affected by a modified string tension. One could
also expect other parameters to be affected, but the ones presented
here are the the most important ones.

\subsection{A pipe-based treatment}
\label{sec:pipe-based-treatment}
In the crude, pipe-based approach, we expect all flux tubes to be directed
dominantly along the rapidity axis.
For a string stretched from parton $(\vec{b_0},y_0)$, via the gluons
$(\vec{b_i},y_i)$, and ending at $(\vec{b_k},y_k)$, the volume of the
corresponding flux tube in (\emph{transverse coord., rapidity}) space is thus
given by
\begin{equation}
  V_{string} = \sum_{i=1}^{k} \pi r_0^2 |y_i - y_{i-1}|.
\label{eq:vstring}
\end{equation}
As the string can go back and forth in rapidity, the separations in
rapidity enters with its absolute value.  To estimate the amount of
overlap between two strings, we must take into account that the string
tubes are not parallel to the rapidity axis, but go up and down in
transverse space. This is approximated by replacing the winding flux
tube by a wider straight pipe, with a correspondingly lower
density. The pipe is parallel to the rapidity axis and stretched
between the partons with the smallest and the largest rapidity,
$y_{min}$ and $y_{max}$. Its center is taken at
$\vec{b}_{cent}=(\vec{b}_{min}+\vec{b}_{max})/2$, and its radius,
$r_{pipe}$, is increased to enclose the major part of a wiggling
string:
\begin{equation}
  r_{pipe}^2 = r_0^2 + \frac{1}{k} \sum_{i=0}^{k} (\vec{b}_i -
   \vec{b}_{cent})^2.
\label{eq:rpipe}
\end{equation}
The volume of the pipe is thus given by $V_{pipe}=\pi
r_{pipe}^2(y_{max}-y_{min})$. Note that since the string can go back and
forth, the ratio $d = V_{string}/V_{pipe}$ can be larger than one. This number
is the relative field density in the pipe, and in terms of overlap, it
represents the string's ``self-overlap''. It is, however, important to keep track of
the direction of the overlap. The relative field density in each pipe is thus assigned both an $m$ and an $n$ component, defined by the sign of its projection on the rapidity axis. For each pipe $i$, we thus have a relative field density with two components $d_{i,n}$ and $d_{i,m}$. To estimate the total $\{m,n\}$ of each string, we must therefore sum over the two components separately, weighting with the geometrical overlap of the pipes, such that:
\begin{equation}
 m = \sum_{i = 0}^N C_id_{i,m} \mathrm{\ \ \ \ and\ \ \ \ }  n = \sum_{i=0}^N C_id_{i,n},
\end{equation}
where $C_i$ is the geometrical overlap with pipe $i$, there are $N$ pipes in
an event. (Note that the geometrical overlap of an object with itself will
always be 1.) These numbers are rounded 
off to integers $m$ and $n$ corresponding to the number of interfering
colour charges and anticharges in the rope. 
To find the relevant colour multiplet $\{p,q\}$ for the rope, we add
$m$ triplets and $n$ antitriplets (parallel and anti-parallel strings)
with random colours, as described in \sectref{sec:ropeformation} and
\appref{sec:pq-calc}.  The $n+m$ strings in the rope should then
fragment in a sequential way in $p+q$ steps, with a gradually
decreasing effective tension.  This is technically difficult to
implement using the \pythia implementation of the Lund fragmentation
model.  In this first study we therefore make an approximation, where
we use the average value for $\kapeff$, given by $\kaprope/(p+q)$ with
$\kaprope$ determined by \eqref{eq:scaledcasimir}. Thus the
enhancement factor $h$ becomes:
\begin{equation}
h = \frac{\langle \kapeff\rangle}{\kappa}=
  \frac{p^2+pq+q^2+3p+3q}{4(p+q)}.
\label{eq:kappaeff_long}
\end{equation}
The averaging described here will not properly take into account the
situation where two triplets become an anti-triplet instead of a
sextet. (Note that the situation where a triplet and an antitriplet
form a singlet is taken care of by the swing described in
\sectref{sec:final-state-swing}.) To account for this, we throw away
strings in a multiplet with probability $1 - \frac{p+q}{m+n}$. A
string that is discarded in this way, is simply not hadronized, and
will not appear in the final state.  Removing strings in this way will
have an effect on total multiplicity, which can largely be tuned away,
but more seriously, it will break energy--momentum conservation. We
have therefore devised a more elaborate scheme in the dipole-based
treatment, which will be presented next.

\subsection{A dipole-based treatment}
\label{sec:dipole-based-treatment}

After the final-state shower in \ariadne, the string can be seen as a
chain of dipoles connected by gluons, and in the string fragmentation
in \pytppp the break-ups of the string basically follow these dipoles
in momentum space. The dipoles, together with their respective
overlaps, are thus the basic structures considered in the more
sophisticated dipole-based treatment. We study one dipole at a time in
its own rest frame, with its two partons along the $z$-axis. All other
dipoles in the event are boosted to the same frame before calculating
their overlap with the dipole under study, using the rapidity span and
transverse distance of each of them. As gluons are massless, the rapidity span
of a dipole can in principle become infinite, and we therefore use a
small gluon mass, $m_0\propto 1/r_0\sim0.2$~GeV, to limit the
rapidities. To allow for a finite formation time of the string pieces
between the partons in a dipole, we let the partons propagate in space
a fixed time before calculating the transverse distances.\footnote{The
  propagation time should be of the order of $r_0$, and the effect of
  this propagation turns out to be very similar to simply reducing the
  transverse thickness of the string pieces.} For a pair of dipoles, we
can now make a linear interpolation between the transverse positions
of their respective partons, and we can thus calculate their overlap
as the region in rapidity where the two string pieces are closer than
$r_0$.

Just as in the pipe-based procedure, the colour charges in the dipoles
are assumed to be random, although they have a definite direction. We
therefore calculate separately the summed overlap of parallel and
antiparallel dipoles as
\begin{equation}
  \label{eq:dipole-nm}
  m_i=\sum_{j_+\ne i}\frac{\delta y_{i(j_+)}}{\Delta y_i}\qquad\mbox{and}\qquad
  n_i=\sum_{j_-\ne i}\frac{\delta y_{i(j_-)}}{\Delta y_i}.
\end{equation}
With these overlaps (rounded to integers) we now perform a random walk
in colour space to arrive at a multiplet $(p_i,q_i)$ for the
dipole. The random walk is, however, somewhat restricted due to the
final-state swing mechanism. If, \eg, we find that $m=0$ and $n=1$ and add a
triplet, we
only allow the step $\{1,0\}\to\{1,1\}$ in colour space, since
the swing is assumed to have taken care of the step
$\{1,0\}\to\{0,0\}$ already.

If we had a proper procedure for the junction swing in
\figref{fig:sextetswing} we could have limited the random walk
further, \eg\ with $m=1$ and $n=0$ we would only allow $\{1,0\}\to\{2,0\}$
and not $\{1,0\}\to\{0,1\}$, and we would always end up in the highest
possible multiplet $\{m,n\}$.  However, since the current version of
\pytppp only can handle a limited number of junctions, we have to allow
such steps and will end up with dipoles with $p_i<m$ or $q_i<n$, which
then corresponds to a partial attenuation of the colour field of the
dipole by other nearby dipoles.

If we consider two completely overlapping dipoles which are in the
multiplet $\{0,1\}$ this would correspond to the right hand side of
\figref{fig:sextetswing}, where we basically only have one string
piece to be hadronized. To avoid junctions, we therefore break one of
the dipoles by replacing the two gluons with a diquark and
antidiquark. In this way we should get approximately the right
multiplicity from the string piece that is left (although its colour
flow is reversed compared to a proper junction treatment) and also get
the same number of (anti-)baryons that would otherwise have come from
the junctions.
However, it should be noted that two connected junctions will not
necessarily result in a baryon--anti-baryon pair. As we have seen in
\sectref{sec:modifiedtension}, the popcorn model for baryon
production assumes that $q\bar{q}$ fluctuations that do not break the
string occur fairly frequently, allowing additional fluctuations to
tunnel out as diquark--antidiquark pairs by locally reversing the
colour flow. In the case of a string piece connecting two nearby
junctions, these fluctuations may actually again turn the colour flow
around, and the $\bar{q}$ ($q$) from the fluctuation may very well
travel along the string and combine with one of the (anti-)quarks in
the nearby junction. Note that the probability for this to happen is
higher than the probability for the corresponding diquark--antidiquark breakup of
a string, since it does not involve the tunneling probability for the
heavier diquarks in \eqref{eq:schwinger2}. Thus the probability of
having a fluctuation, which prevents a baryon--antibaryon pair to
result from two connected junctions, should only be governed by the
$\beta$-parameter in \eqref{eq:hxiscaling}. Therefore, we will not
always break the dipole by introducing diquarks, but with a
probability $\beta$ we will instead use quarks.

This procedure is generalized, so that if a given dipole has overlaps
$m_i$ and $n_i$ resulting in a multiplet $(p_i,q_i)$, the dipole is
broken up with a probability $(1+m_i+n_i-p_i-q_i)/(1+m_i+n_i)$. We
note that it may happen that a dipole cannot break up, \eg\ if the
dipole ends are quarks rather than gluons. In this case the probability
is modified to increase the probability for the other overlapping
dipoles to break by replacing the denominator by $(1+m_i'+n_i')$ where
$m_i'$ and $n_i'$ are calculated as in \eqref{eq:dipole-nm}, but only
summing over breakable dipoles.

For the dipoles that are left, we can now start the hadronization. To
further increase the amount of fluctuations included, we do not
average over all breakups, but hadronize each string piece with a
local effective string tension
\begin{equation}
\label{eq:kappafirst}
\kapeff = \kapeff(p,q) - \kapeff(p-1,q) = \frac{1}{4}\left(2+2p+q\right).
\end{equation}
Note that while the expression in \eqref{eq:scaledcasimir} is
symmetric in $p$ and $q$, \eqref{eq:kappafirst} is not. Hadronizing
one string at a time implies a choice of whether \emph{this} string is
in the $p$ or $q$ direction, and here the implicit choice is taken
towards the $p$ direction. The choice is of course still arbitrary,
with no effect on the result.  The strings are then send to be
hadronized one at a time by \pytppp, utilizing a customization
described in \appref{sec:impl-details}. This customization enables
\pytppp to change hadronization parameters for each string breakup,
according to the calculated value of the effective string tension.

\subsection{Implementation details}
\label{sec:impl-details}
The implementation of the rope production, through estimation of $m$
and $n$ by either an approach based on enclosing cylinders or an
approach based on dipoles, followed by a random walk procedure is
extensively described in \sectref{sec:implementation}, and shall not
be repeated. The use of \pythia for hadronization, and consequently
changing hadronization parameters in a \pythia run, is however not
part of standard usage of the program, and will be described briefly
here. The basic idea is to use \pythia to hadronize one string at a
time, with the ability to change hadronization parameters while the
string is being hadronized, based on the local string tension at that
particular point on the string. Since \pythia normally sets
hadronization parameters once and for all, when the program is
initialized, a feature to intercept the hadronization loop was
introduced. In \pythia such interceptions are done with so-called
\texttt{UserHooks}, which in turn allows for re-initialization of
parameters. The \dipsy program then delivers a single string to
\pythia, which calls back for new hadronization parameters every time
a step is made. Owing to the interpretation of strings as dipoles
connected with gluons introduced in
\sectref{sec:dipole-based-treatment}, the callback needs to include
information about which dipole \pythia has reached, as well as the
position in the dipole. This is done by comparing the invariant mass
of all hadrons made from each string end so far, to the invariant mass
of the dipoles, starting from the same string end. This relies on
\pythia and \dipsy having identical, fixed string ends when the
hadronization procedure begins. This is not the case for gluon loops,
as \pythia will first cut one dipole at random, with probability
proportional to invariant mass squared. As an approximation, gluon
loops are hadronized with the average value of the string tension for
the whole string. Thus, parameters need to be set only once for gluon
loops.

\section{Tuning}
\label{sec:tuning}

To ensure that the observables of main interest for rope formation,
such as the rates of baryons and strange hadrons, are not affected by
global flavour independent effects on particle spectra, we made a
complete tuning of relevant parameters in \dipsy, \ariadne and
\pytppp, with and without inclusion of our new rope models. For this
we have used a selection of data analyses from the Rivet program
\cite{Buckley:2010ar}, and used the Professor framework
\cite{Buckley:2009bj} for the actual tuning.

\subsection{Tuning final-state shower and hadronization}

We followed the standard procedure of first tuning the parameters in
the final-state shower (in \ariadne) and in the hadronization (in
\pytppp) to \tee-observables as measured at LEP. It can be noted that
the old Fortran version of\ariadne has already been tuned to such data
with very good results (see \eg\ ref.~\cite{Abreu:1996na}), and since
the final-state shower in \ariadne is basically unchanged in the new
version, we obtain equally good results for the default version. When
we now add the final-state swing and rope fragmentation we do not
expect the results to change very much, as the number of dipoles
produced are fairly low, and do not allow for many reconnections. We
do expect some changes in parameters, however, since the swing tends
to decrease the total string lengths and therefore also the
multiplicity. Indeed, we find that \eg\ the tuned value for the $a$
parameter in the fragmentation is somewhat increased when the swing is
included.

In \figref{fig:lep-tuning} we show two distributions used in the
tuning. We find that the thrust distribution is equally well described
with and without swing. This is expected since it should be dominated
by effects from the hardest gluon emission, and by construction there
are no effects of the swing for the first two emissions. For the
transverse momentum distribution out of the event plane (defined by
the thrust major and minor axes) we do, however, find some
differences. Here we should be dominated by the two hardest gluon
emissions, and there we can expect larger effects from the swing in
subsequent emissions and in the hadronization. We see that the
description of data is somewhat improved, and in general the total
$\chi^2/$Ndf.\ is also somewhat improved when the swing is included.

Note that we do not expect any effects of the rope hadronization in
\tee$\to$hadrons, as we should be dominated by a single string. In
high multiplicity events there could be some internal overlaps but we
found\footnote{In \tee$\to$hadrons we use the thrust axis rather than
  the beam axis to define the rapidity span of strings for the
  pipe-based treatment.} no change in the tuned observables when
including the ropes.

\FIGURE[ht]{
  \includegraphics[width=0.45\textwidth]{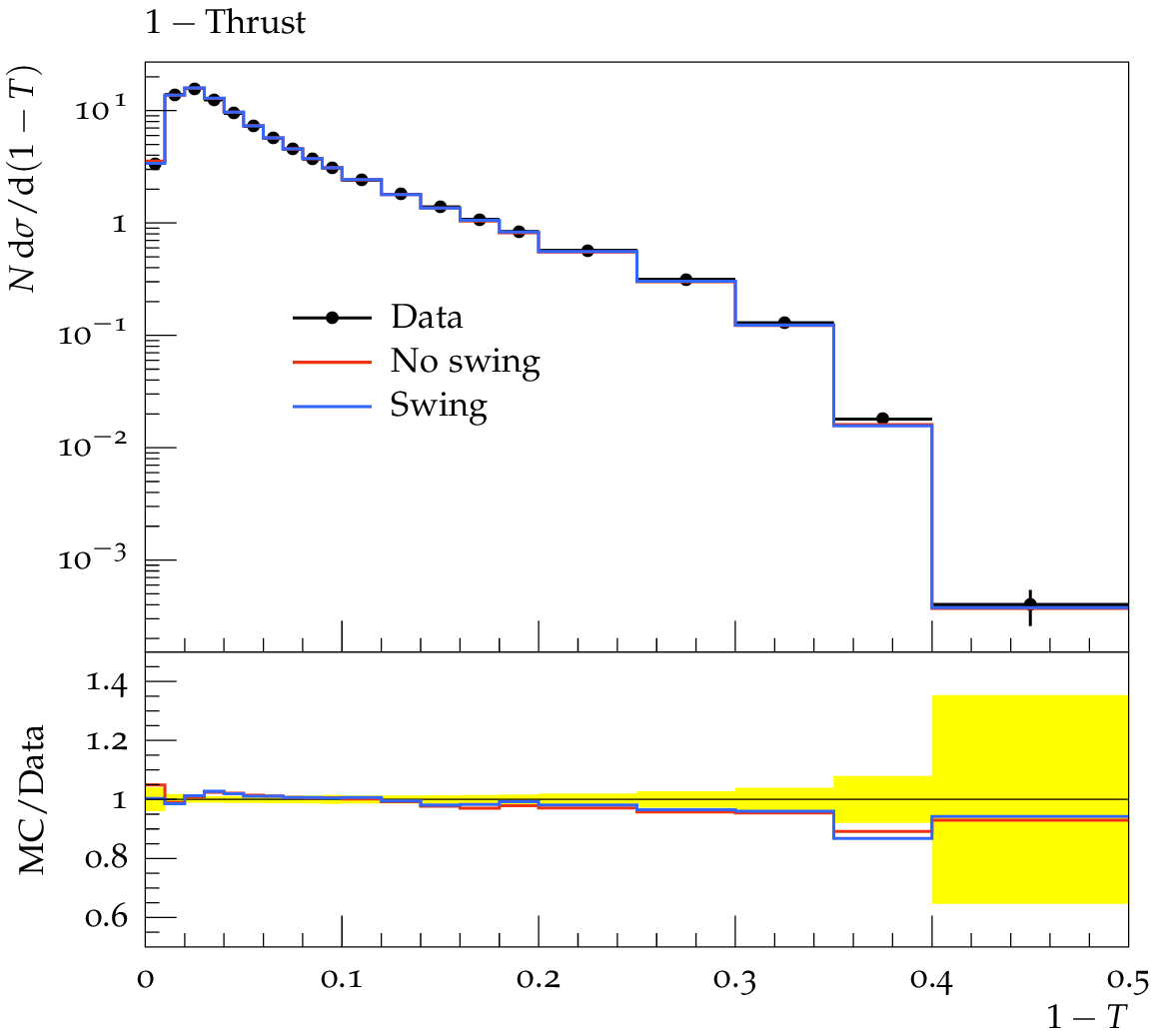}
  \includegraphics[width=0.45\textwidth]{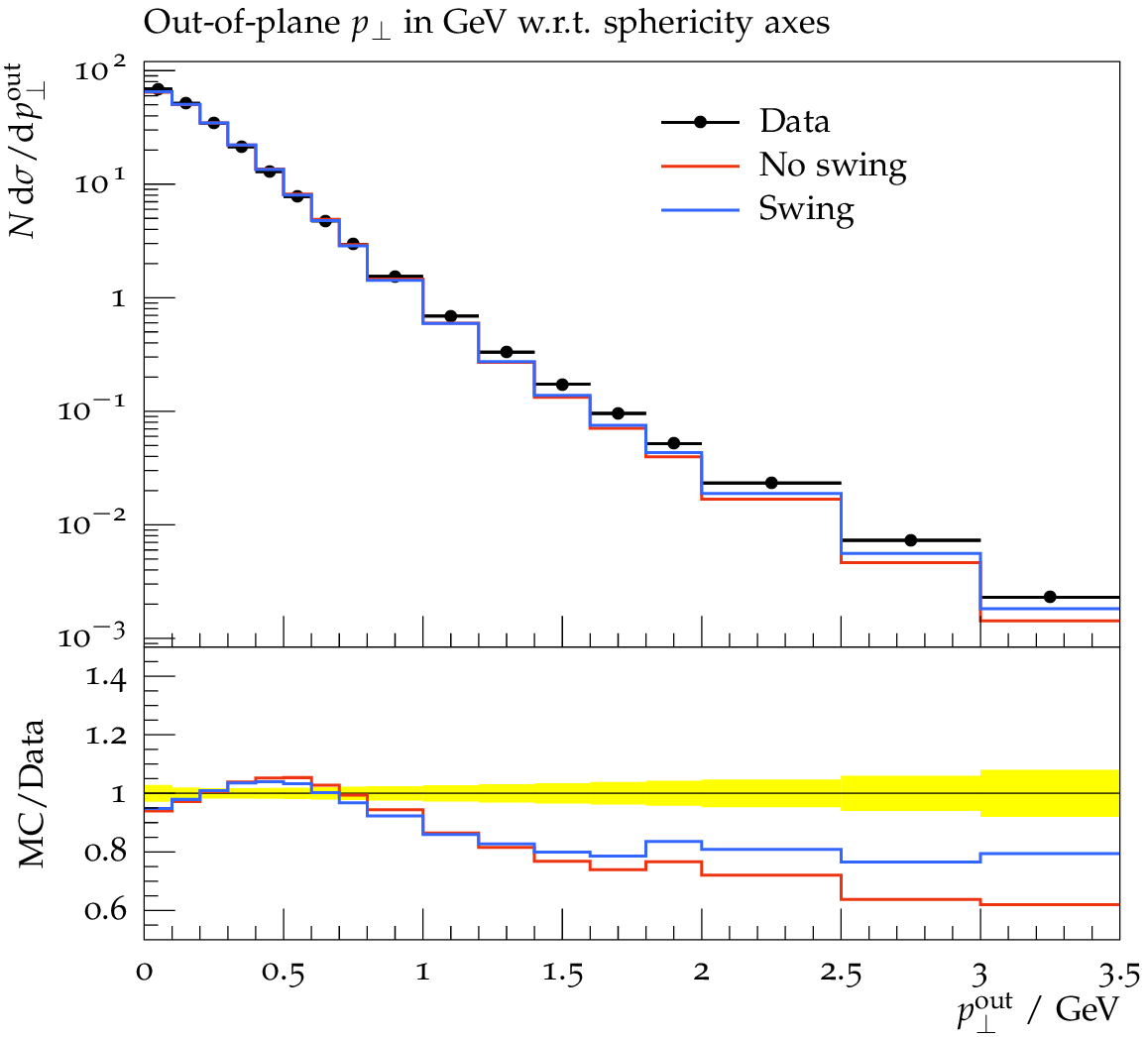}
  \caption[]{\label{fig:lep-tuning} Sample plots from tuning of the
    \ariadne dipole shower with string fragmentation from \pytppp to
    DELPHI data \cite{Abreu:1996na}, with and without final state
    swing. On the left is shown the distribution in thrust, and on the
    right the distribution of transverse momenta out of the event
    plane.}
}

\subsection{Tuning \dipsy}
\label{sec:dipsy-tune}

We then proceed to tune the parameters of the \dipsy model to $pp$
collisions. Here we tune both to the total and elastic cross sections
as well as to final-state observables in minimum bias events. It
should be emphasized that the \dipsy program is not yet ready for
precision description of final-state observables. Although the model
has improvements beyond the leading logarithmic BFKL accuracy, there
are no proper matrix elements included for hard scatterings and there
are no quarks included in the evolution. We therefore do not 
expect it to be able to give a good description of observables
involving high transverse momenta, and indeed we find that the
particle rates above $p_\perp\sim5$~GeV are severely
overestimated. In addition we have found that, although the energy
dependence of the total, elastic, and diffractive cross sections are
well reproduced, the energy dependence of the total multiplicity is a
bit too flat. We therefore decided to make separate tunes for
different collision energies as well as global tunes. As it turned out
that the observables in \sectref{sec:results}, were insensitive to
whether we used separate or global tunes, we there only present
results using the latter.


In \figrefs{fig:900GeV-tuning} and \fig{fig:7000GeV-tuning} we show
examples of observables used in the tuning of 900~GeV and 7~TeV
respectively. The rapidity distribution of charged particles is well
described for both energies and both for the default fragmentation and
for the rope model. While above a couple of GeV, the transverse
momentum distribution in all cases is too hard, the average transverse
momenta are well described as shown in \figref{fig:meanpT-tuning}. In
a future publication we intend to try to cure these deficiencies, but
here we are satisfied that we obtain very similar results with and
without ropes.

\FIGURE[ht]{
  \includegraphics[width=0.45\textwidth]{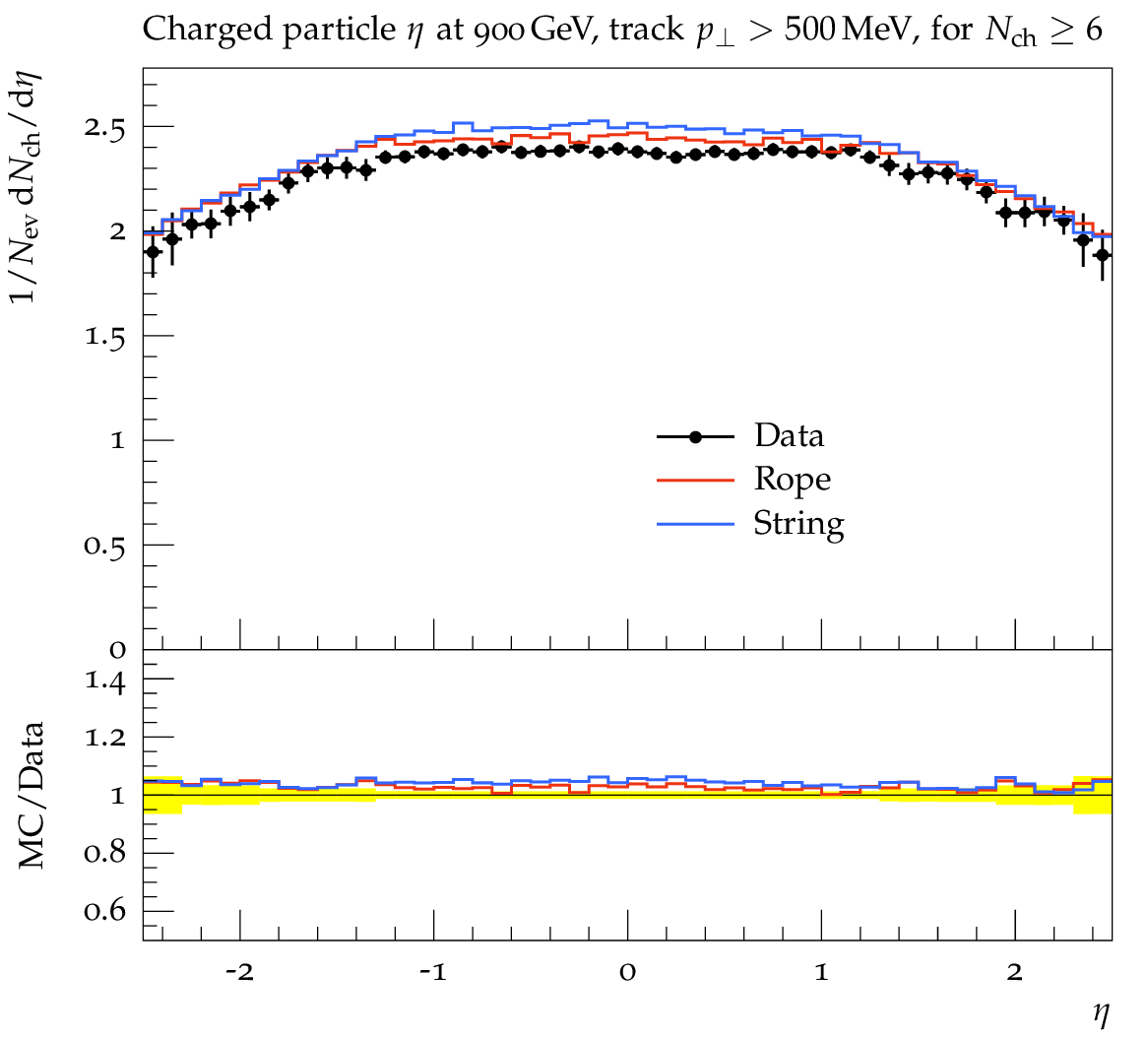}
  \includegraphics[width=0.45\textwidth]{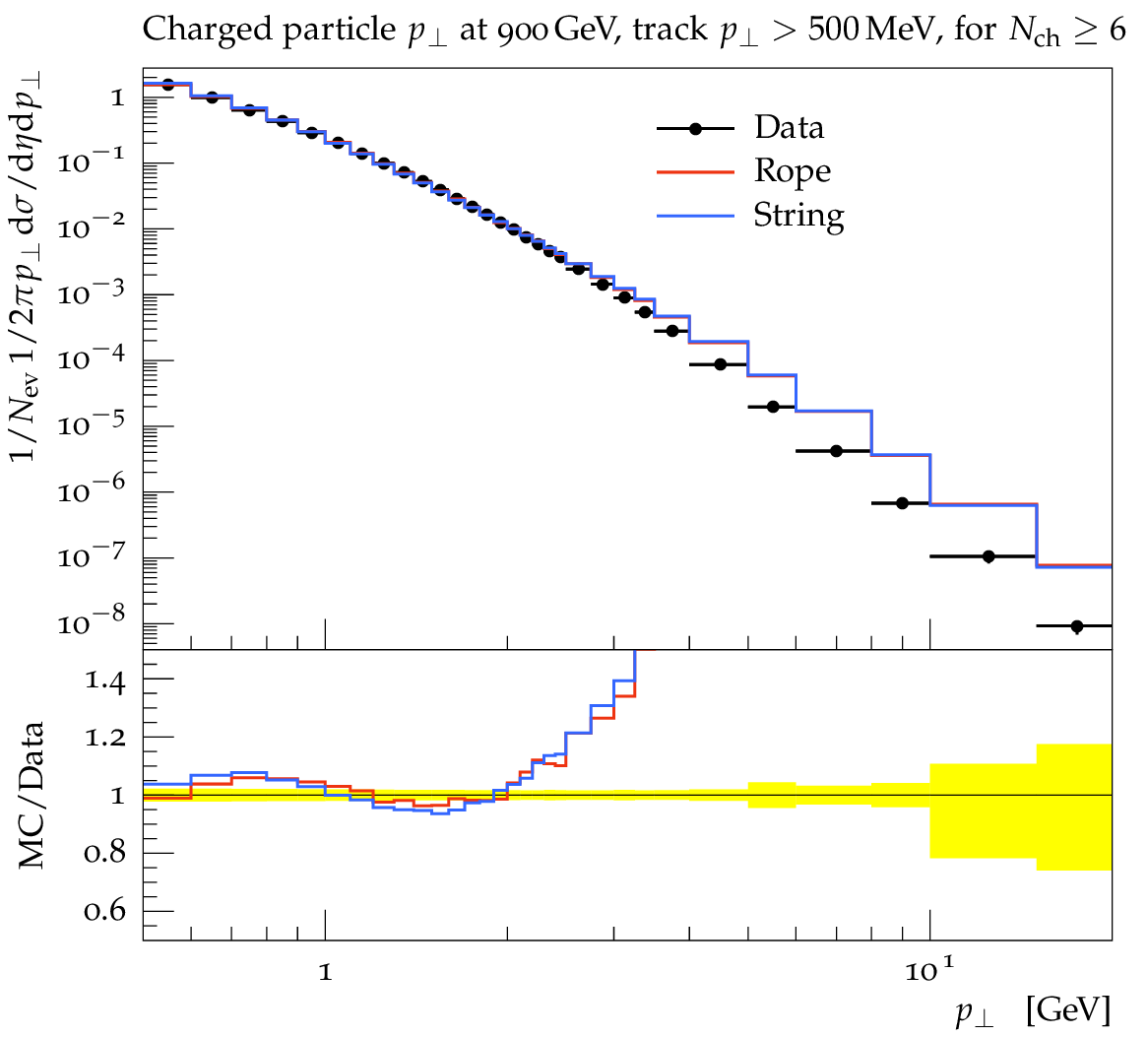}
  \caption{\label{fig:900GeV-tuning} Sample plots from the tuning of
    \dipsy to 900~GeV $pp$ minimum-bias data from ATLAS
    \cite{Aad:2010ac}. On the left is shown the pseudo-rapidity
    distribution of charged particles with transverse momenta larger
    than 500~MeV in events with at least six charged particles. On the
    right is given the transverse momentum distribution of charged
    particles for the same events. In both plots the lines labelled
    \emph{Rope} is \dipsy with the new (dipole-based) rope model, wile
    \emph{String} indicates \dipsy with the standard fragmentation. In
    both cases the final-state swing model is used.}
}
        
\FIGURE[ht]{

  \includegraphics[width=0.45\textwidth]{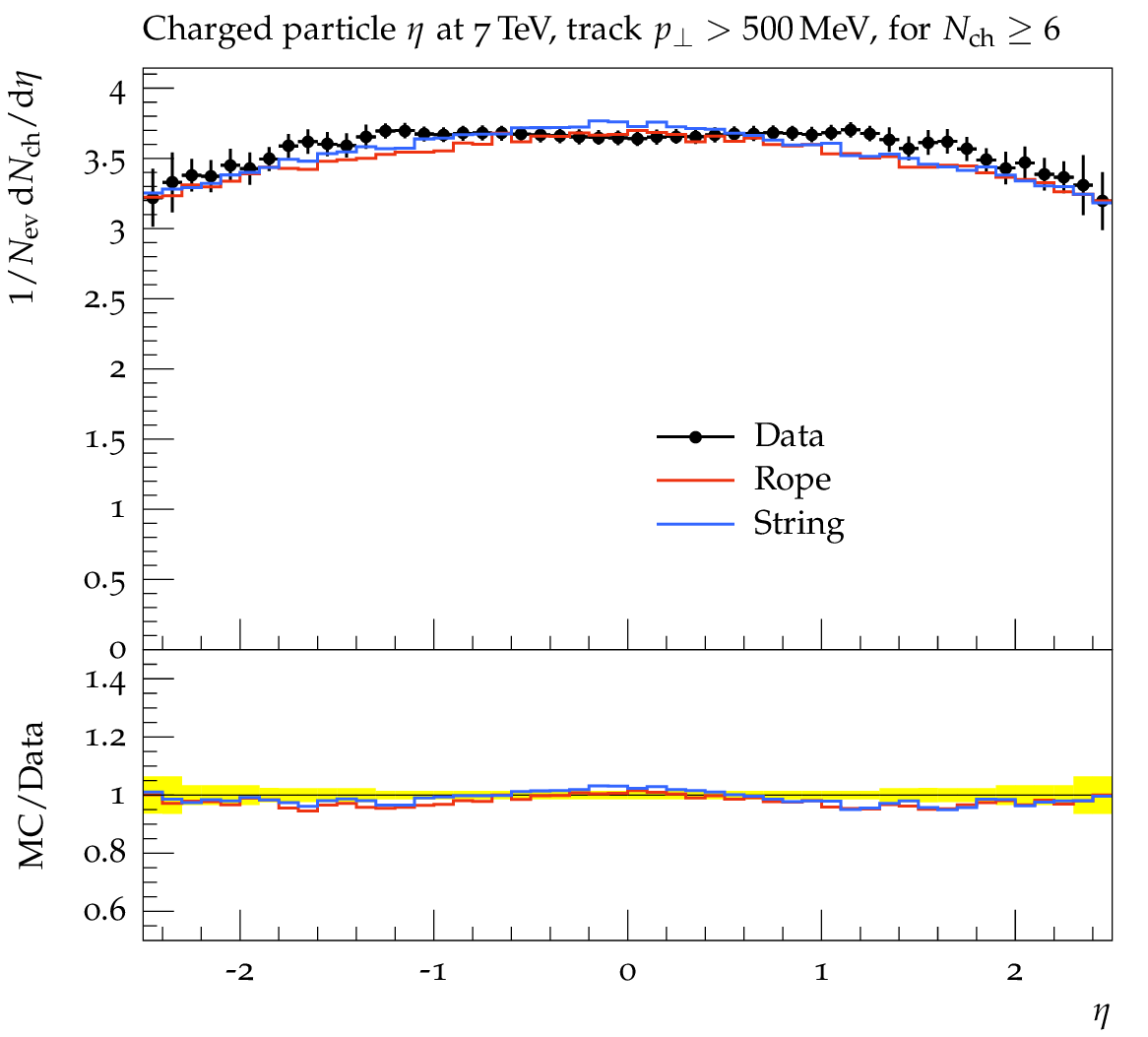}
  \includegraphics[width=0.45\textwidth]{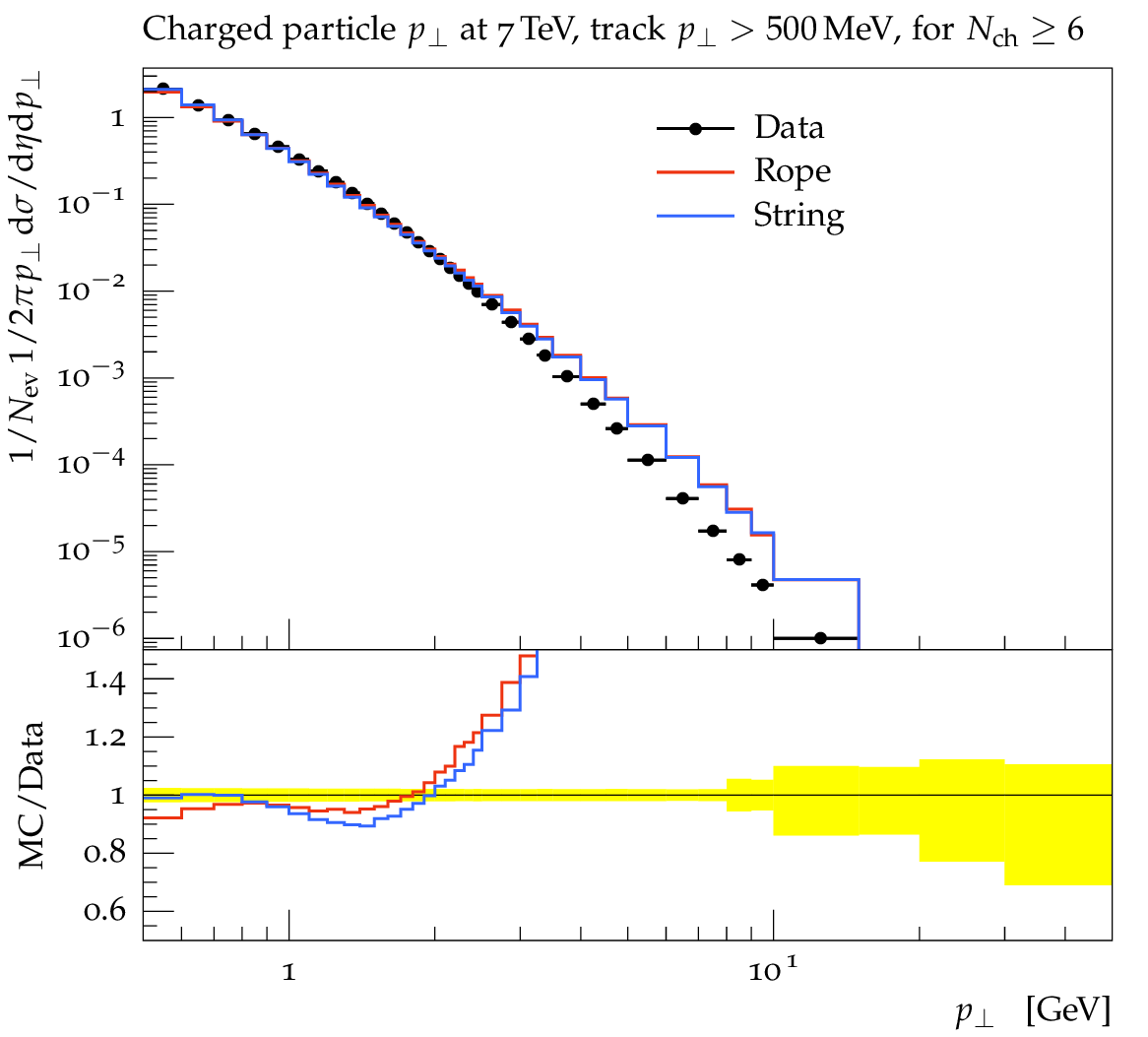}
  \caption{ \label{fig:7000GeV-tuning} Sample plots from the tuning of
    \dipsy to 7~TeV $pp$ minimum-bias data from ATLAS
    \cite{Aad:2010ac}, using the same distributions and models as in
    \figref{fig:900GeV-tuning}.}
}

\FIGURE[ht]{
  \includegraphics[width=0.45\textwidth]{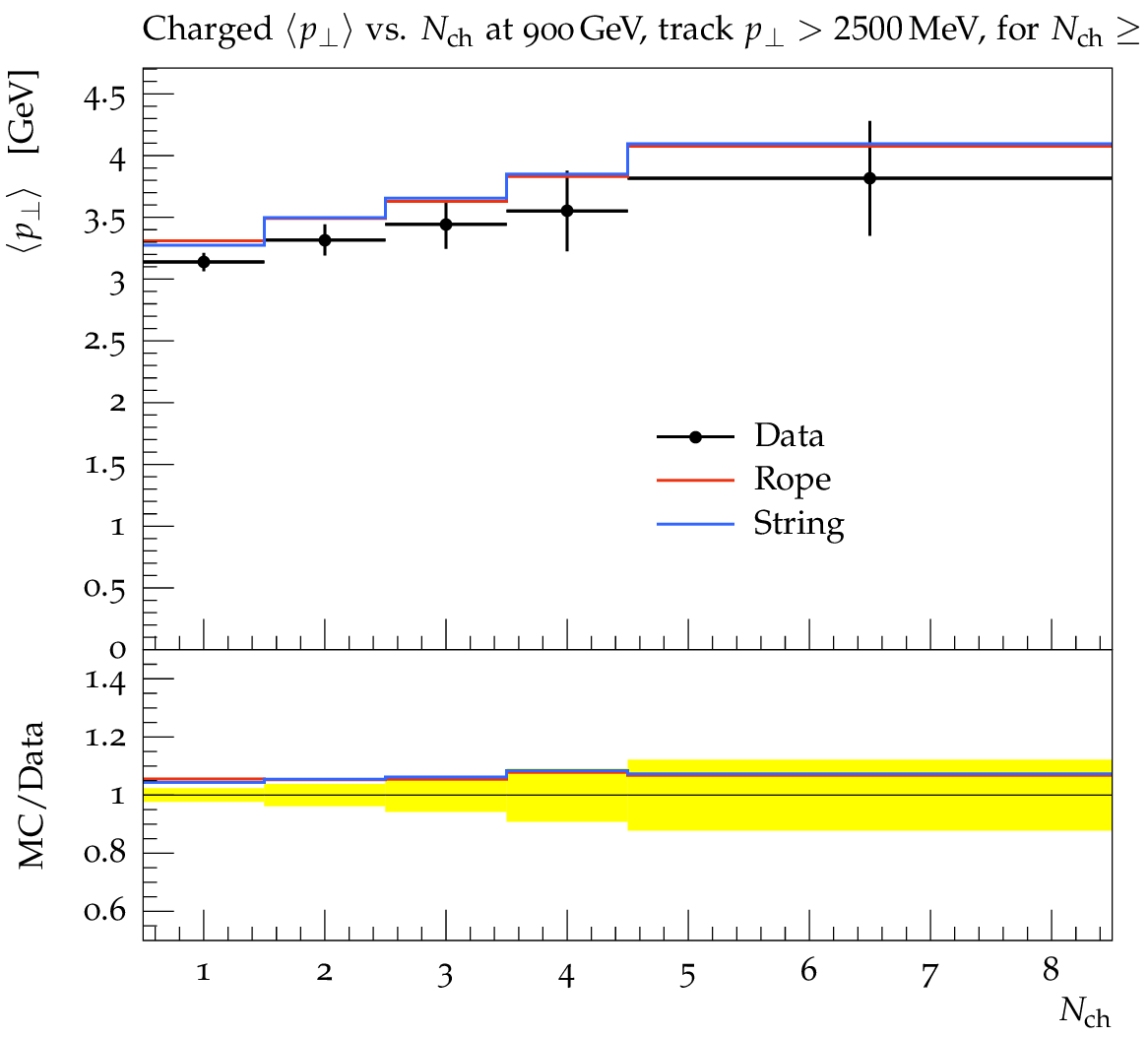}
  \includegraphics[width=0.45\textwidth]{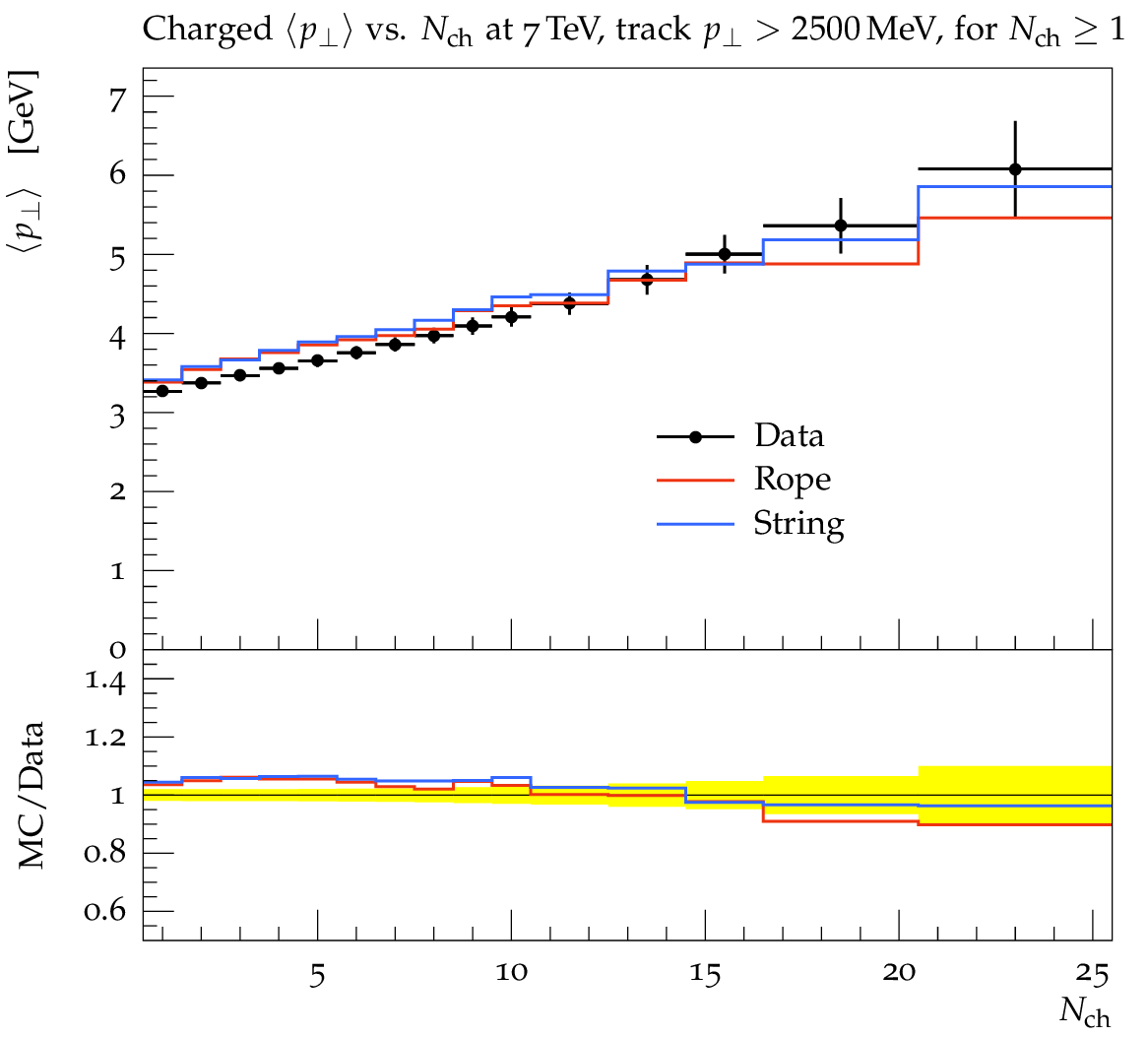}
  \caption{ \label{fig:meanpT-tuning} Sample plots from the tuning of
    \dipsy to 900~GeV (left) and 7~TeV (right) $pp$ minimum-bias data
    from ATLAS \cite{Aad:2010ac}. Both plots show the average
    transverse momenta (above 2.5~GeV) as a function of the number of
    charged particles per event. The models are the same as in
    \figref{fig:900GeV-tuning}.}
}

Since \dipsy in the current version does not describe diffractive
final states (see \cite{Flensburg:2011kk}), it is necessary to choose
observables for comparison which are not greatly affected by the
presence of diffractive events. This is important both for the tuning
and for the observables studied in \sectref{sec:data-cmp}. In \pytppp
it is possible to turn on and off the diffractive contributions, and
such runs can be used to determine which observables should be used
for the present analysis. 
As an example we show in \figref{fig:pytcomp} a comparison between results
from \pytppp and data from CMS for the rapidity 
distribution of $\Lambda$ (\figref{fig:pytcomp} (left)) and the ratio
$K^0_s/\Lambda$ in \figref{fig:pytcomp} (right). It is clear that the
inclusion of diffractive events plays a great 
role when looking at raw per-event distributions. In \pytppp the  diffractive
events are hadronizing in the same way as the non-diffractive, and the
$K^0_s/\Lambda$ ratio is therefore not modified by including diffractive
events. Even if this is not confirmed by experiments at present, we note that
the 
contribution from diffractive events is relatively small. A moderate
difference in particle ratios in diffractive events should therefore not
change predictions for particle ratios in a serious way, and such ratios
should therefore be better observables in comparisons between our model and
data.  

\FIGURE[ht]{
  \includegraphics[width=0.45\textwidth]{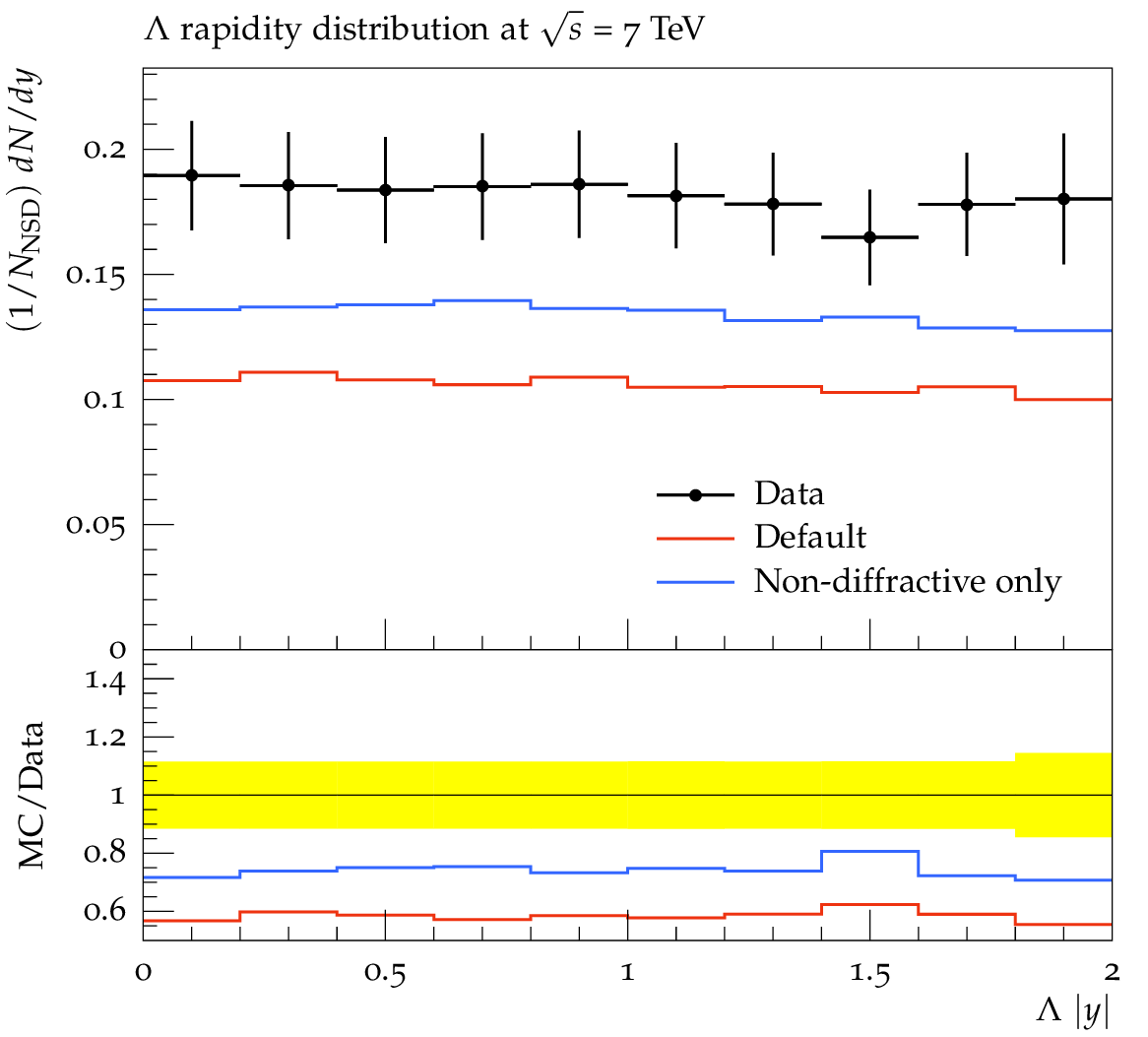}
  \includegraphics[width=0.45\textwidth]{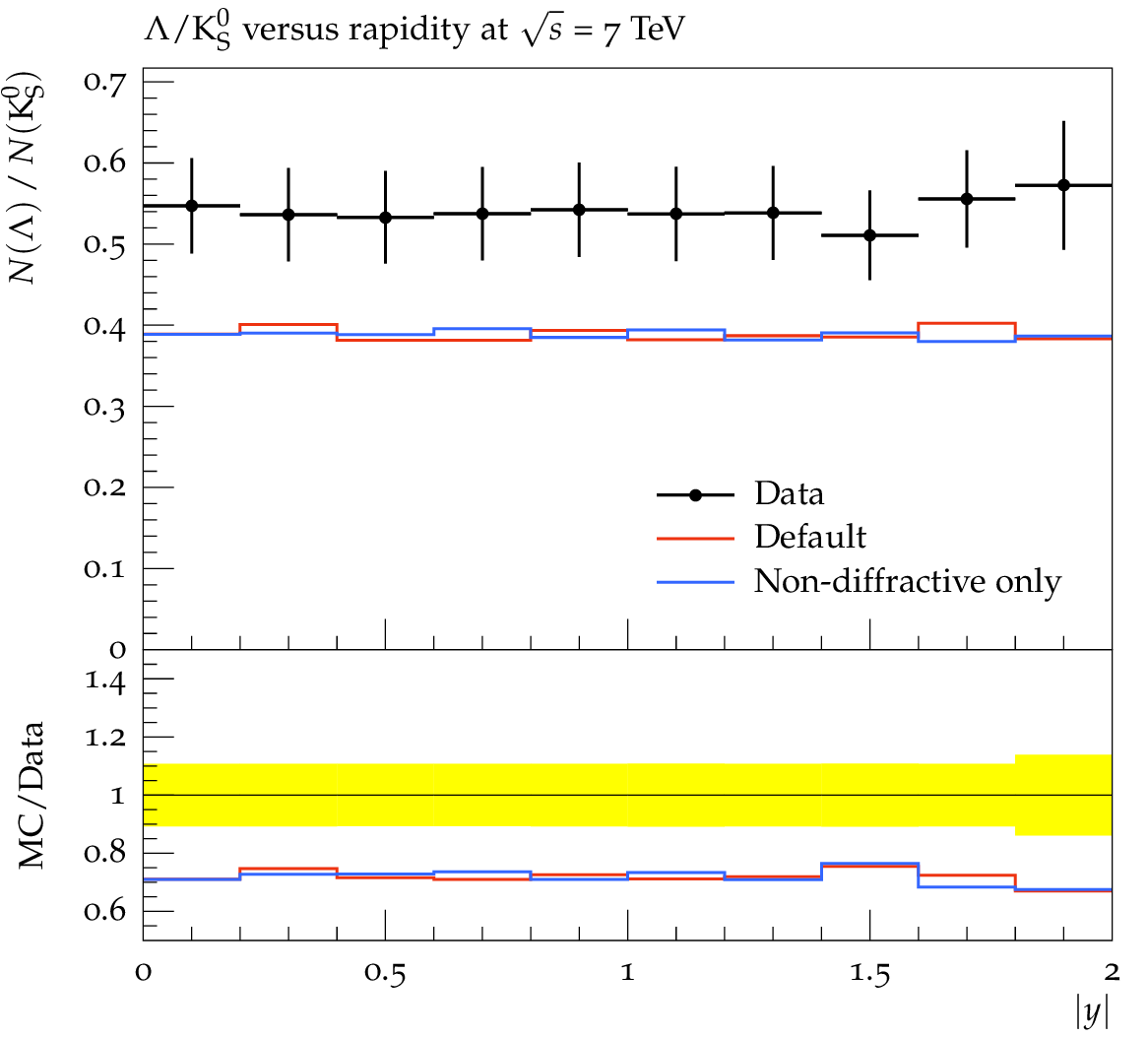}
  \caption{ \label{fig:pytcomp} Pythia non-diffractive only (blue) and
    including diffractive (red), compared to 7000~GeV data from
    CMS for the $\Lambda$ rapidity distribution (left) and the $\Lambda/K_s^0$
    ratio (right). }
}

Looking at \figref{fig:pytcomp} it is also directly visible that
\pytppp has difficulties reproducing these distributions, which is
also the case for almost all contemporary generators (see.\ \eg\
\texttt{\texttt{mcplots.cern.ch}})

\bibliographystyle{utcaps}  
\bibliography{references,refs}

\providecommand{\href}[2]{#2}\begingroup\raggedright\begin{thebibliography}{10}

\bibitem{Sjostrand:2006za}
T.~Sj{\"o}strand, S.~Mrenna, and P.~Skands, {\em JHEP} {\bf 05} (2006)  026,
\href{http://arxiv.org/abs/hep-ph/0603175}{{\tt arXiv:hep-ph/0603175}}.

\bibitem{Bahr:2008pv}
M.~B{\"a}hr {\em et al.},
  \href{http://dx.doi.org/10.1140/epjc/s10052-008-0798-9}{{\em Eur. Phys. J.}
  {\bf C58} (2008)  639--707},
\href{http://arxiv.org/abs/0803.0883}{{\tt arXiv:0803.0883 [hep-ph]}}.

\bibitem{Biro:1984}
T.~S. {Biro}, H.~B. {Nielsen}, and J.~{Knoll},
  \href{http://dx.doi.org/10.1016/0550-3213(84)90441-3}{{\em Nuclear Physics B}
  {\bf 245} (1984)  449--468}.

\bibitem{Bialas:1984ye}
A.~Bia{\l}as and W.~Czyz,
\href{http://dx.doi.org/10.1103/PhysRevD.31.198}{{\em Phys.Rev.} {\bf D31}
  (1985)  198}.

\bibitem{Kerman:1985tj}
A.~K. Kerman, T.~Matsui, and B.~Svetitsky,
\href{http://dx.doi.org/10.1103/PhysRevLett.56.219}{{\em Phys.Rev.Lett.} {\bf
  56} (1986)  219}.

\bibitem{Gyulassy:1986jq}
M.~Gyulassy and A.~Iwazaki,
\href{http://dx.doi.org/10.1016/0370-2693(85)90711-7}{{\em Phys.Lett.} {\bf
  B165} (1985)  157--161}.

\bibitem{Andersson:1991er}
B.~Andersson and P.~Henning,
\href{http://dx.doi.org/10.1016/0550-3213(91)90303-F}{{\em Nucl.Phys.} {\bf
  B355} (1991)  82--105}.

\bibitem{Braun:1991dg}
M.~Braun and C.~Pajares,
\href{http://dx.doi.org/10.1016/0550-3213(93)90467-4}{{\em Nucl.Phys.} {\bf
  B390} (1993)  542--558}.

\bibitem{Braun:1993xw}
M.~Braun and C.~Pajares,
\href{http://dx.doi.org/10.1103/PhysRevD.47.114}{{\em Phys.Rev.} {\bf D47}
  (1993)  114--122}.

\bibitem{Amelin:1994mc}
N.~Amelin, M.~Braun, and C.~Pajares,
\href{http://dx.doi.org/10.1007/BF01580331}{{\em Z.Phys.} {\bf C63} (1994)
  507--516}.

\bibitem{Armesto:1994yg}
N.~Armesto, M.~Braun, E.~Ferreiro, and C.~Pajares,
\href{http://dx.doi.org/10.1016/0370-2693(94)01511-A}{{\em Phys.Lett.} {\bf
  B344} (1995)  301--307}.

\bibitem{Kajantie:1985jh}
K.~Kajantie and T.~Matsui,
\href{http://dx.doi.org/10.1016/0370-2693(85)90343-0}{{\em Phys.Lett.} {\bf
  B164} (1985)  373}.

\bibitem{Gatoff:1987uf}
G.~Gatoff, A.~Kerman, and T.~Matsui,
\href{http://dx.doi.org/10.1103/PhysRevD.36.114}{{\em Phys.Rev.} {\bf D36}
  (1987)  114}.

\bibitem{Braun:1997ch}
M.~Braun, C.~Pajares, and J.~Ranft,
  \href{http://dx.doi.org/10.1142/S0217751X99001354}{{\em Int.J.Mod.Phys.} {\bf
  A14} (1999)  2689--2704},
\href{http://arxiv.org/abs/hep-ph/9707363}{{\tt arXiv:hep-ph/9707363
  [hep-ph]}}.

\bibitem{Avsar:2005iz}
E.~Avsar, G.~Gustafson, and L.~L{\"o}nnblad, {\em JHEP} {\bf 07} (2005)  062,
\href{http://arxiv.org/abs/hep-ph/0503181}{{\tt hep-ph/0503181}}.

\bibitem{Flensburg:2011kk}
C.~Flensburg, G.~Gustafson, and L.~L{\" o}nnblad,
  \href{http://dx.doi.org/10.1007/JHEP08(2011)103}{{\em JHEP} {\bf 1108} (2011)
   103},
\href{http://arxiv.org/abs/1103.4321}{{\tt arXiv:1103.4321 [hep-ph]}}.

\bibitem{Merino:1991nq}
C.~Merino, C.~Pajares, and J.~Ranft,
\href{http://dx.doi.org/10.1016/0370-2693(92)90558-L}{{\em Phys.Lett.} {\bf
  B276} (1992)  168--172}.

\bibitem{Mohring:1992wm}
H.~M{\"o}hring, J.~Ranft, C.~Merino, and C.~Pajares,
\href{http://dx.doi.org/10.1103/PhysRevD.47.4142}{{\em Phys.Rev.} {\bf D47}
  (1993)  4142--4145}.

\bibitem{Sorge:1992ej}
H.~Sorge, M.~Berenguer, H.~Stoecker, and W.~Greiner,
\href{http://dx.doi.org/10.1016/0370-2693(92)91353-B}{{\em Phys.Lett.} {\bf
  B289} (1992)  6--11}.

\bibitem{Bleicher:2000us}
M.~Bleicher, W.~Greiner, H.~Stoecker, and N.~Xu,
  \href{http://dx.doi.org/10.1103/PhysRevC.62.061901}{{\em Phys.Rev.} {\bf C62}
  (2000)  061901},
\href{http://arxiv.org/abs/hep-ph/0007215}{{\tt arXiv:hep-ph/0007215
  [hep-ph]}}.

\bibitem{Soff:2002bn}
S.~Soff, J.~Randrup, H.~Stoecker, and N.~Xu,
  \href{http://dx.doi.org/10.1016/S0370-2693(02)03019-8}{{\em Phys.Lett.} {\bf
  B551} (2003)  115--120},
\href{http://arxiv.org/abs/nucl-th/0209093}{{\tt arXiv:nucl-th/0209093
  [nucl-th]}}.

\bibitem{Andersson:1983jt}
B.~Andersson, G.~Gustafson, and B.~S{\"o}derberg,
{\em Z. Phys.} {\bf C20} (1983)  317.

\bibitem{Andersson:1979ij}
B.~Andersson and G.~Gustafson,
\href{http://dx.doi.org/10.1007/BF01577421}{{\em Z.Phys.} {\bf C3} (1980)
  223}.

\bibitem{Andersson:1983ia}
B.~Andersson, G.~Gustafson, G.~Ingelman, and T.~Sj{\"o}strand,
{\em Phys. Rept.} {\bf 97} (1983)  31.

\bibitem{Buckley:2011ms}
A.~Buckley, J.~Butterworth, S.~Gieseke, D.~Grellscheid, S.~{H\"oche}, {\em et
  al.}, \href{http://arxiv.org/abs/1101.2599}{{\tt arXiv:1101.2599 [hep-ph]}}.

\bibitem{Hamacher:1995df}
K.~Hamacher and M.~Weierstall,
\href{http://arxiv.org/abs/hep-ex/9511011}{{\tt hep-ex/9511011}}.

\bibitem{Sjostrand:2014zea}
T.~Sjöstrand, S.~Ask, J.~R. Christiansen, R.~Corke, N.~Desai, {\em et al.},
\href{http://arxiv.org/abs/1410.3012}{{\tt arXiv:1410.3012 [hep-ph]}}.

\bibitem{Casher:1978wy}
A.~Casher, H.~Neuberger, and S.~Nussinov,
\href{http://dx.doi.org/10.1103/PhysRevD.20.179}{{\em Phys.Rev.} {\bf D20}
  (1979)  179--188}.

\bibitem{Andersson:1980vj}
B.~Andersson, G.~Gustafson, and T.~Sj{\"o}strand,
\href{http://dx.doi.org/10.1007/BF01557774}{{\em Z.Phys.} {\bf C6} (1980)
  235}.

\bibitem{Gurvich:1979nq}
E.~Gurvich,
\href{http://dx.doi.org/10.1016/0370-2693(79)90560-4}{{\em Phys.Lett.} {\bf
  B87} (1979)  386--388}.

\bibitem{Glendenning:1983qq}
N.~Glendenning and T.~Matsui,
\href{http://dx.doi.org/10.1103/PhysRevD.28.2890}{{\em Phys.Rev.} {\bf D28}
  (1983)  2890--2891}.

\bibitem{Schwinger:1951nm}
J.~S. Schwinger,
\href{http://dx.doi.org/10.1103/PhysRev.82.664}{{\em Phys. Rev.} {\bf 82}
  (1951)  664--679}.

\bibitem{Brezin:1970xf}
E.~Brezin and C.~Itzykson,
\href{http://dx.doi.org/10.1103/PhysRevD.2.1191}{{\em Phys. Rev.} {\bf D2}
  (1970)  1191--1199}.

\bibitem{Alner:1985ra}
{\bf UA5 Collaboration} Collaboration, G.~Alner {\em et al.},
\href{http://dx.doi.org/10.1016/0550-3213(85)90624-8}{{\em Nucl.Phys.} {\bf
  B258} (1985)  505}.

\bibitem{Aaron:2008ck}
{\bf H1 Collaboration} Collaboration, F.~Aaron {\em et al.},
  \href{http://dx.doi.org/10.1140/epjc/s10052-009-0995-1}{{\em Eur.Phys.J.}
  {\bf C61} (2009)  185--205},
\href{http://arxiv.org/abs/0810.4036}{{\tt arXiv:0810.4036 [hep-ex]}}.

\bibitem{Khachatryan:2011tm}
{\bf CMS Collaboration} Collaboration, V.~Khachatryan {\em et al.},
  \href{http://dx.doi.org/10.1007/JHEP05(2011)064}{{\em JHEP} {\bf 1105} (2011)
   064},
\href{http://arxiv.org/abs/1102.4282}{{\tt arXiv:1102.4282 [hep-ex]}}.

\bibitem{Ambjorn:1984dp}
J.~Ambj{\o}rn, P.~Olesen, and C.~Peterson,
\href{http://dx.doi.org/10.1016/0550-3213(84)90242-6}{{\em Nucl.Phys.} {\bf
  B240} (1984)  533}.

\bibitem{Bali:2000un}
G.~S. Bali, \href{http://dx.doi.org/10.1103/PhysRevD.62.114503}{{\em Phys.Rev.}
  {\bf D62} (2000)  114503},
\href{http://arxiv.org/abs/hep-lat/0006022}{{\tt arXiv:hep-lat/0006022
  [hep-lat]}}.

\bibitem{Mueller:1993rr}
A.~H. Mueller,
{\em Nucl. Phys.} {\bf B415} (1994)  373--385.

\bibitem{Mueller:1994jq}
A.~H. Mueller and B.~Patel, {\em Nucl. Phys.} {\bf B425} (1994)  471--488,
\href{http://arxiv.org/abs/hep-ph/9403256}{{\tt hep-ph/9403256}}.

\bibitem{Mueller:1994gb}
A.~H. Mueller, {\em Nucl. Phys.} {\bf B437} (1995)  107--126,
\href{http://arxiv.org/abs/hep-ph/9408245}{{\tt hep-ph/9408245}}.

\bibitem{McLerran:1993ni}
L.~D. McLerran and R.~Venugopalan,
  \href{http://dx.doi.org/10.1103/PhysRevD.49.2233}{{\em Phys.Rev.} {\bf D49}
  (1994)  2233--2241},
\href{http://arxiv.org/abs/hep-ph/9309289}{{\tt arXiv:hep-ph/9309289
  [hep-ph]}}.

\bibitem{McLerran:1993ka}
L.~D. McLerran and R.~Venugopalan,
  \href{http://dx.doi.org/10.1103/PhysRevD.49.3352}{{\em Phys.Rev.} {\bf D49}
  (1994)  3352--3355},
\href{http://arxiv.org/abs/hep-ph/9311205}{{\tt arXiv:hep-ph/9311205
  [hep-ph]}}.

\bibitem{Flensburg:2011kj}
C.~Flensburg, G.~Gustafson, L.~L{\"o}nnblad, and A.~Ster,
  \href{http://dx.doi.org/10.1007/JHEP06(2011)066}{{\em JHEP} {\bf 06} (2011)
  066},
\href{http://arxiv.org/abs/1103.4320}{{\tt arXiv:1103.4320 [hep-ph]}}.

\bibitem{Flensburg:2010kq}
C.~Flensburg and G.~Gustafson,
  \href{http://dx.doi.org/10.1007/JHEP10(2010)014}{{\em JHEP} {\bf 1010} (2010)
   014}, \href{http://arxiv.org/abs/1004.5502}{{\tt arXiv:1004.5502 [hep-ph]}}.

\bibitem{Flensburg:2012zy}
C.~Flensburg, G.~Gustafson, and L.~L{\"o}nnblad,
  \href{http://dx.doi.org/10.1007/JHEP12(2012)115}{{\em JHEP} {\bf 1212} (2012)
   115},
\href{http://arxiv.org/abs/1210.2407}{{\tt arXiv:1210.2407 [hep-ph]}}.

\bibitem{Flensburg:2011wx}
C.~Flensburg,
\href{http://arxiv.org/abs/1108.4862}{{\tt arXiv:1108.4862 [nucl-th]}}.

\bibitem{Flensburg:2012zz}
C.~Flensburg,
\href{http://dx.doi.org/10.1143/PTPS.193.172}{{\em Prog.Theor.Phys.Suppl.} {\bf
  193} (2012)  172--175}.

\bibitem{Wilson:1974sk}
K.~G. Wilson,
\href{http://dx.doi.org/10.1103/PhysRevD.10.2445}{{\em Phys. Rev.} {\bf D10}
  (1974)  2445--2459}.

\bibitem{Andersson:1982if}
B.~Andersson and G.~Gustafson, ``{Why are Vector Mesons Suppressed in Jet
  Fragmentation?}.'' Lund preprint, LU-TP-82-5 (1982).

\bibitem{Andersson:1981ce}
B.~Andersson, G.~Gustafson, and T.~Sj{\"o}strand,
\href{http://dx.doi.org/10.1016/0550-3213(82)90153-5}{{\em Nucl.Phys.} {\bf
  B197} (1982)  45}.

\bibitem{Andersson:1984af}
B.~Andersson, G.~Gustafson, and T.~Sj{\"o}strand,
\href{http://dx.doi.org/10.1088/0031-8949/32/6/003}{{\em Phys.Scripta} {\bf 32}
  (1985)  574}.

\bibitem{Jeon:2004rk}
S.~Jeon and R.~Venugopalan,
  \href{http://dx.doi.org/10.1103/PhysRevD.70.105012}{{\em Phys.Rev.} {\bf D70}
  (2004)  105012},
\href{http://arxiv.org/abs/hep-ph/0406169}{{\tt arXiv:hep-ph/0406169
  [hep-ph]}}.

\bibitem{Johnson:1975zp}
K.~Johnson,
{\em Acta Phys.Polon.} {\bf B6} (1975)  865.

\bibitem{Johnson:1975sg}
K.~Johnson and C.~B. Thorn,
\href{http://dx.doi.org/10.1103/PhysRevD.13.1934}{{\em Phys.Rev.} {\bf D13}
  (1976)  1934}.

\bibitem{Semay:2004br}
C.~Semay, \href{http://dx.doi.org/10.1140/epja/i2004-10097-5}{{\em Eur.Phys.J.}
  {\bf A22} (2004)  353--354},
\href{http://arxiv.org/abs/hep-ph/0409105}{{\tt arXiv:hep-ph/0409105
  [hep-ph]}}.

\bibitem{Cardoso:2009kz}
M.~Cardoso, N.~Cardoso, and P.~Bicudo,
  \href{http://dx.doi.org/10.1103/PhysRevD.81.034504}{{\em Phys.Rev.} {\bf D81}
  (2010)  034504},
\href{http://arxiv.org/abs/0912.3181}{{\tt arXiv:0912.3181 [hep-lat]}}.

\bibitem{Cea:2014uja}
P.~Cea, L.~Cosmai, F.~Cuteri, and A.~Papa,
  \href{http://dx.doi.org/10.1103/PhysRevD.89.094505}{{\em Phys.Rev.} {\bf D89}
  (2014)  094505},
\href{http://arxiv.org/abs/1404.1172}{{\tt arXiv:1404.1172 [hep-lat]}}.

\bibitem{Avsar:2006jy}
E.~Avsar, G.~Gustafson, and L.~L{\"o}nnblad, {\em JHEP} {\bf 01} (2007)  012,
\href{http://arxiv.org/abs/hep-ph/0610157}{{\tt hep-ph/0610157}}.

\bibitem{Avsar:2007xg}
E.~Avsar, G.~Gustafson, and L.~L{\"o}nnblad, {\em JHEP} {\bf 12} (2007)  012,
\href{http://arxiv.org/abs/arXiv:0709.1368 [hep-ph]}{{\tt arXiv:0709.1368
  [hep-ph]}}.

\bibitem{Kuraev:1977fs}
E.~A. Kuraev, L.~N. Lipatov, and V.~S. Fadin,
{\em Sov. Phys. JETP} {\bf 45} (1977)  199--204.

\bibitem{Balitsky:1978ic}
I.~I. Balitsky and L.~N. Lipatov,
{\em Sov. J. Nucl. Phys.} {\bf 28} (1978)  822--829.

\bibitem{Gustafson:1986db}
G.~Gustafson,
{\em Phys.~Lett.} {\bf B175} (1986)  453.

\bibitem{Gustafson:1987rq}
G.~Gustafson and U.~Pettersson,
{\em Nucl. Phys.} {\bf B306} (1988)  746.

\bibitem{Lonnblad:1992tz}
L.~L{\"o}nnblad,
{\em Comput.~Phys.~Commun.} {\bf 71} (1992)  15--31.

\bibitem{Sjostrand:2007gs}
T.~Sj{\"o}strand, S.~Mrenna, and P.~Skands,
  \href{http://dx.doi.org/10.1016/j.cpc.2008.01.036}{{\em Comput. Phys.
  Commun.} {\bf 178} (2008)  852--867},
\href{http://arxiv.org/abs/0710.3820}{{\tt arXiv:0710.3820 [hep-ph]}}.

\bibitem{Lonnblad:1995yk}
L.~L{\"o}nnblad,
\href{http://dx.doi.org/10.1007/s002880050087}{{\em Z.Phys.} {\bf C70} (1996)
  107--114}.

\bibitem{Sjostrand:1987su}
T.~Sj{\"o}strand and M.~van Zijl,
{\em Phys. Rev.} {\bf D36} (1987)  2019.

\bibitem{Gustafson:1988fs}
G.~Gustafson, U.~Pettersson, and P.~Zerwas,
\href{http://dx.doi.org/10.1016/0370-2693(88)91836-9}{{\em Phys.Lett.} {\bf
  B209} (1988)  90}.

\bibitem{Sjostrand:1993rb}
T.~Sj{\"o}strand and V.~A. Khoze,
  \href{http://dx.doi.org/10.1103/PhysRevLett.72.28}{{\em Phys.Rev.Lett.} {\bf
  72} (1994)  28--31},
\href{http://arxiv.org/abs/hep-ph/9310276}{{\tt arXiv:hep-ph/9310276
  [hep-ph]}}.

\bibitem{Edin:1995gi}
A.~Edin, G.~Ingelman, and J.~Rathsman,
  \href{http://dx.doi.org/10.1016/0370-2693(95)01391-1}{{\em Phys.Lett.} {\bf
  B366} (1996)  371--378},
\href{http://arxiv.org/abs/hep-ph/9508386}{{\tt arXiv:hep-ph/9508386
  [hep-ph]}}.

\bibitem{Enberg:2001vq}
R.~Enberg, G.~Ingelman, and N.~Timneanu,
  \href{http://dx.doi.org/10.1103/PhysRevD.64.114015}{{\em Phys.Rev.} {\bf D64}
  (2001)  114015},
\href{http://arxiv.org/abs/hep-ph/0106246}{{\tt arXiv:hep-ph/0106246
  [hep-ph]}}.

\bibitem{Andersson:1985qr}
B.~Andersson, G.~Gustafson, and B.~S{\"o}derberg,
\href{http://dx.doi.org/10.1016/0550-3213(86)90471-2}{{\em Nucl.Phys.} {\bf
  B264} (1986)  29}.

\bibitem{Andersson:1988ee}
B.~Andersson, P.~Dahlkvist, and G.~Gustafson,
\href{http://dx.doi.org/10.1016/0370-2693(88)90128-1}{{\em Phys.Lett.} {\bf
  B214} (1988)  604--608}.

\bibitem{Lonnblad:2006pt}
L.~{L\"onnblad}, \href{http://dx.doi.org/10.1016/j.nima.2005.11.143}{{\em
  Nucl.Instrum.Meth.} {\bf A559} (2006)  246--248}.

\bibitem{Karneyeu:2013aha}
A.~Karneyeu, L.~Mijovic, S.~Prestel, and P.~Skands,
  \href{http://dx.doi.org/10.1140/epjc/s10052-014-2714-9}{{\em Eur.Phys.J.}
  {\bf C74} (2014)  2714},
\href{http://arxiv.org/abs/1306.3436}{{\tt arXiv:1306.3436 [hep-ph]}}.

\bibitem{Adams:2006nd}
{\bf STAR Collaboration} Collaboration, J.~Adams {\em et al.},
  \href{http://dx.doi.org/10.1016/j.physletb.2006.04.032}{{\em Phys.Lett.} {\bf
  B637} (2006)  161--169},
\href{http://arxiv.org/abs/nucl-ex/0601033}{{\tt arXiv:nucl-ex/0601033
  [nucl-ex]}}.

\bibitem{Werner:2014xoa}
K.~Werner, B.~Guiot, I.~Karpenko, and T.~Pierog,
\href{http://arxiv.org/abs/1411.1048}{{\tt arXiv:1411.1048 [nucl-th]}}.

\bibitem{Abreu:2007kv}
N.~Armesto, N.~Borghini, S.~Jeon, U.~Wiedemann, S.~Abreu, {\em et al.},
  \href{http://dx.doi.org/10.1088/0954-3899/35/5/054001}{{\em J.Phys.} {\bf
  G35} (2008)  054001},
\href{http://arxiv.org/abs/0711.0974}{{\tt arXiv:0711.0974 [hep-ph]}}.

\bibitem{Ortiz:2013yxa}
A.~Ortiz~Velasquez, P.~Christiansen, E.~Cuautle~Flores, I.~Maldonado~Cervantes,
  and G.~Paić, \href{http://dx.doi.org/10.1103/PhysRevLett.111.042001}{{\em
  Phys.Rev.Lett.} {\bf 111} (2013)  042001},
\href{http://arxiv.org/abs/1303.6326}{{\tt arXiv:1303.6326 [hep-ph]}}.

\bibitem{Abelev:2014laa}
{\bf ALICE Collaboration} Collaboration, B.~B. Abelev {\em et al.},
  \href{http://dx.doi.org/10.1016/j.physletb.2014.07.011}{{\em Phys.Lett.} {\bf
  B736} (2014)  196--207},
\href{http://arxiv.org/abs/1401.1250}{{\tt arXiv:1401.1250 [nucl-ex]}}.

\bibitem{Engel:1995yda}
R.~Engel and J.~Ranft, \href{http://dx.doi.org/10.1103/PhysRevD.54.4244}{{\em
  Phys.Rev.} {\bf D54} (1996)  4244--4262},
\href{http://arxiv.org/abs/hep-ph/9509373}{{\tt arXiv:hep-ph/9509373
  [hep-ph]}}.

\bibitem{Butterworth:1996zw}
J.~M. Butterworth, J.~R. Forshaw, and M.~H. Seymour, {\em Z. Phys.} {\bf C72}
  (1996)  637--646,
\href{http://arXiv.org/abs/hep-ph/9601371}{{\tt hep-ph/9601371}}.

\bibitem{Fadin:1998py}
V.~S. Fadin and L.~N. Lipatov, {\em Phys. Lett.} {\bf B429} (1998)  127--134,
\href{http://arXiv.org/abs/hep-ph/9802290}{{\tt hep-ph/9802290}}.

\bibitem{Ciafaloni:1998gs}
M.~Ciafaloni and G.~Camici, {\em Phys. Lett.} {\bf B430} (1998)  349--354,
\href{http://arXiv.org/abs/hep-ph/9803389}{{\tt hep-ph/9803389}}.

\bibitem{Salam:1999cn}
G.~P. Salam, {\em Acta Phys. Polon.} {\bf B30} (1999)  3679--3705,
\href{http://arxiv.org/abs/hep-ph/9910492}{{\tt hep-ph/9910492}}.

\bibitem{Kwiecinski:1996td}
J.~Kwiecinski, A.~D. Martin, and P.~J. Sutton, {\em Z. Phys.} {\bf C71} (1996)
  585--594,
\href{http://arXiv.org/abs/hep-ph/9602320}{{\tt hep-ph/9602320}}.

\bibitem{Andersson:1996ju}
B.~Andersson, G.~Gustafson, and J.~Samuelsson,
{\em Nucl. Phys.} {\bf B467} (1996)  443--478.

\bibitem{Balitsky:1995ub}
I.~Balitsky, {\em Nucl. Phys.} {\bf B463} (1996)  99--160,
\href{http://arxiv.org/abs/hep-ph/9509348}{{\tt hep-ph/9509348}}.

\bibitem{Kovchegov:1999yj}
Y.~V. Kovchegov, {\em Phys. Rev.} {\bf D60} (1999)  034008,
\href{http://arxiv.org/abs/hep-ph/9901281}{{\tt hep-ph/9901281}}.

\bibitem{Buckley:2010ar}
A.~Buckley {\em et al.},
\href{http://arxiv.org/abs/1003.0694}{{\tt arXiv:1003.0694 [hep-ph]}}.

\bibitem{Buckley:2009bj}
A.~Buckley, H.~Hoeth, H.~Lacker, H.~Schulz, and J.~E. von Seggern,
  \href{http://dx.doi.org/10.1140/epjc/s10052-009-1196-7}{{\em Eur.Phys.J.}
  {\bf C65} (2010)  331--357},
\href{http://arxiv.org/abs/0907.2973}{{\tt arXiv:0907.2973 [hep-ph]}}.

\bibitem{Abreu:1996na}
{\bf DELPHI} Collaboration, P.~Abreu {\em et al.},
{\em Z. Phys.} {\bf C73} (1996)  11--60.

\bibitem{Aad:2010ac}
{\bf ATLAS Collaboration} Collaboration, G.~Aad {\em et al.},
  \href{http://dx.doi.org/10.1088/1367-2630/13/5/053033}{{\em New J.Phys.} {\bf
  13} (2011)  053033},
\href{http://arxiv.org/abs/1012.5104}{{\tt arXiv:1012.5104 [hep-ex]}}.

\end{thebibliography}\endgroup

\end{document}